%% file: pear-depletion_pmdg_2020.tex
\documentclass[graphicx,twocolumn,jcp]{revtex4}
\usepackage{amssymb,mathtools}

\usepackage[varg]{txfonts}
\usepackage{color}
\usepackage{setspace}
\usepackage{relsize}

\usepackage{tikz}
\usepackage{filecontents,pgfplots}
\usetikzlibrary{calc,fadings,decorations.pathreplacing,hobby}
\usetikzlibrary{arrows,positioning}
\usetikzlibrary{patterns}
\usetikzlibrary{fit}
\usepackage{MnSymbol}

\pgfplotsset{select coords between index/.style 2 args={
    x filter/.code={
        \ifnum\coordindex<#1\fi
        \ifnum\coordindex>#2\fi
    }
}}

\usepgfplotslibrary{fillbetween}
\usetikzlibrary{shapes}
\tikzset{>=latex}

\newcommand{\fig}[1]{FIG.\,\ref{#1}}

\newcommand{\sect}[1]{Sec.\,\ref{#1}}
\newcommand{\app}[1]{App.\,\ref{#1}}

\newcommand\orangeline[1][]{%
   \,\tikz[baseline]\draw[#1,very thick,orange](0,0.35*\ht\strutbox)--(1.5*\ht\strutbox,0.35*\ht\strutbox);}

\begin{document}

\title{Self-assembly and entropic effects in pear-shaped colloid systems:\\ II. Depletion attraction of pear-shaped particles in a hard sphere solvent} 

\author{Philipp W. A. Sch\"onh\"ofer}
\email[]{Philipp.Schoenhoefer@fau.de}
\affiliation{College of Science, Health, Engineering and Education, Mathematics and Statistics, Murdoch University, 90 South Street, 6150 Murdoch, WA, Australia}
\affiliation{Institut f\"ur Theoretische Physik I, Friedrich-Alexander-Universit\"at Erlangen-N\"urnberg, Staudtstra\ss{}e 7, 91058 Erlangen, Germany}
\author{Matthieu Marechal}
\affiliation{Institut f\"ur Theoretische Physik I, Friedrich-Alexander-Universit\"at Erlangen-N\"urnberg, Staudtstra\ss{}e 7, 91058 Erlangen, Germany}
\author{Douglas J. Cleaver}
\affiliation{Materials and Engineering Research Institute, Sheffield Hallam University, Sheffield S1 1WB, UK}
\author{Gerd E. Schr\"oder-Turk}
\email[]{G.Schroeder-Turk@murdoch.edu.au}
\affiliation{College of Science, Health, Engineering and Education, Mathematics and Statistics, Murdoch University, 90 South Street, 6150 Murdoch, WA, Australia}
\affiliation{Department of Applied Mathematics, Research School of Physical Sciences and Engineering, The Australian National University, 0200 Canberra, ACT, Australia}
\affiliation{Department of Food Science, University of Copenhagen, Rolighedsvej 26, 1958 Frederiksberg C, Denmark}
\affiliation{Physical Chemistry, Center for Chemistry and Chemical Engineering, Lund University, Lund 22100, Sweden}
\date{\today}

\begin{abstract}
We consider depletion effects of a pear-shaped colloidal particle in a hard-sphere solvent, for two different model realisations of the pear-shaped colloidal particle. The two models are the pear hard Gaussian overlap (PHGO) particles and the hard pears of revolution (HPR). The motivation for this study is to provide a microscopic understanding for the substantially different mesoscopic self-assembly properties of these pear-shaped colloids, in dense suspensions, that have been reported in the first part of this series. This is done by determining the differing depletion attraction via MC simulations of PHGO and HPR particles in a pool of a hard spherical solvent and comparing them with excluded volume calculations of numerically obtained ideal configurations on the microscopic level. While the HPR model behaves as predicted by the analysis of excluded volumes, the PHGO model showcases a preference for splay between neighbouring particles, which can be attributed to the special non-additive characteristics of the PHGO contact function. Lastly, we propose a potentially experimentally realisable pear-shaped particle model, the non-additive hard pear of revolution (NAHPR) model, which is based on the HPR model but also features non-additive traits similar to those of PHGO particles to mimic their depletion behaviour.
\end{abstract}

\maketitle 

This article's concern is the equilibrium self-assembly process, where by self-organisation relatively simple hard-core particles spontaneously adopt complex three-dimensionally ordered mesoscopic structures. On the one hand, particle shape is the sole parameter that tunes structure formation in this process and many simple shape characteristics (such as particle elongation) have been identified as determinants of structure formation \cite{IMN1976,HBLLNAL1996,KLW2006,PRB-LDR2006,AWGHE-S1996,THY2008,CKM2007,GS2007}. On the other hand, the self-assembly often depends in a drastic, non-linear way on details of the particle shape. Even though some shape features of particles can be related to specific global order, like the aspect ratio to nematic-like orientational order \cite{VF1990,FMMcT1984} or close-packed structures to a high isoperimetric quotient \cite{DEG2012} these correlations are often a rule of thumb and specific multi-particle behaviours can hardly be targeted in this straight-forward fashion. Small changes to the shape can have major repercussions for the structure formation.\\

In recent years, various reversed engineered approaches successfully circumvented this issue and opened the door to design self-assembled materials more precisely. Specifically in purely entropic systems, where the potentials are reduced to hard-core interactions and, therefore, are determined by the shape of the colloids, an iterative technique called \textit{digital alchemy} made it possible to create specific polyhedral building blocks for the formation of targeted structures \cite{vAKKDG2015}. Despite this remarkable achievement, those kind of strategies can still not pinpoint concrete relations between microscopic particle features and mesoscopic order.\\

Hence, the question which particle properties are necessary and which are sufficient for specific structure formation remains unanswered. To highlight the complexity of this question, this second paper in our study of pear-shaped colloids addresses this question through a depletion study of pear-shaped particles.\\

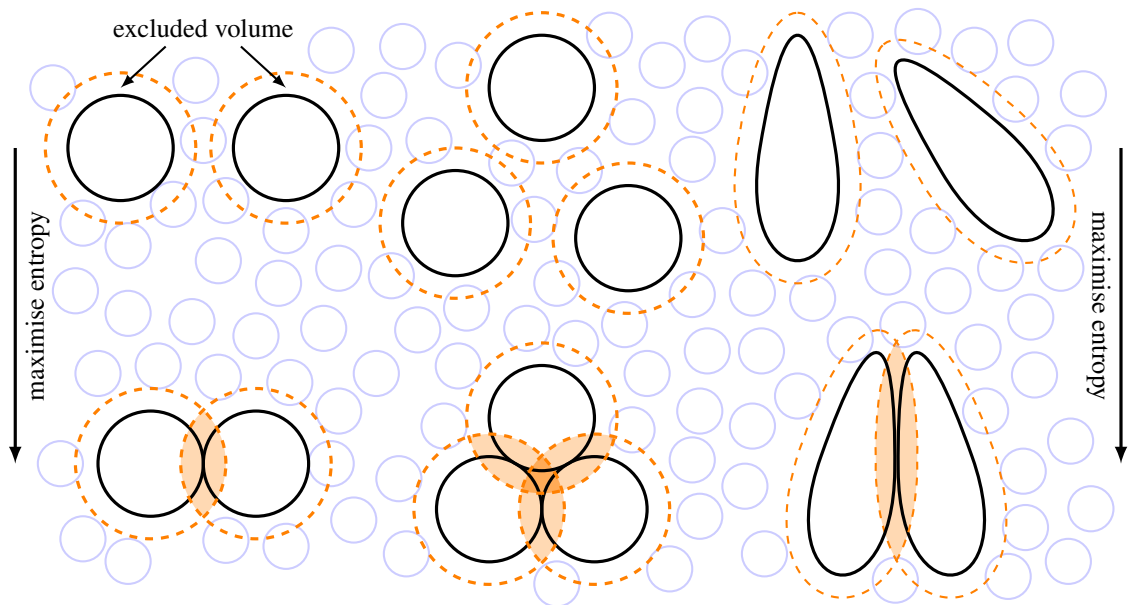
\begin{figure*}[t!]
\centering
\input{images/depletion_sketch.tex}
\caption{The concept of depletion is sketched by the example of two hard-core spherical colloids (left), three hard-core spherical colloids (centre) and two hard-core pear-shaped colloids (right) dissolved in a liquid of smaller hard spheres (indicated in light blue). The system is driven mainly by the entropy of the solvent particles and maximises the free energy by minimising the excluded volume of the bigger colloidal particles. The excluded volume (\orangeline[dashed]) cannot be penetrated by the depletant due to the presence of the colloid. Thus, the larger objects pack together such that their excluded volumes maximally overlap (indicated in orange) and more space is provided for the depletants. Overall this mechanism can be interpreted as an effective, entropically driven attraction between the colloids.}
\label{fig:DSketch}
\end{figure*}

Pear-shaped colloids, or rather their contact function, have been modelled using the self-non-addivitive pear hard Gaussian overlap (PHGO) model which is a computationally much faster approximation than the proper hard pears of revolution (HPR) model. We showed in part 1 \cite{SMCS-T2020_1} and other earlier studies that pear-shaped particles, which contact is approximated by the PHGO potential \cite{BRZC2003}, spontaneously form cubic, bicontinuous phases, like the double gyroid \cite{EMBC2006,SEMCS-T2017} or, when diluted with a small amount of hard-sphere solvent, the double diamond \cite{SCS-T2018}. We define pear-shaped particles by the B\'ezier-curve which, when extended to a solid of revolution, yields the pear-shaped particle shape with a smooth bounding surface \cite{BRZC2003} (see also \fig{fig:DSketch} for the outline of a pear-shaped particle). This description has two parameters, $k$ and $k_\theta$ which tune the aspect ratio and the degree of tapering of the pears, respectively. Even though PHGO particles are best illustrated by the B\'ezier pear-shape, the computational PHGO model does not represent hard interactions between those B\'ezier objects perfectly. In particular, PHGO pear-shaped particles partially overestimate or underestimate the inter-particle distance compared to the B\'ezier curve representation, which leads to small overlaps and gaps depending on relative particle orientations \footnote{Note that these ``overlaps'' do not enable the pear-shaped particles to invade the space occupied by other pears according to the PHGO potential. The interactions are governed by a hard-core potential.}. These inaccuracies, despite small, affect the phase behaviour of the pears and have previously been -- incorrectly -- believed not to be important for the self-assembly processes \cite{EMBC2006,SEMCS-T2017}.\\

A more accurate, but computationally substantially more expensive model is based on triangulated meshes of the pear-surface, denoted as the hard pears of revolution (HPR) model. Here, the contact coincides with the B\'ezier description to much higher accuracy (essentially only limited by the discretisation used for the mesh). Even though the difference between the PHGO and the HPR model is small (see in-depth discussion about the differences in the contact function in part 1), the first part of this study shows that the gyroid phase is not formed by HPR particle \cite{SMCS-T2020_1}.\\

We show that also the excluded volume interactions of pears in a solvent of hard spheres are impacted by these distinctions. This depletion behaviour enables us to explain some of the differences between the PHGO and HPR self-assembly behaviour of the pure systems, without solvent which were discussed in part 1 of this series \cite{SMCS-T2020_1}.\\

Depletion forces, which arise from the osmotic pressure on neighbouring colloids by the surrounding small depletants, lead to effective short range-attraction \cite{AO1954,AO1958,V1976,MCL1995,RED2000} or repulsion \cite{CNW1996,BBF1996,DAS1997,GED1998} between colloidal particles. Already 70 years ago, these depletion forces has been predicted as a purely entropically driven effect similar to the entropic self-assembly of colloids into liquid crystal phases. More specifically, Asakura and Oosawa \cite{AO1954,AO1958} argued that, as the free energy of the system is predominantly governed by the degrees of freedom of the solvent particles, the minimisation of free energy induces the colloids to arrange in the most compact arrangement such that their excluded volume, which can not be penetrated by the solvent, is minimised (see \fig{fig:DSketch}). Since then depletion forces of spherical particles have been studied extensively both in theory for different solvent models, like the penetrable hard spheres model \cite{WR1970,LT2011} polymers based on the ideal chain-model \cite{EHD1996,ATL2002}, hard-core spheres \cite{MCL1995,RED2000,DAS1997}, hard-core rods \cite{MCL1997,R2002,LYM2008}, or hard-core disks \cite{PW2000,HD2004}, and also experimentally \cite{EN1988,CMDY1999,KFL1994,DWPY1997,KBCS2007,BH1991,LS1993,R1994,R1995,ID1995,DvRE1999,TRDeK2003}.\\

The study of depletion effects between two pear-shaped particles in a solvent of hard spheres can help understand the collective self-assembly mechanisms in a one-component pear particle system, for the following two reasons: Firstly, the simulation process is numerically less expensive (especially for the HPR model) by dealing with only two particles with complicated contact functions. Secondly and more importantly, in all liquid crystal phases, obtained for the PHGO system so far \cite{EMBC2006,SEMCS-T2017,SCS-T2018}, the arrangement of each pear is highly affected by a multitude of next nearest neighbours. This elaborate interplay of particles coupled with the aspherical pear-shape, which features a significant degree of complexity, makes a more detailed analysis of the direct influence between adjacent particles in one-component systems impracticable. Hence, we reduce the complexity of our simulations and shift our focus to the depletion systems which encapsulate the fundamental features of pure two-particle interactions.\\

This article is structured as follows: We first identify the optimal arrangement of pears in terms of minimal collective excluded volume using numerical tools in \sect{sec:Excluded}. Next (\sect{sec:MonteCarlo}) we perform MC simulations of two large pear-shaped particles within a solution of smaller hard spheres; This is done for both the PHGO and HPR particle models to compare the computational results with the previous predictions of the ideal excluded volume, obtained by the numerical technique. These allow us to pinpoint the specific differences between the two models more efficiently. We show that the PHGO particles favour the formation of bilayer phases (including the bilayer smectic and gyroid phases) in contrast to the HPR particle. Finally in \sect{sec:Outlook}, we will give a short outlook, how bilayer phases could possibly be stabilised in monodisperse systems based on the HPR interactions by introducing non-additivity to the contact function.\\

\section{Excluded volume of two pear-shaped particles}
\label{sec:Excluded}

Similar to other self-assembly processes, the shape of the molecules/colloids naturally impacts how a pair of two colloidal particles in a solvent eventually arranges under the influence of depletion. By changing colloids from a simple sphere to objects with more complicated shapes, the excluded volume does not only depend on the separation but also the relative orientation of the particles (see \fig{fig:DSketch}). Consequently, depletion does not only induce attraction between colloids but also an orientational rearrangement of the particles. For instance, it has been shown that by adding dimples to one of the spheres the other colloid preferentially attaches to these concavities \cite{SICP2010,WWZYSPW2014}. This ``lock-and-key'' mechanism can be used as a tool to control the depletion of particles. Another sort of directionality can be introduced by creating elongated colloids. At a wall, hard prolate ellipsoids \cite{EBM2003,LM2003} and spherocylinders \cite{RvRAMD2002} align with their long axis along the flat interface due to depletion. Moreover, it is known theoretically \cite{ADKF1998,KVKvDPL1999} and from experiments \cite{SYI1989,AF1998} that rod-like colloids self-assemble into clusters with nematic order when non-absorbing polymers are added. Excluded volume mechanisms provide access to rich phase behaviours for various mixtures of hard aspherical particles and depletant particles \cite{ADKF1998,BF2000,DPGAF2004,BDvR2012,AMV2016,GOT2018,GTMOWL2018}, including fascinating effects like depletion induced shape-selective separation in colloidal mixtures by the addition of non-adsorbing polymers \cite{C1942,M2002,BFvHGFLYYCM2010,PKV2010}.\\

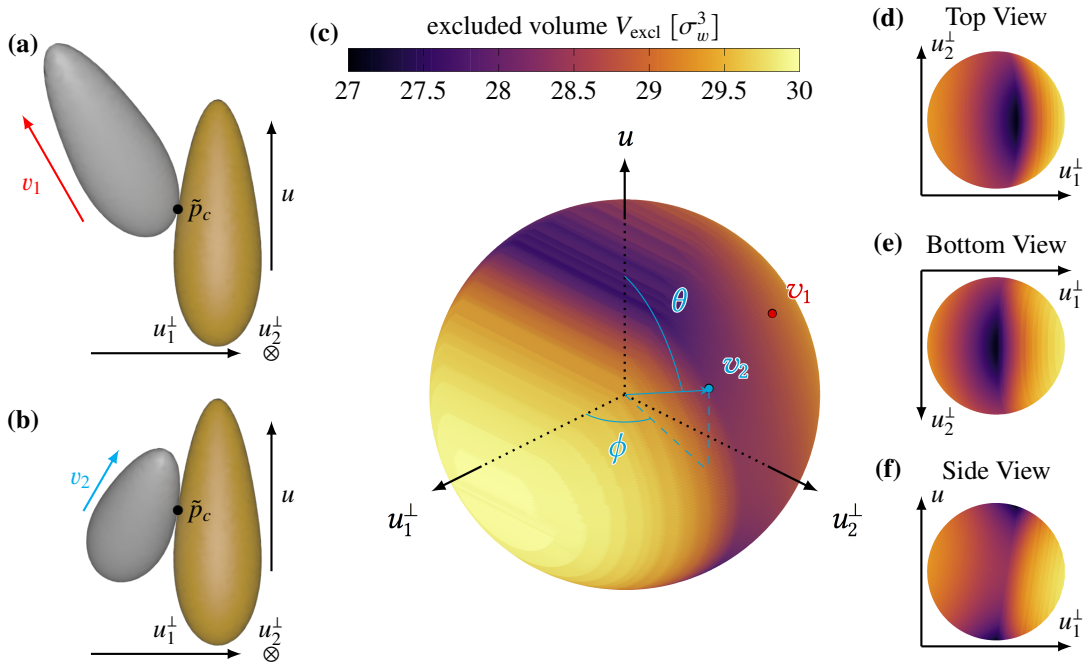
\begin{figure*}[t!]
\centering
\input{images/ExclVolSph.tex}
\caption{The excluded volume of two pear-shaped particles with $k{=}3$, $\theta_k{=}15^{\circ}$ and $r_\text{depl}{=}0.31\sigma_w$ in relation to the relative orientation of the pears on the unit sphere. The algorithm to calculate the excluded volumes is described in \app{app:Appendix}. The contact point $p_c$ is fixed for the reference pear and chosen such that the configuration with the global minimum can be adopted. In the centre (c), the orientation of the free pear $\mathbf{v}$ is given in spherical coordinates dependent on the orientation of the reference pear $\mathbf{u}$ and the direction towards $p_c$ $\mathbf{u}_1^{\perp}$ . On the right, the unit sphere is viewed from the top (d), bottom (e) and side (f) perspective. On the left (a)+(b) two exemplary configurations are shown. The locations of their corresponding orientations $\mathbf{v}_1$ and $\mathbf{v}_2$ on the unit sphere are indicated.}
\label{fig:ExclVolSph}
\end{figure*}

Here we perform numerical calculations to predict the ideal configuration of two pear-shaped particles in a solvent. The ideal configuration is defined as the arrangement of two pears such that the excluded volume caused by these two particles is minimised (that is, that the overlap of the 'halos' of these particles representing the volume that solvent particles cannot enter is maximised). This is achieved by a computational algorithm (see \app{app:Appendix}) that generates mesh representations of these halos and that, using computational geometry ''Boolean'' algorithms to compute the set intersection of these halos, then computes the intersection volumes for all possible relative orientations of the pear-shaped particles. For rotationally symmetric particles like pears defined by B\'ezier-curves, three degrees of freedom have to be considered in addition to the particle separation to define a specific constellation between two pears. Two of these degrees of freedom relate to the relative orientations of the particles $\mathbf{u}$ and $\mathbf{v}$. The last one relates to the flexibility to select the contact point $p_c$ on the surface of one colloid, in the case both particles are in touch and the separation is 0. The choices of $\mathbf{u}$, $\mathbf{v}$ and $p_c$, automatically determine the contact point on the surface of the other object (see \fig{fig:ExclVolSph}a+b). Theoretically, we are able to sweep the whole configurational space of the two-pear-depletion-problem and identify the configuration with the largest excluded volume overlap. Therefore, we apply the sampling algorithm to pears with aspect ratio $k{=}3$ and tapering angle $\theta_k{=}15^{\circ}$, which lie well within the gyroid phase for the PHGO model \cite{SEMCS-T2017} but does not form cubic phases for the HPR-model.\\

To sample the configuration space as efficiently as possible, we first show that the three-dimensional excluded volume problem can be narrowed down to its two-dimensional counterpart. In more mathematical terms, we only consider arrangements of pears where the orientation vectors of the two pears $\mathbf{u}$ and $\mathbf{v}$ and their relative position vector $\mathbf{R}$ are linearly dependent. Only those positions are in line to find the ideal placement of pears. It seems intuitive that, due to the pear's rotational symmetry, the configuration which occupies the least amount of space falls into the category of those linearly dependent arrangements rather than of asymmetric configurations. Moreover, any expansions of the excluded volume in the form of dilatations into the third dimension (like those indicated in \fig{fig:ExclVolAlg}) can be prevented by restricting the particles to a plane. This guess is confirmed by computing the excluded volume for different relative orientations with a fixed contact point $p_c$ of one of the pears as plotted in \fig{fig:ExclVolSph}c. Here, the pear with constant $p_c$ acts as a reference (see \fig{fig:ExclVolSph}a+b) such that $\mathbf{v}$ can be written in spherical coordinates with respect to the frame defined by $\mathbf{u}$ and $p_c$. The azimuthal angle $\phi{=}0$ of the spherical coordinate system is defined by the direction from the contact point $p_c$ to the centre of the reference pear. For all the tested values of $p_c$, the extremal values in $V_\text{excl}$, and hence both its global maximum and minimum, are attained by linearly dependent configurations, that is where the polar angle of $\mathbf{v}$ is either $\phi{=}0$ or $\phi{=}\pi$.\\

\begin{figure*}[t!]
\centering
\input{images/excluded_V.tex}
\caption{The excluded volume of two pear-shaped particles with $k{=}3$, $\theta_k{=}15^{\circ}$ and $r_\text{depl}{=}0.31\sigma_w$ along the ``rol'' (blue) and ``slide'' (red) route, where the particles share the same contact point pc, in terms of the angle $\beta$ between the orientation of the pears and the normal direction into the pear at $p_c$ . The algorithm to calculate the excluded volumes is described in \app{app:Appendix} and both sampling pathways are sketched above. The plots show a minimum of the same value which can be identified as the global minimum of the system. The corresponding optimal configurations are highlighted in the small coloured boxes.}
\label{fig:excluded_V}
\end{figure*}
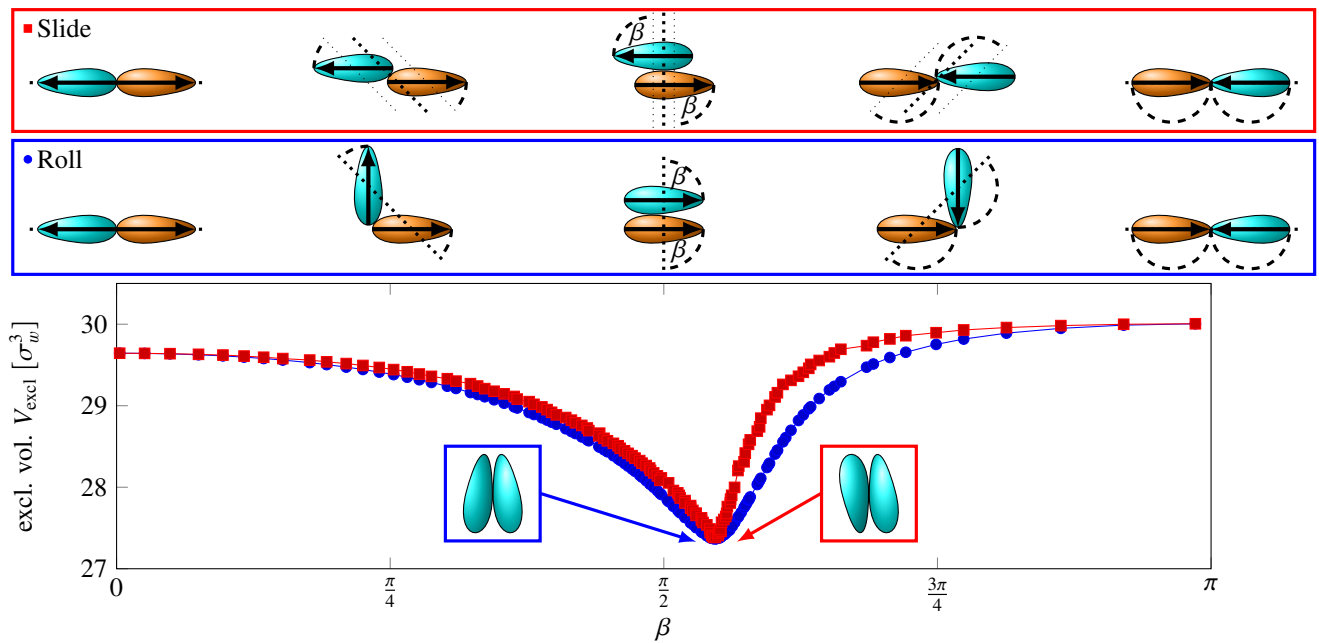

To reduce the configurational space even further, we can utilise another argument about the symmetry of the system. Specifically, the contact, which leads to the maximal or minimal excluded volume, has to be at the same point on both pear surfaces as the choice of the reference pear is arbitrary. Otherwise, the system would have two solutions with the same relative orientations, which is not possible for convex particles. Overall this leaves us with a sampling domain which practically only depends on one degree of freedom, namely on the shared $p_c$. By adding the constraint of linearly dependent orientations with $\phi{=}0/\phi{=}\pi$ the polar angle, $\theta$ is restricted to maximally two possible orientations. The excluded volume calculations for the ``roll'' and ``slide'' sampling of the different contact points $p_c$ are plotted in \fig{fig:excluded_V}.
\begin{itemize}
\item \textbf{Roll route}: The particles start from an antiparallel configuration, when the pears touch with their blunt ends, pass through a parallel alignment next to each other and eventually end up antiparallel again where their pointy ends meet. This sampling can be interpreted as one pear is rolled over the other.
\item \textbf{Slide route}: During the ``slide'' sampling the pears are perfectly antiparallel for all $p_c$ which resembles a slide of one pear along the surface of the other.
\end{itemize}
Hence, the duality of $\theta$ is covered by those two computational pathways. The contact $p_c$ is given by the angle $\beta$ between $\mathbf{u}$ and the normal vector into the pear at $p_c$.\\

Interestingly, the different paths reveal two distinct relative configurations with the same contact point $p_c {=}\tilde{p}_c$ , which both can be associated with the global minimum of the excluded volume $V_\text{excl}$. In one solution the pears are placed side-by-side and oriented perfectly antiparallel towards one another: $\mathbf{u}{\cdot} \mathbf{v} {=} -1$ (see \fig{fig:excluded_V}). The minimum, however, does not occur for $\beta{=}\frac{\pi}{2}$ when the pears are at the same height. The particles are rather shifted towards their blunt ends by a small distance. The second ideal configuration exists due to the broken inversion symmetry of the pear-shape and is found when the two pears point roughly in the same general direction (see \fig{fig:excluded_V}). However, here the colloids are not perfectly aligned but slightly tilted towards each other. This tilt also becomes apparent by looking at the excluded volume plot of different orientations at $\tilde{p}_c$ in \fig{fig:ExclVolSph}d-f. The top, bottom and especially side view of the unit-sphere clearly show that the minimum at the northern hemisphere is shifted away from the north pole. The tilt can be related directly to the pear-shape. In particular, the angle between the pear-shaped solids is identified as their tapering angle of $\theta_k {=} 15^{\circ}$ . Hence, $\theta_k$ also defines the shift in the antiparallel domain, as both optimal configurations are attained for $\tilde{p}_c$.\\

Furthermore, the computations show that configurations, where the blunt ends touch ($\beta<\frac{\pi}{2}$ in \fig{fig:excluded_V}), tend to be often more favourable than arrangements where the pears come together with their pointy ends ($\beta{>}\frac{\pi}{2}$). Also in \fig{fig:ExclVolSph}c a similar observation can be made. If the particle is oriented away from the reference pear and comes in contact with the blunt end, the excluded volume is smaller than if the pear points directly towards $\tilde{p}_c$. This general behaviour indicates that during the rearrangement of inversion asymmetric particles from a configuration where the colloids are separated to one where they are in contact due to depletion interactions, the colloids are likely to first approach each other with their bigger ends before eventually equilibrating into the most compact formation. Note that an indication of this blunt-end-attraction can be seen in the gyroid-phase self-assembly \cite{SMCS-T2020_1} where the blunt ends form the network-like domains of the bicontinuous cubic phase \cite{EMBC2006,SEMCS-T2017,SCS-T2018}. This indicates that also the hard HPR pears has a tendency to cluster with their blunt ends.\\

\section{Monte Carlo simulations of depletion effects of pear-shaped particles}
\label{sec:MonteCarlo}

Having determined the most favourable configuration of pairs of pear-shaped particles in regards to their excluded volume, we compare the computational predictions to results obtained by computer simulations. Our goal is to replicate the behaviour of pear-shaped colloids due to depletion and, moreover, to study if the pears indeed prefer the states calculated in \sect{sec:Excluded}. Therefore, we apply simple Metropolis Monte Carlo methods below. Before doing so, we review alternative methods for calculating depletion forces and free energies and describe why, sadly, these methods are too complex to implement for the pear-shaped particle system.\\

One successful theoretical approach to describe depletion is density functional theory (DFT). Roth introduced a so-called ``insertion approach'' \cite{RED2000,KRD2006} within DFT, where the depletion potential is calculated from the solvent density distribution close to one fixed colloid by insertion of a second colloidal particle and use of the potential distribution theorem \cite{H1983}. The interactions in a mixture of hard spheres \cite{RED2000}, a system of a spherocylinder close to a hard wall \cite{RvRAMD2002}, and a mixture of aspherical, but inversion symmetric particles \cite{KRD2006} have been derived with this ansatz. Also, other theories have been applied to calculate depletion interactions \cite{BBF1996,DAS1997,K2004} but have shown to be less efficient as every single configuration has to be treated individually. However, all of those theoretical approaches only cover a set of particles with simple shapes and have not yet been applied to pear-shaped particles. Even though a density functional for hard pear-shaped particles representing the HPR model has been derived \cite{SS-TM2018}, the difficulty is enhanced even more as we would have to develop a functional of orientational-dependent contact functions like for PHGO particles as well.\\

Alternatively, depletion interactions have been obtained with Monte Carlo simulation techniques. A typical procedure to calculate the depletion forces between various particles is the ``acceptance'' approach where the free energies between two different configuration states are compared. During these simulations, the positions and orientations of both colloids are fixed, and only the hard sphere solvent is displaced in the process of the Monte Carlo step. Finally, the free energy difference between two states can be related to the acceptance rate to jump from one colloid particle's relative position to the other and vice versa without causing particles to overlap \cite{LM2003,B1976,AT1991,LM2002}. This procedure has been advanced using Wang-Landau Monte Carlo approaches \cite{WL2001,SDP2002,MLM2014,HKHS2016}. Also, a hybrid of simulation and DFT has been suggested \cite{JW2011}. Those approaches are, however, very complicated for the pear shape (in case of the hybrid approach) or very time inefficient, as for every configuration state a separate MC run has to be performed in the acceptance approach. Combining these issues with the already computationally demanding overlap check between two meshes for the HPR particles and hard spheres, the mentioned techniques are all impracticable. However, in general, we are not necessarily interested in the specific free energy-calculations of the different states but merely want to clarify the distinctions between the HPR and PHGO model. Therefore, the question of depletion is tackled by applying Monte-Carlo simulations in the following and more straightforward fashion.\\

\subsection{Depletion interactions between HPR particles}
\label{sec:DepletionHPR}

\begin{figure*}[t!]
\centering
\input{images/separation.tex}
\caption{Representative progressions of the separation $R$ of two pear-shaped particles (red: HPR, blue: PHGO, orange: NAHPR.) surrounded by 1498 hard spheres, acting as a solvent during the Monte-Carlo simulations. The simulations are performed at a global density of $\rho_g{=}0.45$. All models show an effective attraction into the zone of influence, where the excluded volumes of the pears can be considered overlapping, induced by depletion effects. The shaded area approximates this zone of influence where the particles can be considered in contact.}
\label{fig:separation}
\end{figure*}
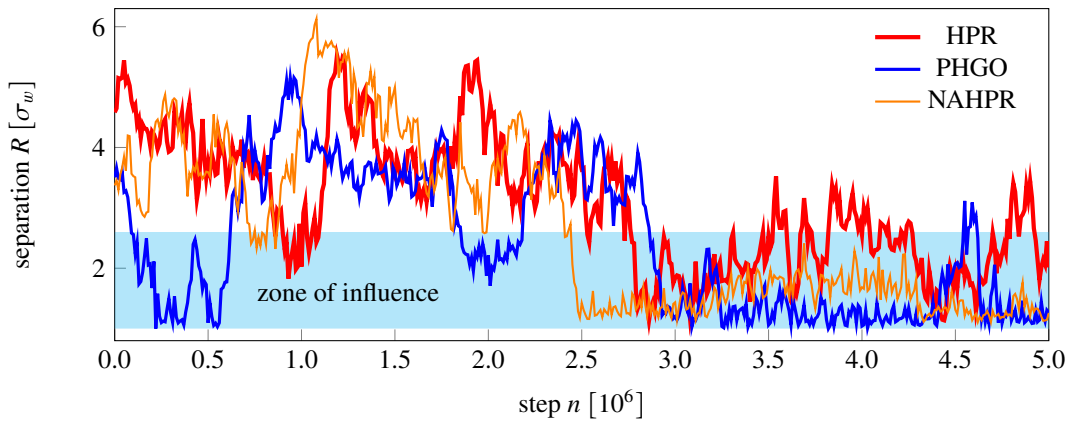

Monte Carlo simulations are performed on systems with $N_\text{pear}{=}2$ hard-core pear-shaped particles within a solvent, which is approximated by a large number $N_\text{sph}{=}1498$ of surrounding smaller hard spheres, within a cubic box with periodic boundary conditions in all three dimensions. The aspect ratio $k{=}3$ and tapering parameter $\theta_k {=} 15^{\circ}$ of the pear-shaped particles are chosen to easily compare the simulation results with the calculations of \fig{fig:ExclVolSph}. For the same reason the sphere radii of the solvent $r_\text{depl}$ is set to $0.31\sigma_w$, which corresponds to the volume ratio between the spheres and pears $v{=} \frac{V_\text{depl}}{V_\text{pear}} {=} 0.08$. An acceptance rate of roughly 50 \% has been achieved by setting the maximal translation $\Delta_{q,\text{max}}{=}0.085\sigma_w$ and the maximal orientational displacement $\Delta_{u,\text{max}}{=}0.085\sigma_w$ per step. The greater number of depletants assures that the simulations are not affected by the boundary conditions and the system can indeed be interpreted as two pear-shaped colloids surrounded by hard sphere solvent. Furthermore, the sphere size is small enough to see depletion interactions between the particles occurring at higher densities. All sets are performed in the \textit{NVT}-ensemble starting from different diluted initial states at
\begin{equation}
\rho_g=\frac{N_\text{pear}\cdot V_\text{pear} + N_\text{sph}\cdot V_\text{sph}}{V_\text{box}}=0.1.
\end{equation}

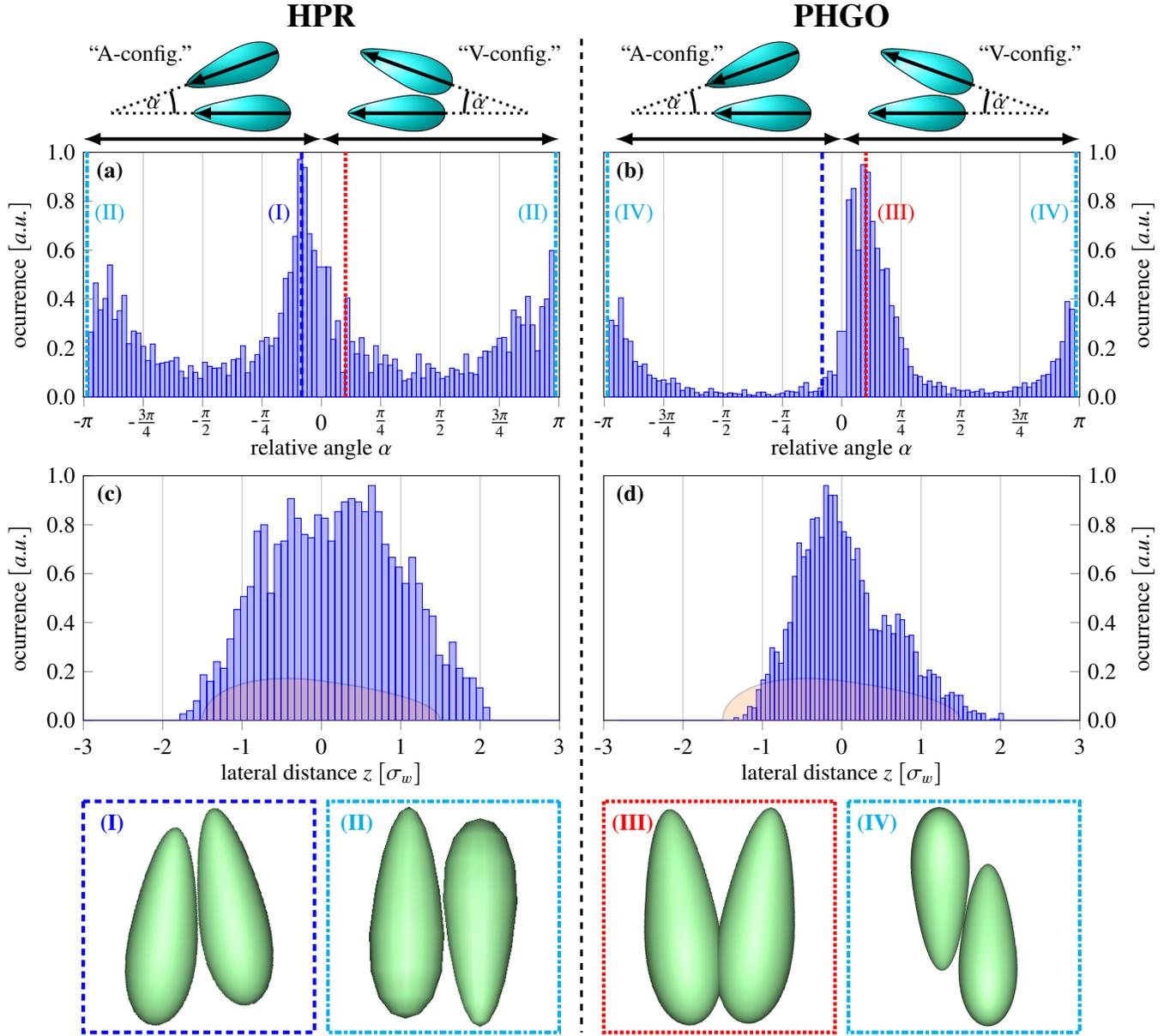
\begin{figure*}[t!]
\centering
\input{images/ori_hard_go.tex}
\caption{The relative orientation (a \& b) and lateral distance distribution (c \& d) of two HPR/PHGO particles surrounded by 1498 hard spheres, acting as a solvent at global density $\rho_g {=} 0.45$, on the left. The particle parameters are set to $k{=}3$, $\theta_k{=}15^{\circ}$ and $r_\text{depl}{=}0.31\sigma_w$ ($\frac{V_\text{depl}}{V_\text{pear}} {=} 0.08$). Only pair-configurations are considered if the pear-shaped particles are close to each other such that the excluded volumes overlap. Positive
angles $\alpha$ indicate V-configurations (blunt ends together), whereas negative $\alpha$ values describe A-configurations (pointy ends together). On the bottom, typical arrangements of the HPR (I+II) and PHGO (III-IV) depletion systems, extracted from both type of simulations, are shown. The left snapshot (dashed line, (I)) corresponds to the indicated peak in (a) and coincides with the parallel solution for maximal excluded volume overlap. The centre left configuration (dash-dotted line, (II)) contributes to the second peak of (a) and matches the anti-parallel solution in terms of minimised excluded volume. The centre right snapshot (dotted line, (III)) shows a V-configuration, which corresponds to the indicated peak in (b). This configuration does not coincide with the parallel solution for maximal excluded volume overlap of B\'ezier pears. The right configuration (dash-dotted line, (IV)) contributes to the second peak in (b) and matches the anti-parallel solution in terms of minimised excluded volume.}
\label{fig:ori_hard_go}
\end{figure*}

After a sequence of compressions to the final density $\rho_g{=}0.45$ the system is studied for $5.0{\cdot}10^6$ steps. This density turned out to be sufficiently high to observe considerable entropic forces between the pear-shaped colloids and low enough to prevent crystallisation in the surrounding hard sphere liquid.\\

We first simulate HPR pears in a hard sphere fluid, where the overlap of two particles is determined by checking for intersections of two meshes representing the surfaces of the pears \cite{HLCGM1997,JTT2001}. For every simulation run, the entropic depletion attraction between the pear-particles is determined when the colloids are in each other's vicinity, which means that their excluded volumes overlap. More precisely, the particles stay together for a considerable number of MC steps (see \fig{fig:separation}), which leads to the conclusion that the system indeed favours the particles coming in contact. However, the entropic attraction seems to be short range and rather weak. This can be seen in \fig{fig:separation}, where, during a typical MC simulation run, the particles repeatedly separate prior to reaching a seemingly steady state where they remain in contact \footnote{Note that the MC dynamics described here do, of course, not represent true particle dynamics or trajectories.}. Nevertheless, the preferred sampling of close pear arrangements is a strong indication for depletion interactions.\\

Even though the particles are affected by the presence of the second colloid, the determination of the relative arrangements of the colloid pair faces some difficulties. The main issue which has to be overcome is poor statistics. As we are studying a two-particle problem, it is hardly feasible to gather enough data for a detailed combined analysis of the possible states due to computational time constraints. Therefore, we decouple the degrees of freedom and only investigate one between two close HPR particles is plotted. For these plots, relative parameter at a time. In \fig{fig:ori_hard_go}a the relative polar angle only configurations are considered if the excluded volumes overlap. This ensures that the sampled relative orientations are actually influenced by the close distance between the particles. The relative angle $\alpha$ between the orientation vectors of the pears $\mathbf{u}$ and $\mathbf{v}$ is split into two domains to characterise the orientational states further. For positive angles, the pears point away from each other such that their blunt ends are in contact. A negative angle indicates that the pears face towards one another and that their pointy ends are closer together. In the following, we will refer to these two domains ``V''-configurations ($\alpha{>}0$) and ``A''-configuration ($\alpha<0$).\\

The histogram of the relative pear orientations shows two distinct peaks which match perfectly with the ideal configurations predicted in \fig{fig:ExclVolSph}c and \fig{fig:excluded_V}. The first preferred orientation is measured at $\alpha{=}-0.26{=}-15^{\circ}$, and hence categorised as an A-configuration. This relative angle corresponds directly to the parallel solution for minimal excluded volume as it coincides with the tapering angle $\theta_k {=} 15^{\circ}$. The configuration can also be extracted from the simulations directly (see a snapshot in \fig{fig:ori_hard_go}I). The second peak at $\alpha{=} \pm \pi{=} \pm 180^{\circ}$ is identified as a single characteristic orientation due to the duality of the A- and V-configuration for $\cos(\alpha){=} -1$. Moreover, this orientation also coincides with the predictions as it fits the second solution of the excluded volume calculations, where the particles are aligned anti-parallelly next to each other. A snapshot from the MC simulation of this particular configuration is depicted in \fig{fig:ori_hard_go}II.\\

The observations are corroborated by the lateral distance distribution between two particles when in contact. \fig{fig:ori_hard_go}c highlights that the neighbouring pears are not distributed around the centre point of the reference particles. The distribution is rather slightly shifted towards the pointy end. The inversion asymmetric shape of the HPR particle consequently introduces a move of the optimal contact point above the centre-point. Hence, the HPR particles behave precisely as expected according to \sect{sec:Excluded} and according to the solutions of the ideal configurations to maximise the available space for the hard spheres.\\

\subsection{Depletion interactions between PHGO particles}
\label{sec:DepletionPHGO}

We established that the HPR particle model describes the imposed pear-shape very well in terms of depletion and reproduces the analytical predictions. The HPR model now serves as a reference for other models which describe pear-shaped particles, such as the PHGO model. This gives us an opportunity to study the ramifications to describe a pear using a hard Gaussian overlap approach. Thus, the depletion MC simulations are repeated. The same parameters are applied except that the HPR contact function is replaced with the hard PHGO potential to approximate the particle overlap \cite{BRZC2003}.\\

The first distinction between the PHGO and HPR system becomes apparent during the MC sampling already. By tracking the distances between both particles for every MC step in \fig{fig:separation} the depletion attraction between two PHGO pears seems to be much stronger than in the equivalent HPR case. This can be explained by the development of the separation once the two PHGO pears are close together. After the pears pass a sequence of arbitrary displacements and eventually approach each other, the touching configuration stays stable for a significantly longer time (see \fig{fig:separation}). This is in contrast to the split-ups of the HPR particles where very short periods of configurations close together alternate with stages of separation and subsequent recombination. The repeated attachment/detachment of the pear colloids in the HPR model indicates that the depletion attraction is comparable to thermal energies, that is, it is of the order of $k_BT$. The greater propensity of the PHGO pear colloids to remain in contact (rather than to detach again) is a clear indication that the depletion effects are stronger for PHGO particles than for HPR particles. The increased strength of the entropic force, however, can be related to the contact function of the PHGO pear. Presuming the particles are in the optimal state, an attempted translational step and especially an attempted rotational step is much more strongly penalised for PHGO than for HPR particles. This is manifested in the contact distance of roughly perpendicular arrangements (see Fig. 1 of part 1 \cite{SMCS-T2020_1}). Here, the pear size is overestimated, and a particle pair is accounted as overlapping even though they are not in contact according to the B\'ezier-curve depiction. The effect is comparable to the PHGO pears and HGO ellipsoids \cite{P2008} entering orientationally ordered phases already for low densities. The depth of the effective potential does not necessarily indicate that the two models differ qualitatively, but suggests that the depletion is more guided towards the equilibrium states.\\

The relative orientation distribution between two PHGO particles in close contact is plotted in \fig{fig:ori_hard_go}b. Two distinct peaks emerge similar to the equivalent HPR system. The smaller peak is found at $\alpha{=} \pm \pi$ which again corresponds to an antiparallel configuration. Therefore, the orientation distribution suggests that the PHGO pear model reproduces the antiparallel solution sufficiently. In this domain, the HPR and PHGO differ the least from each other, such that it is quite intuitive that in the anti-parallel case both models share the same solution. Additionally, we find many configurations as depicted in \fig{fig:ori_hard_go}IV, which contribute to the pronounced peak at $\alpha{=} \pm \pi$ and coincide with the ideal solution to a sufficient degree. By focusing on the second larger peak, however, we observe two major differences compared to the HPR system. Firstly the peak is significantly more intense. This indicates that for PHGO particles the parallel configuration is more beneficial than the antiparallel solution. This is explained by the ability of PHGO particles coming closer together than HPR particles when parallelly aligned. By changing the relative angle between the pear-shaped particles, the overlap tends to be underestimated by the PHGO model which consequently leads to a lower excluded volume. Thus, the duality of the ideal configuration is broken by the particular angle dependence of the PHGO contact function and weighted to the benefit of parallel arrangements. This observation is in accordance to the pair correlation functions of the monodisperse pear-shaped particle systems, obtained in the first part. Also these plots indicated a pronounced polar alignment between neighbouring PHGO particles within the bilayer architecture of the gyroid structure compared to the HPR particles.\\

The second difference is the position of the peak, which is shifted from $\alpha{=}-15^{\circ}$ to a positive value close to $\alpha{=}20^{\circ}$. Hence, the particles form V-configurations rather than the expected A-configurations. To clarify the reason behind this transition we take a closer look at those V-configurations which can be obtained from the simulations directly. A representative pair is portrayed in \fig{fig:ori_hard_go}III. It becomes apparent that the pears slightly overlap. Here, the term ``overlap'' might be misleading as the particles do not technically overlap in terms of their PHGO contact function but according to the best possible illustration using the B\'ezier representation. However, it also has to be mentioned that the spheres interact with the pear according to this B\'ezier shape. Thus, the solvent particles interact with the PHGO particles in terms of a different effective shape than two PHGO particles with each other. Furthermore, the underlying underestimation of the PHGO-contact function enables the pear-shaped particles to occupy space, which by design cannot be reached by hard spheres and would also be prohibited for HPR particles. This effect is known as pairwise \textit{non-additivity} and is well studied for hard binary sphere mixtures \cite{BPGG1986,LALA1996,RE2001,HS2010,ZFLSSO'H2015}, which successfully model the behaviour of binary alloys \cite{KA1995,SKCFC2011} or organic mixtures \cite{M1977,PM2006}.\\

The V-configurations also can be associated with a special kind of non-additivity effect between two PHGO pears, which we called \textit{self-non-additivity} in the first part \cite{SMCS-T2020_1} of this series. Due to the self-non-additivity between the blunt ends of PHGO particles, the excluded volume is decreased instead of simple alignment by an alternative route, namely by increasing the overlap of the two particles. For pears with $k{=}3$ and $\theta_k {=} 15^{\circ}$ the maximal overlap according to the B\'ezier shape occurs roughly at an angle of $\alpha_\text{overlap}{\approx} 30^{\circ}$. This is considerably higher than the measured angle between the pears in the V-configuration observed in the simulations. However, we can argue that the adopted angle results from the intricate interplay of reducing excluded volume via overlap and alignment and the sphere radius of the solvent. For small volume ratios the overlap is more dominant and the V-arrangement more favourable, whereas for large ratios the contribution of the overlap becomes negligible and the aligned A-configuration will be adopted.\\

To complete the comparison between the HPR and PHGO particles, we investigate the lateral distance of the PHGO pears to its fellow pear in close contact in \fig{fig:ori_hard_go}d. Compared to \fig{fig:ori_hard_go}c the distribution is much narrower and shifted towards the blunt end which leads the impression that the HPR particles are more flexible to obtain the equilibrium state whereas the PHGO pears are more restricted in terms of fluctuations from the ideal configuration. The emergence of the shifted peaks can again be attributed to the non-additive characteristics of the PHGO model. Furthermore, the two maxima at lateral distance $z{=}-0.17$ and $z{=}0.70$ indicate the existence of two different contact points. One is associated with the V position ($z {<} 0$), the other peak can be identified as the contact for the antiparallel solution $z{>}0$.

\section{Conclusion and Outlook}
\label{sec:Outlook}

In this article, we studied depletion effects on pear-shaped particles due to a solvent of hard spheres. To this end, we investigated the depletion interactions of a pair of pear-shaped particles surrounded by a hard sphere solvent. In the course of this study, we first determined the optimal pear configurations in terms of minimised total excluded volume based on the B\'ezier curves to predict the equilibrated particle formation. Using numerical calculation techniques, we identified two configurations that both correspond to two global minima; a parallel and antiparallel solution, which both share the same contact point on the pear surface. Both configurations could be related directly to the taper of the particle. Afterwards, the predicted states could be obtained in Monte Carlo simulations of two HPR pear particles dissolved in a hard sphere solvent. However, depletion attraction is weak for the chosen parameters.\\

In comparison, the PHGO pear particles revealed differences to the predictions in \sect{sec:Excluded}. Even though the antiparallel configuration was also reproduced for PHGO pears, the parallel solution was found to be more dominant and shifted from an A- to a V-configuration with a different contact point. We argue that the V-configuration is governed by the PHGO contact function which underestimates the pear contact distance slightly and causes overlaps according to the B\'ezier representation. Moreover, it has been shown that the depletion attraction between two PHGO particles is much stronger than between HPR particles.\\

\begin{figure*}[t!]
\centering
\input{images/pricklypear.tex}
\caption{Possible design of a ``prickly'' pear-shaped colloid which copies the properties of the PHGO and NAHPR model. The self-non-additivity is modelled by a region of spikes (blue) which is pervious for spikes of other pear-shaped colloids but not for their hard body (black). (b) The procedure to obtain the second mesh in the NAHPR model which determined the overlap between the blunt ends of two pears with $k{=}3$ and $\theta_k{=}15^{\circ}$. First, two pears are placed symmetrically at an angle $\alpha{=} 30^{\circ}$ such that the pears are exactly in contact according to the PHGO contact function. The distance is decreased by $-0.035\sigma_w$ also to compensate the contact overestimation for A-configurations. Afterwards, the overlap is cut from the initial contour (dashed) such that a concavity occurs (dotted line). The equivalent non-additive contour is obtained from its convex hull (dash-dotted). This procedure is repeated for different angles between $\alpha{=} 30^{\circ}{\pm}10^{\circ}$. The final contour (solid line) is the basis of the solid of revolution from which the mesh is generated.}
\label{fig:pricklypear}
\end{figure*}
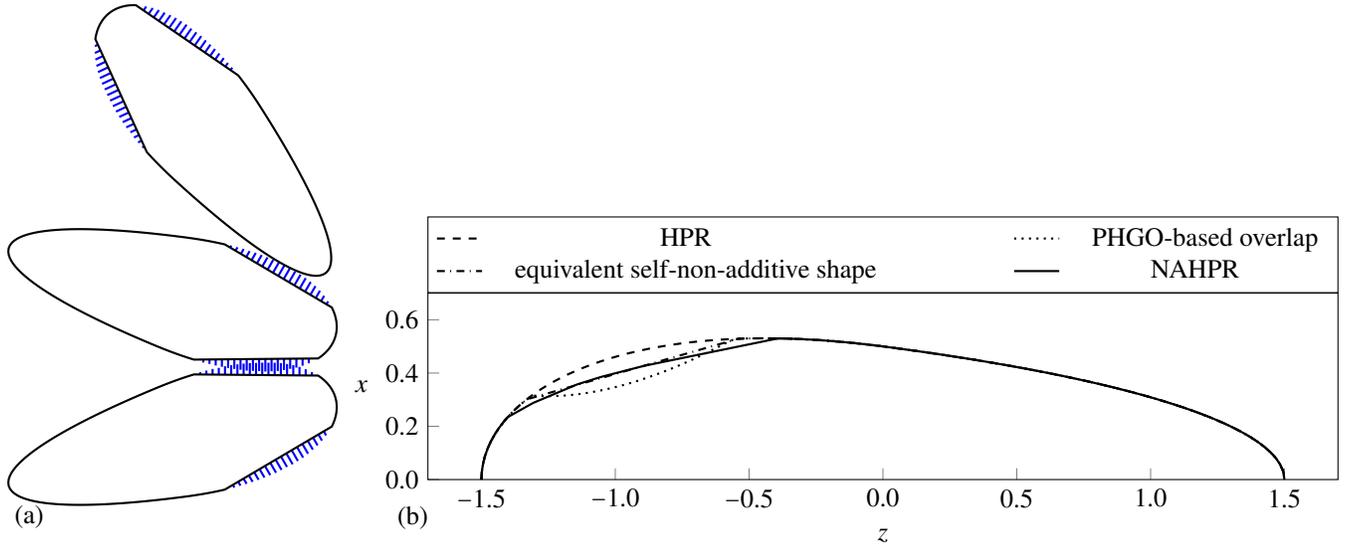

The discrepancies in the depletion behaviour also give improved insight into why the PHGO model has a propensity to forming interdigitated bilayer phases and why such bilayers are absent in the phase diagram of HPR particles. It is more than likely that specific details of the relative positions between neighbouring pear-shaped particles are varied due to the enhanced complexity of the excluded volume effects in one-component assemblies. Nevertheless, based also on the pair correlation functions in part 1 we can reason that the non-interdigitating quality of the arrangements would not change and hence, general statements about the local formations can be made. Especially three contributions to the stabilisation mechanisms of bilayer configurations \cite{SEMCS-T2017} are identified.\\
\begin{enumerate}
\item By breaking the duality of the optimal configurations (parallel and anti-parallel), the systems introduce a local polar order. In the PHGO model, this leads to a dominant formation of parallel alignments between adjacent pears. Hence, the system is guided towards the formation of sheets, which are a prerequisite of interdigitated bilayers.
\item The interdigitation is enhanced by the preferred parallel order into V- rather than A-configurations. It is quite intuitive to imagine that sheets, which consist of an array of V-aligned pears, interlock analogous to a zip mechanism in an ``zig-zag''-pattern and subsequently develop bilayers.
\item The greater fluctuations of the contact point in HPR systems hinder a targeted alignment of particles. This consequently leads to an increased susceptibility for defects within the bilayers, and a weaker correlation of translational order as those observed in typical smectics let alone gyroid or lamellar phases.
\end{enumerate}

\begin{figure*}[t!]
\centering
\input{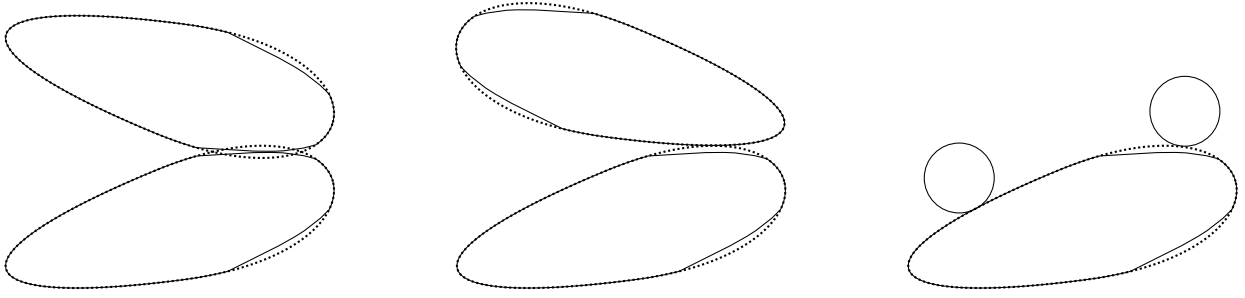}
\caption{The concept of the overlap determination for the NAHPR model. The pear consists of an inner contour (solid line, non-additive part) and an outer contour (dotted line, similar to the HPR model). If the pears coms together with their blunt ends (left) the particles are considered in contact if their inner contours touch. Otherwise (centre) the outer contours determine the overlap. The interactions with hard spheres are also according to the outer contour (right).}
\label{fig:nonadd}
\end{figure*}

These three factors raise the question of how pear-shaped particles adopting bilayer phases can be realised in experiments. In the first paper of this series, we have discussed aspects of whether the HPR or the PHGO model is closer to potential experimentally synthetised colloidal particles \cite{SEMCS-T2017}. In the following, we pursue a different question. Namely, we analyse some concepts of how a non-additive pear-shaped particle with a contact function of the PHGO particle would need to be designed and, more precisely, how the HPR contact profile has to be modified to obtain the key characteristics of the PHGO contact function. Therefore, we propose a promising approach as an outlook and introduce non-additive features to the mesh-description of pears as well .\\

To mimic the behaviour of PHGO particles non-additive features have to be added to the blunt ends of the pear particles. Using this approach we specifically try to engineer an HPR potential which favours the formation of V-configurations due to depletion interactions. One idea is to introduce a ``prickly'' pear-shaped colloid. Here non-additivity is modelled by a region of spikes, which is pervious by thorns of other colloids, leading to an effective ``overlap'' of the pear shapes, but cannot be penetrated by their hard bodies (see \fig{fig:pricklypear}a). Here, we have to consider that the spikes should not be too dense, which would prevent the full penetration of spikes or causes the particles to wedge. On the downside, if the spikes are distributed only sparsely, also the hard body can enter the non-additive region. Nevertheless, it seems feasible that we can effectively replicate the self-non-additive properties of the PHGO model by colloids with spikes in appropriate distances and optimised angles of the thorns.\\

To avoid optimising the prickly pear-shaped colloids in terms of spike distance and angle, we describe in our simulations the semi-penetrable region of the colloid by a second mesh in addition to the one used for calculating the HPR interactions. This mesh which describes the interaction between two blunt ends is based on the distance of two PHGO particles with the largest overlap. As mentioned this occurs for $\alpha_\text{overlap}{\approx} 30^{\circ}$. However, the distance is decreased even further by $-0.035\sigma_w$ to additionally compensate for the contact overestimation for A-configurations which otherwise would not be considered. The contour of the non-additive shape is created by introducing a flat line between the two points where both B\'ezier curves meet (see \fig{fig:pricklypear}b). Taking this new contour as a basis, we repeat the procedure for different angles $\alpha{=}30^{\circ}{\pm}10^{\circ}$ to allow some flexibility of the adopted orientations. Afterwards, a triangulated mesh of the solid of revolution of the resulting contour is generated. The mesh is implemented into the MC algorithm such that only the blunt ends of the pears are allowed to overlap according to the B\'ezier shape. To put it differently, the particles interact via the non-additive mesh exclusively if the particles come together with their blunt ends. Otherwise, the overlap is determined by the regular mesh describing the pear surface (see \fig{fig:nonadd}). Furthermore, the pear-sphere interactions stay unmodified such that the hard solvent still experiences the HPR pear. We will refer to this model as the non-additive hard pears of revolution (NAHPR) model. In experiments, the underlying contact function might be realised by preparing pear colloids with a rougher surface at the pointy than at the blunt ends. By using different roughness, the strength between different parts of a colloid can be controlled, and therefore an effective entropic attraction between specific moieties of the colloid can be introduced \cite{ZM2007,KNSHYWvBGDK2012}.\\

\begin{figure*}[t!]
\centering
\input{images/ori_nonadd.tex}
\caption{
The relative orientation (a) and lateral distance distribution (b) of two non-additive HPR particles surrounded by 1498 hard spheres, acting as a solvent at global density $\rho_g {=} 0.45$, on the left. The particle parameter are set to $k{=}3$, $\theta_k{=}15^{\circ}$ and $r_\text{depl}{=}0.31\sigma_w$ ($\frac{V_\text{depl}}{V_\text{pear}} {=} 0.08$). Only pair-configurations are considered if the pear-shaped particles are close to each other and the excluded volumes overlap. Positive angles $\alpha$ indicate V-configurations (blunt ends together). Negative $\alpha$ values describe A-configurations (pointy ends together). This is also indicated above the plot. On the right two typical arrangements, extracted from the simulations, are shown. The top snapshot (dotted line, (c)) corresponds to the indicated peak and shows the engineered V-configuration. The bottom configuration (dash-dotted line, (d)) is a defect of the non-additive mesh and contributes next to the anti-parallel solution also to the second indicated peak.}
\label{fig:ori_nonadd}
\end{figure*}

After implementing the non-additive contact function, the depletion MC simulations are again repeated with the same parameters. Both \fig{fig:separation} and \fig{fig:ori_nonadd} reveal that many of the features of the PHGO model have been adopted by the NAHPR model. By investigating the separation during the MC simulation in \fig{fig:separation} it becomes apparent that the depletion interaction increases. Even though the PHGO particles show slightly weaker attraction, the NAHPR particles remain in the zone of influence similarly as soon as they are within their vicinities. More interesting, however, is the orientation distribution for NAHPR particles in contact (see \fig{fig:ori_nonadd}a). The non-additivity at the blunt ends indeed stabilises the desired V-configurations creating a dominant peak at around $\alpha{=}20^{\circ}$. Nevertheless, by taking a close look, a small peak at the A-configurations can be observed as well. This leads to the conclusion that two minima for the excluded volume can be obtained within the parallel configurations. The global one is attributed to the V-configuration and the non-additivity, the second minor one can be ascribed to the A-position and the parallel alignment of the pears according to their tapering parameter.\\

The NAHPR model can also reproduce roughly the lateral distance distribution of the PHGO particle. Even though the distribution in \fig{fig:ori_nonadd}b is broader than the one in \fig{fig:ori_hard_go}b, most of the contact points are located underneath the centre point of the pear-shaped particle as well. However, the NAHPR model still does not reproduce all feature of the PHGO particles. For instance, some of the simulations end up in configurations which contribute to the preferred antiparallel alignment but do not coincide with the prediction. Although the predicted anti-parallel arrangement, where thin and blunt ends of the pear-shaped particles are next to each other, is still the dominant configuration, the non-additivity allows the particles also to overlap with the blunt ends in an antiparallel configuration (S-configuration, see \fig{fig:ori_nonadd}d) and also introduces in the antiparallel case a secondary minimum.\\

In a nutshell, the NAHPR particles can recreate some of the features of the PHGO contact function, like the formation of V-configurations, the enhanced depletion attraction or the shift of the contact point towards the blunt ends. Some other features like the symmetry breaking into heavily favoured anti-parallel configuration could not be resolved by the modified model yet. Unfortunately we could not determine if the NAHPR particles indeed do form bilayer phases, due to the very time-consuming calculations of the contact function and, hence, major equilibration issues. However, the introduction of non-additivity between blunt ends seems to be a pivotal factor to enable bilayer formation. The present issues might be resolved by further alternations of the NAHPR interactions. One solution might be to add additional angle dependence to the non-additivity, such that blunt ends are only able to overlap if the particles are pointing roughly in the same direction. This would probably diminish the formation of S-configurations. This, however, is in contrast with the original idea of prickly pear-shaped colloids, where this asymmetry seems hardly achievable. Another approach might be to replace the rounded pear surface with a partially flat surface. This would allow us to control not only the non-additivity attraction but also the depletion attraction via alignment by introducing more or less curvature to the surfaces.\\

As a final note of this paper series, we have to mention the importance of detail in self-assembly processes of complex structures again. Not only have we shown in the first part, based on the presence and absence of the gyroid phase in the PHGO and HPR model, respectively, that already small variations in particle shape can alter the phase behaviour of colloids drastically. We also shed light on the formation of bilayer-like gyroid structures in this paper. The depletion interactions reported here indicate that the bilayers are a result of a delicate interplay between the taper of the pear-shape and the self-non-additive features of the PHGO contact function. Therefore we argue that solely particle asymmetry is not sufficient but, in addition to self-non-additivity, necessary to create gyroid-like configurations.\\

\begin{acknowledgments}
We thank Universities Australia and the German Academic Exchange Service (DAAD) for funds through a collaboration funding scheme, through the grant ``Absorption and confinement of complex fluids''. We also thank the DFG through the ME1361/11-2 grant and through the research group ``Geometry and Physics of Spatial Random Systems'' (GPSRS) for funding. We gratefully acknowledge Klaus Mecke's support and advice in useful discussions. P.W.A.S. acknowledges a Murdoch University Postgraduate Research Scholarship. G.E.S-T is grateful to the Food Science Department at the University of Copenhagen and the Physical Chemistry group at Lund University for their hospitality and to Copenhagen University, the Camurus Lipid Research Foundation and the Danish National Bank for enabling a sabbatical stay in Denmark and Sweden.
\end{acknowledgments}

\subsection*{Data availability}
The data that supports the findings of this study are available within the article. Data set lists are available from the corresponding authors upon reasonable request.

\appendix
\section{Sampling algorithm}
\label{app:Appendix}

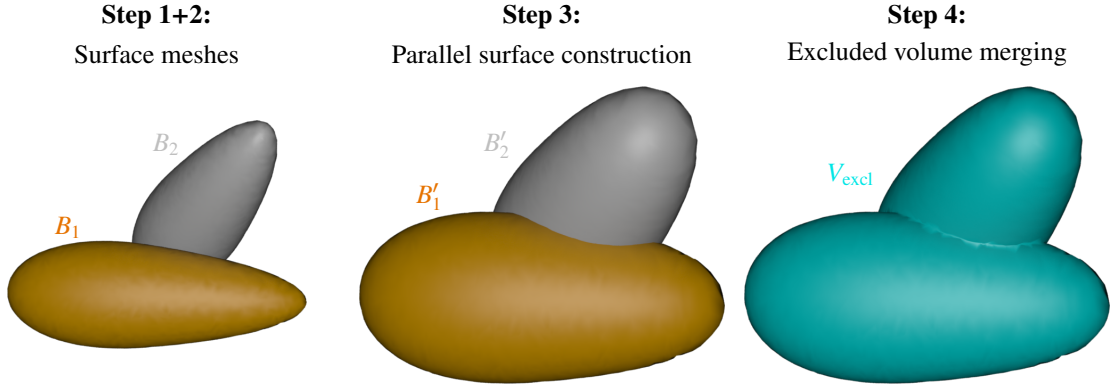
\begin{figure*}[t!]
\centering
\input{images/ExclVolAlg.tex}
\caption{The main steps of the algorithm to predict the ideal two pear-shaped particle arrangement in terms of excluded volume. In the first and second step (left) a configuration is chosen, and the surface meshes B1 and B2 of the pear-shaped particles are created. In the third step (centre) the individual excluded volumes of the pears $B'_1$ and $B'_2$ are created by constructing the parallel surface of $B_1$ and $B_2$. Afterwards, (right) the two meshes are merged and the total excluded volume $V_\text{excl}$ is computed. The steps are repeated until enough configurations are sampled.}
\label{fig:ExclVolAlg}
\end{figure*}

This appendix describes the sampling algorithm to determine the most compact arrangement between two pear-shaped particles. The most important steps are both sketched in \fig{fig:ExclVolAlg} and itemised below:
\begin{enumerate}
\item In the first step, an initial arrangement of two pear-shaped particles is chosen. We only consider arrangements where the two pears are in contact, as those configurations provide the minimal excluded volume for convex particles in terms of separation.
\item Afterwards, the surfaces of the particles are triangulated to create two separate meshes ($B_1$ and $B_2$) representing the pear shape.
\item In the next step, the parallel surfaces of the triangulations are generated. The vertices pt of the triangulations are translated in normal direction $\mathbf{\hat{n}}$ by $r_\text{depl}$.
\begin{equation}
\begin{aligned}
f_{\parallel,r_\text{depl}} &: B \rightarrow B' \\
f_{\parallel,r_\text{depl}}(p_t) &= p_t+r_\text{depl}\cdot \mathbf{\hat{n}}(p_t).
\end{aligned}
\end{equation}
The resulting new meshes $B'_1(r_\text{depl})$ and $B'_2(r_\text{depl})$ correspond to the interface separating the impenetrable and available space of virtual hard spheres with radius $r_\text{depl}$ caused by the first and second pear, respectively.
\item Subsequently, $B'_1(r_\text{depl})$ and $B'_2(r_\text{depl})$ are merged to calculate the collective excluded volume defined by
\begin{equation}
V_\text{excl}(r_\text{depl}) = B'_1(r_\text{depl}) \cup B'_2(r_\text{depl}).
\end{equation}
\item Another configuration, which has not been observed yet, is chosen and the algorithm returns to step 2. This procedure is repeated until the configuration space is sampled sufficiently densely.
\end{enumerate}

In this article this algorithm is applied to pears with aspect ratio $k{=}3$ and tapering parameter $\theta_k {=} 15^{\circ}$. Moreover, we use $r_\text{depl} {=}0.31\sigma_w$ , which corresponds to spheres with $V\text{sph} {=}0.08{\cdot} V_\text{pear}$ to create the data for \fig{fig:ExclVolSph} and \fig{fig:excluded_V}. The computations are performed using the ''Boolean operator'' of the 3D animation software tool Houdini \cite{Houdini} for creating intersections between mesh-representations of two pear-shaped particles.


\bibliographystyle{unsrt}
\bibliography{reference}
\end{document}

%% file: images/depletion_sketch.tex
\begin{tikzpicture}

\draw[very thick,->] (-2.5,0.2) -- (-2.5,-4) node[midway,rotate=90,anchor=north] {maximise entropy};
\draw[very thick,->] (12.2,0.2) -- (12.2,-4) node[midway,rotate=-90,anchor=north] {maximise entropy};

\begin{scope}[shift={(0,0.2)}]

	\draw[blue,thick,opacity=0.2] (-2,0.8) circle (0.3);		
	\draw[blue,thick,opacity=0.2] (-1.6,-1) circle (0.3);
	\draw[blue,thick,opacity=0.2] (-1,-1.3) circle (0.3);
	\draw[blue,thick,opacity=0.2] (-0.4,-0.75) circle (0.3);
	\draw[blue,thick,opacity=0.2] (-0.1,0.9) circle (0.3);
	\draw[blue,thick,opacity=0.2] (0,0.1) circle (0.3);
	\draw[blue,thick,opacity=0.2] (0.2,-1.5) circle (0.3);
	\draw[blue,thick,opacity=0.2] (0.3,-0.7) circle (0.3);
	\draw[blue,thick,opacity=0.2] (1.1,-1.1) circle (0.3);
	\draw[blue,thick,opacity=0.2] (1.7,1.3) circle (0.3);
	\draw[blue,thick,opacity=0.2] (1.8,-1.6) circle (0.3);
	\draw[blue,thick,opacity=0.2] (2,-0.75) circle (0.3);
	\draw[blue,thick,opacity=0.2] (2.2,0) circle (0.3);
	\draw[blue,thick,opacity=0.2] (2.4,0.7) circle (0.3);
	\draw[blue,thick,opacity=0.2] (2.65,1.35) circle (0.3);
	
	\node at (0,1.6) {excluded volume};
	\draw[very thick] (1.1,0) circle (0.7);	
	\draw[very thick] (-1.1,0) circle (0.7);	
	\draw[dashed,orange,very thick] (1.1,0) circle (1);	
	\draw[dashed,orange,very thick] (-1.1,0) circle (1);	
	
	\draw[thick,->] (0.5,1.35) -- (1.1,0.8);	
	\draw[thick,->] (-0.5,1.35) -- (-1.1,0.8);	
\end{scope}
\begin{scope}[shift={(0,-4)}]
	\begin{scope}
	
	\draw[blue,thick,opacity=0.2] (-1.9,0) circle (0.3);	
	\draw[blue,thick,opacity=0.2] (-1.7,2.3) circle (0.3);		
	\draw[blue,thick,opacity=0.2] (-1.6,-1) circle (0.3);
	\draw[blue,thick,opacity=0.2] (-1.4,1.2) circle (0.3);
	\draw[blue,thick,opacity=0.2] (-1,-1.3) circle (0.3);
	\draw[blue,thick,opacity=0.2] (-1,2) circle (0.3);
	\draw[blue,thick,opacity=0.2] (-0.6,1.3) circle (0.3);
	\draw[blue,thick,opacity=0.2] (-0.15,1.89) circle (0.3);	
	\draw[blue,thick,opacity=0.2] (0.2,1.15) circle (0.3);
	\draw[blue,thick,opacity=0.2] (0.3,-1.02) circle (0.3);
	\draw[blue,thick,opacity=0.2] (0.7,1.7) circle (0.3);
	\draw[blue,thick,opacity=0.2] (1,2.4) circle (0.3);	
	\draw[blue,thick,opacity=0.2] (1.1,-1.1) circle (0.3);
	\draw[blue,thick,opacity=0.2] (1.15,1.08) circle (0.3);
	\draw[blue,thick,opacity=0.2] (1.6,1.8) circle (0.3);	
	\draw[blue,thick,opacity=0.2] (1.7,0) circle (0.3);
	\draw[blue,thick,opacity=0.2] (1.8,0.7) circle (0.3);
	\draw[blue,thick,opacity=0.2] (2,-0.75) circle (0.3);
	\draw[blue,thick,opacity=0.2] (2.3,1.2) circle (0.3);
	\draw[blue,thick,opacity=0.2] (2.85,1.78) circle (0.3);

	\clip (-0.7,0) circle (1);	
	\draw[draw=none,fill=orange,opacity=0.3] (0.7,0) circle (1);
	\end{scope}
	\draw[very thick] (0.7,0) circle (0.7);	
	\draw[very thick] (-0.7,0) circle (0.7);	
	\draw[dashed,orange,very thick] (0.7,0) circle (1);	
	\draw[dashed,orange,very thick] (-0.7,0) circle (1);	
	
\end{scope}

\begin{scope}[shift={(4.5,0)}]

	\draw[blue,thick,opacity=0.2] (-2,-1.5) circle (0.3);
	\draw[blue,thick,opacity=0.2] (-1.5,0.3) circle (0.3);
	\draw[blue,thick,opacity=0.2] (-1.11,1.3) circle (0.3);
	\draw[blue,thick,opacity=0.2] (-1,-1.93) circle (0.3);
	\draw[blue,thick,opacity=0.2] (-0.4,0) circle (0.3);
	\draw[blue,thick,opacity=0.2] (-0.1,-0.75) circle (0.3);
	\draw[blue,thick,opacity=0.2] (0.2,-1.6) circle (0.3);
	\draw[blue,thick,opacity=0.2] (0.5,-0.1) circle (0.3);
	\draw[blue,thick,opacity=0.2] (0.9,1.7) circle (0.3);
	\draw[blue,thick,opacity=0.2] (1.1,-2.15) circle (0.3);
	\draw[blue,thick,opacity=0.2] (1.2,0.55) circle (0.3);
	\draw[blue,thick,opacity=0.2] (1.6,1.3) circle (0.3);
	\draw[blue,thick,opacity=0.2] (2,-1.75) circle (0.3);
	\draw[blue,thick,opacity=0.2] (2,0.05) circle (0.3);
	\draw[blue,thick,opacity=0.2] (2.1,0.7) circle (0.3);
	\draw[blue,thick,opacity=0.2] (2.2,1.65) circle (0.3);
	\draw[blue,thick,opacity=0.2] (2.4,-0.9) circle (0.3);

	\draw[very thick] (1.15,-1) circle (0.7);	
	\draw[very thick] (-1.15,-0.8) circle (0.7);	
	\draw[very thick] (0,1) circle (0.7);	
	\draw[dashed,orange,very thick] (1.15,-1) circle (1);	
	\draw[dashed,orange,very thick] (-1.15,-0.8) circle (1);	
	\draw[dashed,orange,very thick] (0,1) circle (1);	
\end{scope}

\begin{scope}[shift={(4.5,-4)}]

	\draw[blue,thick,opacity=0.2] (-1.9,-0.2) circle (0.3);
	\draw[blue,thick,opacity=0.2] (-1.8,-1.2) circle (0.3);
	\draw[blue,thick,opacity=0.2] (-1.3,0.5) circle (0.3);
	\draw[blue,thick,opacity=0.2] (-1.11,1.3) circle (0.3);
	\draw[blue,thick,opacity=0.2] (-0.2,1.8) circle (0.3);
	\draw[blue,thick,opacity=0.2] (0.2,-1.6) circle (0.3);
	\draw[blue,thick,opacity=0.2] (0.45,1.6) circle (0.3);
	\draw[blue,thick,opacity=0.2] (1.2,0.45) circle (0.3);
	\draw[blue,thick,opacity=0.2] (1.4,1.2) circle (0.3);
	\draw[blue,thick,opacity=0.2] (1.7,-1.4) circle (0.3);
	\draw[blue,thick,opacity=0.2] (2,0.05) circle (0.3);
	\draw[blue,thick,opacity=0.2] (2.1,-0.7) circle (0.3);
	\draw[blue,thick,opacity=0.2] (2.1,0.7) circle (0.3);
	\draw[blue,thick,opacity=0.2] (2.2,1.5) circle (0.3);

	\begin{scope}
	\clip (0.7,-0.60622) circle (1);		
	\draw[draw=none,fill=orange,opacity=0.3] (-0.7,-0.60622) circle (1);
	\end{scope};
	\begin{scope}
	\clip (-0.7,-0.60622) circle (1);
	\draw[draw=none,fill=orange,opacity=0.3] (0,0.60622) circle (1);	
	\end{scope};
	\begin{scope}
	\clip (0.7,-0.60622) circle (1);
	\draw[draw=none,fill=orange,opacity=0.3] (0,0.60622) circle (1);	
	\end{scope};
		
	\draw[very thick] (0.7,-0.60622) circle (0.7);	
	\draw[very thick] (-0.7,-0.60622) circle (0.7);	
	\draw[very thick] (0,0.60622) circle (0.7);	
	\draw[dashed,orange,very thick] (0.7,-0.60622) circle (1);	
	\draw[dashed,orange,very thick] (-0.7,-0.60622) circle (1);	
	\draw[dashed,orange,very thick] (0,0.60622) circle (1);	
\end{scope}

\begin{scope}[shift={(7.9,0.2)}]

	\draw[blue,thick,opacity=0.2] (-0.7,1) circle (0.3);
	\draw[blue,thick,opacity=0.2] (-0.6,-1.93) circle (0.3);
	\draw[blue,thick,opacity=0.2] (0.2,-2) circle (0.3);
	\draw[blue,thick,opacity=0.2] (0.7,1.5) circle (0.3);
	\draw[blue,thick,opacity=0.2] (1,0.7) circle (0.3);
	\draw[blue,thick,opacity=0.2] (1.08,-0.85) circle (0.3);
	\draw[blue,thick,opacity=0.2] (1.1,-0.1) circle (0.3);	
	\draw[blue,thick,opacity=0.2] (1.2,-1.65) circle (0.3);
	\draw[blue,thick,opacity=0.2] (1.6,1.55) circle (0.3);
	\draw[blue,thick,opacity=0.2] (1.8,-1.1) circle (0.3);
	\draw[blue,thick,opacity=0.2] (2,-1.9) circle (0.3);
	\draw[blue,thick,opacity=0.2] (2.35,1.3) circle (0.3);
	\draw[blue,thick,opacity=0.2] (2.7,-1.6) circle (0.3);
	\draw[blue,thick,opacity=0.2] (2.9,0.65) circle (0.3);
	\draw[blue,thick,opacity=0.2] (3.1,1.4) circle (0.3);
	\draw[blue,thick,opacity=0.2] (3.1,-2.25) circle (0.3);
	\draw[blue,thick,opacity=0.2] (3.5,-1.6) circle (0.3);
	\draw[blue,thick,opacity=0.2] (3.6,0) circle (0.3);
	\draw[blue,thick,opacity=0.2] (3.8,0.72) circle (0.3);

	\draw[very thick, shift={(2.3,0)},rotate=40] (-0.5,0) .. controls (-0.166666666,2) and (0.166666666,2) .. (0.5,0) .. controls (0.833333333,-2) and (-0.8333333333,-2) .. (-0.5,0) ;
    \draw[orange,dashed,thick,shift={(2.3,0)},rotate=40] (0,1.79959)--
(0.0563487,1.79304)--
(0.111487,1.77204)--
(0.164498,1.74021)--
(0.21447,1.69771)--
(0.261053,1.64823)--
(0.303952,1.59337)--
(0.343053,1.53473)--
(0.378782,1.47526)--
(0.411322,1.41578)--
(0.440731,1.35643)--
(0.467551,1.29867)--
(0.491973,1.24258)--
(0.514192,1.18823)--
(0.534402,1.13566)--
(0.552796,1.08492)--
(0.569569,1.03604)--
(0.584913,0.989034)--
(0.59902,0.943904)--
(0.61168,0.90006)--
(0.62344,0.858092)--
(0.634484,0.817972)--
(0.64499,0.779659)--
(0.654665,0.742572)--
(0.66363,0.706694)--
(0.672,0.672)--
(0.680405,0.638943)--
(0.68794,0.606501)--
(0.695754,0.575578)--
(0.702875,0.545206)--
(0.709928,0.515793)--
(0.716422,0.48688)--
(0.722406,0.458453)--
(0.728538,0.430856)--
(0.733657,0.403331)--
(0.738403,0.376235)--
(0.743458,0.349845)--
(0.747591,0.323512)--
(0.751467,0.297527)--
(0.755784,0.272099)--
(0.759923,0.246914)--
(0.763909,0.221936)--
(0.767764,0.197128)--
(0.772197,0.172606)--
(0.776543,0.148133)--
(0.780812,0.123669)--
(0.785715,0.0992589)--
(0.789853,0.0746631)--
(0.794631,0.049994)--
(0.798636,0.0250982)--
(0.802566,-1.02009e-15)--
(0.805704,-0.0253203)--
(0.808745,-0.0508819)--
(0.811676,-0.076726)--
(0.814478,-0.102892)--
(0.817129,-0.129421)--
(0.818912,-0.156216)--
(0.821192,-0.183558)--
(0.82324,-0.211372)--
(0.825701,-0.239888)--
(0.827846,-0.268983)--
(0.829633,-0.298686)--
(0.831676,-0.329284)--
(0.833901,-0.360861)--
(0.835591,-0.393199)--
(0.836689,-0.426314)--
(0.837757,-0.460561)--
(0.837484,-0.495287)--
(0.837633,-0.531578)--
(0.836313,-0.568358)--
(0.834638,-0.6064)--
(0.832499,-0.645752)--
(0.829239,-0.686006)--
(0.825846,-0.728081)--
(0.821126,-0.771088)--
(0.8155,-0.8155)--
(0.80836,-0.860816)--
(0.800562,-0.90806)--
(0.791476,-0.956731)--
(0.780537,-1.00626)--
(0.768079,-1.05717)--
(0.753571,-1.10885)--
(0.736935,-1.16122)--
(0.718094,-1.21423)--
(0.696291,-1.26655)--
(0.672215,-1.3193)--
(0.644895,-1.37047)--
(0.615009,-1.4212)--
(0.582038,-1.47006)--
(0.546114,-1.51689)--
(0.507376,-1.56154)--
(0.465967,-1.60387)--
(0.421689,-1.64237)--
(0.375137,-1.67827)--
(0.326079,-1.70937)--
(0.275102,-1.73692)--
(0.222269,-1.75944)--
(0.168025,-1.77752)--
(0.112642,-1.79039)--
(0.0565264,-1.7987)--
(8.21861e-15,-1.801)--
(-0.0565264,-1.7987)--
(-0.112642,-1.79039)--
(-0.168025,-1.77752)--
(-0.222269,-1.75944)--
(-0.275102,-1.73692)--
(-0.326079,-1.70937)--
(-0.375137,-1.67827)--
(-0.421689,-1.64237)--
(-0.465967,-1.60387)--
(-0.507376,-1.56154)--
(-0.546114,-1.51689)--
(-0.582038,-1.47006)--
(-0.615009,-1.4212)--
(-0.644895,-1.37047)--
(-0.672215,-1.3193)--
(-0.696291,-1.26655)--
(-0.718094,-1.21423)--
(-0.736935,-1.16122)--
(-0.753571,-1.10885)--
(-0.768079,-1.05717)--
(-0.780537,-1.00626)--
(-0.791476,-0.956731)--
(-0.800562,-0.90806)--
(-0.80836,-0.860816)--
(-0.8155,-0.8155)--
(-0.821126,-0.771088)--
(-0.825846,-0.728081)--
(-0.829239,-0.686006)--
(-0.832499,-0.645752)--
(-0.834638,-0.6064)--
(-0.836313,-0.568358)--
(-0.837633,-0.531578)--
(-0.837484,-0.495287)--
(-0.837757,-0.460561)--
(-0.836689,-0.426314)--
(-0.835591,-0.393199)--
(-0.833901,-0.360861)--
(-0.831676,-0.329284)--
(-0.829633,-0.298686)--
(-0.827846,-0.268983)--
(-0.825701,-0.239888)--
(-0.82324,-0.211372)--
(-0.821192,-0.183558)--
(-0.818912,-0.156216)--
(-0.817129,-0.129421)--
(-0.814478,-0.102892)--
(-0.811676,-0.076726)--
(-0.808745,-0.0508819)--
(-0.805704,-0.0253203)--
(-0.802566,-1.08398e-14)--
(-0.798636,0.0250982)--
(-0.794631,0.049994)--
(-0.789853,0.0746631)--
(-0.785715,0.0992589)--
(-0.780812,0.123669)--
(-0.776543,0.148133)--
(-0.772197,0.172606)--
(-0.767764,0.197128)--
(-0.763909,0.221936)--
(-0.759923,0.246914)--
(-0.755784,0.272099)--
(-0.751467,0.297527)--
(-0.747591,0.323512)--
(-0.743458,0.349845)--
(-0.738403,0.376235)--
(-0.733657,0.403331)--
(-0.728538,0.430856)--
(-0.722406,0.458453)--
(-0.716422,0.48688)--
(-0.709928,0.515793)--
(-0.702875,0.545206)--
(-0.695754,0.575578)--
(-0.68794,0.606501)--
(-0.680405,0.638943)--
(-0.672,0.672)--
(-0.66363,0.706694)--
(-0.654665,0.742572)--
(-0.64499,0.779659)--
(-0.634484,0.817972)--
(-0.62344,0.858092)--
(-0.61168,0.90006)--
(-0.59902,0.943904)--
(-0.584913,0.989034)--
(-0.569569,1.03604)--
(-0.552796,1.08492)--
(-0.534402,1.13566)--
(-0.514192,1.18823)--
(-0.491973,1.24258)--
(-0.467551,1.29867)--
(-0.440731,1.35643)--
(-0.411322,1.41578)--
(-0.378782,1.47526)--
(-0.343053,1.53473)--
(-0.303952,1.59337)--
(-0.261053,1.64823)--
(-0.21447,1.69771)--
(-0.164498,1.74021)--
(-0.111487,1.77204)--
(-0.0563487,1.79304)--
(-4.03996e-14,1.79959);

	\draw[very thick] (-0.5,0) .. controls (-0.166666666,2) and (0.166666666,2) .. (0.5,0) .. controls (0.833333333,-2) and (-0.8333333333,-2) .. (-0.5,0) ;
    \draw[orange,dashed,thick] (0,1.79959)--
(0.0563487,1.79304)--
(0.111487,1.77204)--
(0.164498,1.74021)--
(0.21447,1.69771)--
(0.261053,1.64823)--
(0.303952,1.59337)--
(0.343053,1.53473)--
(0.378782,1.47526)--
(0.411322,1.41578)--
(0.440731,1.35643)--
(0.467551,1.29867)--
(0.491973,1.24258)--
(0.514192,1.18823)--
(0.534402,1.13566)--
(0.552796,1.08492)--
(0.569569,1.03604)--
(0.584913,0.989034)--
(0.59902,0.943904)--
(0.61168,0.90006)--
(0.62344,0.858092)--
(0.634484,0.817972)--
(0.64499,0.779659)--
(0.654665,0.742572)--
(0.66363,0.706694)--
(0.672,0.672)--
(0.680405,0.638943)--
(0.68794,0.606501)--
(0.695754,0.575578)--
(0.702875,0.545206)--
(0.709928,0.515793)--
(0.716422,0.48688)--
(0.722406,0.458453)--
(0.728538,0.430856)--
(0.733657,0.403331)--
(0.738403,0.376235)--
(0.743458,0.349845)--
(0.747591,0.323512)--
(0.751467,0.297527)--
(0.755784,0.272099)--
(0.759923,0.246914)--
(0.763909,0.221936)--
(0.767764,0.197128)--
(0.772197,0.172606)--
(0.776543,0.148133)--
(0.780812,0.123669)--
(0.785715,0.0992589)--
(0.789853,0.0746631)--
(0.794631,0.049994)--
(0.798636,0.0250982)--
(0.802566,-1.02009e-15)--
(0.805704,-0.0253203)--
(0.808745,-0.0508819)--
(0.811676,-0.076726)--
(0.814478,-0.102892)--
(0.817129,-0.129421)--
(0.818912,-0.156216)--
(0.821192,-0.183558)--
(0.82324,-0.211372)--
(0.825701,-0.239888)--
(0.827846,-0.268983)--
(0.829633,-0.298686)--
(0.831676,-0.329284)--
(0.833901,-0.360861)--
(0.835591,-0.393199)--
(0.836689,-0.426314)--
(0.837757,-0.460561)--
(0.837484,-0.495287)--
(0.837633,-0.531578)--
(0.836313,-0.568358)--
(0.834638,-0.6064)--
(0.832499,-0.645752)--
(0.829239,-0.686006)--
(0.825846,-0.728081)--
(0.821126,-0.771088)--
(0.8155,-0.8155)--
(0.80836,-0.860816)--
(0.800562,-0.90806)--
(0.791476,-0.956731)--
(0.780537,-1.00626)--
(0.768079,-1.05717)--
(0.753571,-1.10885)--
(0.736935,-1.16122)--
(0.718094,-1.21423)--
(0.696291,-1.26655)--
(0.672215,-1.3193)--
(0.644895,-1.37047)--
(0.615009,-1.4212)--
(0.582038,-1.47006)--
(0.546114,-1.51689)--
(0.507376,-1.56154)--
(0.465967,-1.60387)--
(0.421689,-1.64237)--
(0.375137,-1.67827)--
(0.326079,-1.70937)--
(0.275102,-1.73692)--
(0.222269,-1.75944)--
(0.168025,-1.77752)--
(0.112642,-1.79039)--
(0.0565264,-1.7987)--
(8.21861e-15,-1.801)--
(-0.0565264,-1.7987)--
(-0.112642,-1.79039)--
(-0.168025,-1.77752)--
(-0.222269,-1.75944)--
(-0.275102,-1.73692)--
(-0.326079,-1.70937)--
(-0.375137,-1.67827)--
(-0.421689,-1.64237)--
(-0.465967,-1.60387)--
(-0.507376,-1.56154)--
(-0.546114,-1.51689)--
(-0.582038,-1.47006)--
(-0.615009,-1.4212)--
(-0.644895,-1.37047)--
(-0.672215,-1.3193)--
(-0.696291,-1.26655)--
(-0.718094,-1.21423)--
(-0.736935,-1.16122)--
(-0.753571,-1.10885)--
(-0.768079,-1.05717)--
(-0.780537,-1.00626)--
(-0.791476,-0.956731)--
(-0.800562,-0.90806)--
(-0.80836,-0.860816)--
(-0.8155,-0.8155)--
(-0.821126,-0.771088)--
(-0.825846,-0.728081)--
(-0.829239,-0.686006)--
(-0.832499,-0.645752)--
(-0.834638,-0.6064)--
(-0.836313,-0.568358)--
(-0.837633,-0.531578)--
(-0.837484,-0.495287)--
(-0.837757,-0.460561)--
(-0.836689,-0.426314)--
(-0.835591,-0.393199)--
(-0.833901,-0.360861)--
(-0.831676,-0.329284)--
(-0.829633,-0.298686)--
(-0.827846,-0.268983)--
(-0.825701,-0.239888)--
(-0.82324,-0.211372)--
(-0.821192,-0.183558)--
(-0.818912,-0.156216)--
(-0.817129,-0.129421)--
(-0.814478,-0.102892)--
(-0.811676,-0.076726)--
(-0.808745,-0.0508819)--
(-0.805704,-0.0253203)--
(-0.802566,-1.08398e-14)--
(-0.798636,0.0250982)--
(-0.794631,0.049994)--
(-0.789853,0.0746631)--
(-0.785715,0.0992589)--
(-0.780812,0.123669)--
(-0.776543,0.148133)--
(-0.772197,0.172606)--
(-0.767764,0.197128)--
(-0.763909,0.221936)--
(-0.759923,0.246914)--
(-0.755784,0.272099)--
(-0.751467,0.297527)--
(-0.747591,0.323512)--
(-0.743458,0.349845)--
(-0.738403,0.376235)--
(-0.733657,0.403331)--
(-0.728538,0.430856)--
(-0.722406,0.458453)--
(-0.716422,0.48688)--
(-0.709928,0.515793)--
(-0.702875,0.545206)--
(-0.695754,0.575578)--
(-0.68794,0.606501)--
(-0.680405,0.638943)--
(-0.672,0.672)--
(-0.66363,0.706694)--
(-0.654665,0.742572)--
(-0.64499,0.779659)--
(-0.634484,0.817972)--
(-0.62344,0.858092)--
(-0.61168,0.90006)--
(-0.59902,0.943904)--
(-0.584913,0.989034)--
(-0.569569,1.03604)--
(-0.552796,1.08492)--
(-0.534402,1.13566)--
(-0.514192,1.18823)--
(-0.491973,1.24258)--
(-0.467551,1.29867)--
(-0.440731,1.35643)--
(-0.411322,1.41578)--
(-0.378782,1.47526)--
(-0.343053,1.53473)--
(-0.303952,1.59337)--
(-0.261053,1.64823)--
(-0.21447,1.69771)--
(-0.164498,1.74021)--
(-0.111487,1.77204)--
(-0.0563487,1.79304)--
(-4.03996e-14,1.79959);
			
\end{scope}

\begin{scope}[shift={(8.7,-4)}]

	\draw[blue,thick,opacity=0.2] (-1.5,-0.3) circle (0.3);
	\draw[blue,thick,opacity=0.2] (-1.4,0.5) circle (0.3);
	\draw[blue,thick,opacity=0.2] (-1.28,1.5) circle (0.3);	
	\draw[blue,thick,opacity=0.2] (-1.24,-1.2) circle (0.3);
	\draw[blue,thick,opacity=0.2] (-0.8,0.9) circle (0.3);
	\draw[blue,thick,opacity=0.2] (-0.1,1.65) circle (0.3);
	\draw[blue,thick,opacity=0.2] (0.5,-1.55) circle (0.3);
	\draw[blue,thick,opacity=0.2] (0.6,1.85) circle (0.3);
	\draw[blue,thick,opacity=0.2] (1.6,1.55) circle (0.3);	
	\draw[blue,thick,opacity=0.2] (1.7,0.7) circle (0.3);
	\draw[blue,thick,opacity=0.2] (2,-1.5) circle (0.3);
	\draw[blue,thick,opacity=0.2] (2.2,-0.1) circle (0.3);	
	\draw[blue,thick,opacity=0.2] (2.23,-0.85) circle (0.3);
	\draw[blue,thick,opacity=0.2] (2.35,1.18) circle (0.3);
	\draw[blue,thick,opacity=0.2] (2.8,-1.3) circle (0.3);
	\draw[blue,thick,opacity=0.2] (2.83,0.3) circle (0.3);
	\draw[blue,thick,opacity=0.2] (3.1,-0.6) circle (0.3);
\begin{scope}
    \clip[shift={(1.03,0)},rotate=10] (0,1.79959)--
(0.0563487,1.79304)--
(0.111487,1.77204)--
(0.164498,1.74021)--
(0.21447,1.69771)--
(0.261053,1.64823)--
(0.303952,1.59337)--
(0.343053,1.53473)--
(0.378782,1.47526)--
(0.411322,1.41578)--
(0.440731,1.35643)--
(0.467551,1.29867)--
(0.491973,1.24258)--
(0.514192,1.18823)--
(0.534402,1.13566)--
(0.552796,1.08492)--
(0.569569,1.03604)--
(0.584913,0.989034)--
(0.59902,0.943904)--
(0.61168,0.90006)--
(0.62344,0.858092)--
(0.634484,0.817972)--
(0.64499,0.779659)--
(0.654665,0.742572)--
(0.66363,0.706694)--
(0.672,0.672)--
(0.680405,0.638943)--
(0.68794,0.606501)--
(0.695754,0.575578)--
(0.702875,0.545206)--
(0.709928,0.515793)--
(0.716422,0.48688)--
(0.722406,0.458453)--
(0.728538,0.430856)--
(0.733657,0.403331)--
(0.738403,0.376235)--
(0.743458,0.349845)--
(0.747591,0.323512)--
(0.751467,0.297527)--
(0.755784,0.272099)--
(0.759923,0.246914)--
(0.763909,0.221936)--
(0.767764,0.197128)--
(0.772197,0.172606)--
(0.776543,0.148133)--
(0.780812,0.123669)--
(0.785715,0.0992589)--
(0.789853,0.0746631)--
(0.794631,0.049994)--
(0.798636,0.0250982)--
(0.802566,-1.02009e-15)--
(0.805704,-0.0253203)--
(0.808745,-0.0508819)--
(0.811676,-0.076726)--
(0.814478,-0.102892)--
(0.817129,-0.129421)--
(0.818912,-0.156216)--
(0.821192,-0.183558)--
(0.82324,-0.211372)--
(0.825701,-0.239888)--
(0.827846,-0.268983)--
(0.829633,-0.298686)--
(0.831676,-0.329284)--
(0.833901,-0.360861)--
(0.835591,-0.393199)--
(0.836689,-0.426314)--
(0.837757,-0.460561)--
(0.837484,-0.495287)--
(0.837633,-0.531578)--
(0.836313,-0.568358)--
(0.834638,-0.6064)--
(0.832499,-0.645752)--
(0.829239,-0.686006)--
(0.825846,-0.728081)--
(0.821126,-0.771088)--
(0.8155,-0.8155)--
(0.80836,-0.860816)--
(0.800562,-0.90806)--
(0.791476,-0.956731)--
(0.780537,-1.00626)--
(0.768079,-1.05717)--
(0.753571,-1.10885)--
(0.736935,-1.16122)--
(0.718094,-1.21423)--
(0.696291,-1.26655)--
(0.672215,-1.3193)--
(0.644895,-1.37047)--
(0.615009,-1.4212)--
(0.582038,-1.47006)--
(0.546114,-1.51689)--
(0.507376,-1.56154)--
(0.465967,-1.60387)--
(0.421689,-1.64237)--
(0.375137,-1.67827)--
(0.326079,-1.70937)--
(0.275102,-1.73692)--
(0.222269,-1.75944)--
(0.168025,-1.77752)--
(0.112642,-1.79039)--
(0.0565264,-1.7987)--
(8.21861e-15,-1.801)--
(-0.0565264,-1.7987)--
(-0.112642,-1.79039)--
(-0.168025,-1.77752)--
(-0.222269,-1.75944)--
(-0.275102,-1.73692)--
(-0.326079,-1.70937)--
(-0.375137,-1.67827)--
(-0.421689,-1.64237)--
(-0.465967,-1.60387)--
(-0.507376,-1.56154)--
(-0.546114,-1.51689)--
(-0.582038,-1.47006)--
(-0.615009,-1.4212)--
(-0.644895,-1.37047)--
(-0.672215,-1.3193)--
(-0.696291,-1.26655)--
(-0.718094,-1.21423)--
(-0.736935,-1.16122)--
(-0.753571,-1.10885)--
(-0.768079,-1.05717)--
(-0.780537,-1.00626)--
(-0.791476,-0.956731)--
(-0.800562,-0.90806)--
(-0.80836,-0.860816)--
(-0.8155,-0.8155)--
(-0.821126,-0.771088)--
(-0.825846,-0.728081)--
(-0.829239,-0.686006)--
(-0.832499,-0.645752)--
(-0.834638,-0.6064)--
(-0.836313,-0.568358)--
(-0.837633,-0.531578)--
(-0.837484,-0.495287)--
(-0.837757,-0.460561)--
(-0.836689,-0.426314)--
(-0.835591,-0.393199)--
(-0.833901,-0.360861)--
(-0.831676,-0.329284)--
(-0.829633,-0.298686)--
(-0.827846,-0.268983)--
(-0.825701,-0.239888)--
(-0.82324,-0.211372)--
(-0.821192,-0.183558)--
(-0.818912,-0.156216)--
(-0.817129,-0.129421)--
(-0.814478,-0.102892)--
(-0.811676,-0.076726)--
(-0.808745,-0.0508819)--
(-0.805704,-0.0253203)--
(-0.802566,-1.08398e-14)--
(-0.798636,0.0250982)--
(-0.794631,0.049994)--
(-0.789853,0.0746631)--
(-0.785715,0.0992589)--
(-0.780812,0.123669)--
(-0.776543,0.148133)--
(-0.772197,0.172606)--
(-0.767764,0.197128)--
(-0.763909,0.221936)--
(-0.759923,0.246914)--
(-0.755784,0.272099)--
(-0.751467,0.297527)--
(-0.747591,0.323512)--
(-0.743458,0.349845)--
(-0.738403,0.376235)--
(-0.733657,0.403331)--
(-0.728538,0.430856)--
(-0.722406,0.458453)--
(-0.716422,0.48688)--
(-0.709928,0.515793)--
(-0.702875,0.545206)--
(-0.695754,0.575578)--
(-0.68794,0.606501)--
(-0.680405,0.638943)--
(-0.672,0.672)--
(-0.66363,0.706694)--
(-0.654665,0.742572)--
(-0.64499,0.779659)--
(-0.634484,0.817972)--
(-0.62344,0.858092)--
(-0.61168,0.90006)--
(-0.59902,0.943904)--
(-0.584913,0.989034)--
(-0.569569,1.03604)--
(-0.552796,1.08492)--
(-0.534402,1.13566)--
(-0.514192,1.18823)--
(-0.491973,1.24258)--
(-0.467551,1.29867)--
(-0.440731,1.35643)--
(-0.411322,1.41578)--
(-0.378782,1.47526)--
(-0.343053,1.53473)--
(-0.303952,1.59337)--
(-0.261053,1.64823)--
(-0.21447,1.69771)--
(-0.164498,1.74021)--
(-0.111487,1.77204)--
(-0.0563487,1.79304)--
(-4.03996e-14,1.79959);

    \draw[fill=orange,opacity=0.3,draw=none,rotate=-10] (0,1.79959)--
(0.0563487,1.79304)--
(0.111487,1.77204)--
(0.164498,1.74021)--
(0.21447,1.69771)--
(0.261053,1.64823)--
(0.303952,1.59337)--
(0.343053,1.53473)--
(0.378782,1.47526)--
(0.411322,1.41578)--
(0.440731,1.35643)--
(0.467551,1.29867)--
(0.491973,1.24258)--
(0.514192,1.18823)--
(0.534402,1.13566)--
(0.552796,1.08492)--
(0.569569,1.03604)--
(0.584913,0.989034)--
(0.59902,0.943904)--
(0.61168,0.90006)--
(0.62344,0.858092)--
(0.634484,0.817972)--
(0.64499,0.779659)--
(0.654665,0.742572)--
(0.66363,0.706694)--
(0.672,0.672)--
(0.680405,0.638943)--
(0.68794,0.606501)--
(0.695754,0.575578)--
(0.702875,0.545206)--
(0.709928,0.515793)--
(0.716422,0.48688)--
(0.722406,0.458453)--
(0.728538,0.430856)--
(0.733657,0.403331)--
(0.738403,0.376235)--
(0.743458,0.349845)--
(0.747591,0.323512)--
(0.751467,0.297527)--
(0.755784,0.272099)--
(0.759923,0.246914)--
(0.763909,0.221936)--
(0.767764,0.197128)--
(0.772197,0.172606)--
(0.776543,0.148133)--
(0.780812,0.123669)--
(0.785715,0.0992589)--
(0.789853,0.0746631)--
(0.794631,0.049994)--
(0.798636,0.0250982)--
(0.802566,-1.02009e-15)--
(0.805704,-0.0253203)--
(0.808745,-0.0508819)--
(0.811676,-0.076726)--
(0.814478,-0.102892)--
(0.817129,-0.129421)--
(0.818912,-0.156216)--
(0.821192,-0.183558)--
(0.82324,-0.211372)--
(0.825701,-0.239888)--
(0.827846,-0.268983)--
(0.829633,-0.298686)--
(0.831676,-0.329284)--
(0.833901,-0.360861)--
(0.835591,-0.393199)--
(0.836689,-0.426314)--
(0.837757,-0.460561)--
(0.837484,-0.495287)--
(0.837633,-0.531578)--
(0.836313,-0.568358)--
(0.834638,-0.6064)--
(0.832499,-0.645752)--
(0.829239,-0.686006)--
(0.825846,-0.728081)--
(0.821126,-0.771088)--
(0.8155,-0.8155)--
(0.80836,-0.860816)--
(0.800562,-0.90806)--
(0.791476,-0.956731)--
(0.780537,-1.00626)--
(0.768079,-1.05717)--
(0.753571,-1.10885)--
(0.736935,-1.16122)--
(0.718094,-1.21423)--
(0.696291,-1.26655)--
(0.672215,-1.3193)--
(0.644895,-1.37047)--
(0.615009,-1.4212)--
(0.582038,-1.47006)--
(0.546114,-1.51689)--
(0.507376,-1.56154)--
(0.465967,-1.60387)--
(0.421689,-1.64237)--
(0.375137,-1.67827)--
(0.326079,-1.70937)--
(0.275102,-1.73692)--
(0.222269,-1.75944)--
(0.168025,-1.77752)--
(0.112642,-1.79039)--
(0.0565264,-1.7987)--
(8.21861e-15,-1.801)--
(-0.0565264,-1.7987)--
(-0.112642,-1.79039)--
(-0.168025,-1.77752)--
(-0.222269,-1.75944)--
(-0.275102,-1.73692)--
(-0.326079,-1.70937)--
(-0.375137,-1.67827)--
(-0.421689,-1.64237)--
(-0.465967,-1.60387)--
(-0.507376,-1.56154)--
(-0.546114,-1.51689)--
(-0.582038,-1.47006)--
(-0.615009,-1.4212)--
(-0.644895,-1.37047)--
(-0.672215,-1.3193)--
(-0.696291,-1.26655)--
(-0.718094,-1.21423)--
(-0.736935,-1.16122)--
(-0.753571,-1.10885)--
(-0.768079,-1.05717)--
(-0.780537,-1.00626)--
(-0.791476,-0.956731)--
(-0.800562,-0.90806)--
(-0.80836,-0.860816)--
(-0.8155,-0.8155)--
(-0.821126,-0.771088)--
(-0.825846,-0.728081)--
(-0.829239,-0.686006)--
(-0.832499,-0.645752)--
(-0.834638,-0.6064)--
(-0.836313,-0.568358)--
(-0.837633,-0.531578)--
(-0.837484,-0.495287)--
(-0.837757,-0.460561)--
(-0.836689,-0.426314)--
(-0.835591,-0.393199)--
(-0.833901,-0.360861)--
(-0.831676,-0.329284)--
(-0.829633,-0.298686)--
(-0.827846,-0.268983)--
(-0.825701,-0.239888)--
(-0.82324,-0.211372)--
(-0.821192,-0.183558)--
(-0.818912,-0.156216)--
(-0.817129,-0.129421)--
(-0.814478,-0.102892)--
(-0.811676,-0.076726)--
(-0.808745,-0.0508819)--
(-0.805704,-0.0253203)--
(-0.802566,-1.08398e-14)--
(-0.798636,0.0250982)--
(-0.794631,0.049994)--
(-0.789853,0.0746631)--
(-0.785715,0.0992589)--
(-0.780812,0.123669)--
(-0.776543,0.148133)--
(-0.772197,0.172606)--
(-0.767764,0.197128)--
(-0.763909,0.221936)--
(-0.759923,0.246914)--
(-0.755784,0.272099)--
(-0.751467,0.297527)--
(-0.747591,0.323512)--
(-0.743458,0.349845)--
(-0.738403,0.376235)--
(-0.733657,0.403331)--
(-0.728538,0.430856)--
(-0.722406,0.458453)--
(-0.716422,0.48688)--
(-0.709928,0.515793)--
(-0.702875,0.545206)--
(-0.695754,0.575578)--
(-0.68794,0.606501)--
(-0.680405,0.638943)--
(-0.672,0.672)--
(-0.66363,0.706694)--
(-0.654665,0.742572)--
(-0.64499,0.779659)--
(-0.634484,0.817972)--
(-0.62344,0.858092)--
(-0.61168,0.90006)--
(-0.59902,0.943904)--
(-0.584913,0.989034)--
(-0.569569,1.03604)--
(-0.552796,1.08492)--
(-0.534402,1.13566)--
(-0.514192,1.18823)--
(-0.491973,1.24258)--
(-0.467551,1.29867)--
(-0.440731,1.35643)--
(-0.411322,1.41578)--
(-0.378782,1.47526)--
(-0.343053,1.53473)--
(-0.303952,1.59337)--
(-0.261053,1.64823)--
(-0.21447,1.69771)--
(-0.164498,1.74021)--
(-0.111487,1.77204)--
(-0.0563487,1.79304)--
(-4.03996e-14,1.79959);
\end{scope}
	
	\draw[very thick, shift={(1.03,0)},rotate=10] (-0.5,0) .. controls (-0.166666666,2) and (0.166666666,2) .. (0.5,0) .. controls (0.833333333,-2) and (-0.8333333333,-2) .. (-0.5,0) ;
	    \draw[orange,dashed,thick,shift={(1.03,0)},rotate=10] (0,1.79959)--
(0.0563487,1.79304)--
(0.111487,1.77204)--
(0.164498,1.74021)--
(0.21447,1.69771)--
(0.261053,1.64823)--
(0.303952,1.59337)--
(0.343053,1.53473)--
(0.378782,1.47526)--
(0.411322,1.41578)--
(0.440731,1.35643)--
(0.467551,1.29867)--
(0.491973,1.24258)--
(0.514192,1.18823)--
(0.534402,1.13566)--
(0.552796,1.08492)--
(0.569569,1.03604)--
(0.584913,0.989034)--
(0.59902,0.943904)--
(0.61168,0.90006)--
(0.62344,0.858092)--
(0.634484,0.817972)--
(0.64499,0.779659)--
(0.654665,0.742572)--
(0.66363,0.706694)--
(0.672,0.672)--
(0.680405,0.638943)--
(0.68794,0.606501)--
(0.695754,0.575578)--
(0.702875,0.545206)--
(0.709928,0.515793)--
(0.716422,0.48688)--
(0.722406,0.458453)--
(0.728538,0.430856)--
(0.733657,0.403331)--
(0.738403,0.376235)--
(0.743458,0.349845)--
(0.747591,0.323512)--
(0.751467,0.297527)--
(0.755784,0.272099)--
(0.759923,0.246914)--
(0.763909,0.221936)--
(0.767764,0.197128)--
(0.772197,0.172606)--
(0.776543,0.148133)--
(0.780812,0.123669)--
(0.785715,0.0992589)--
(0.789853,0.0746631)--
(0.794631,0.049994)--
(0.798636,0.0250982)--
(0.802566,-1.02009e-15)--
(0.805704,-0.0253203)--
(0.808745,-0.0508819)--
(0.811676,-0.076726)--
(0.814478,-0.102892)--
(0.817129,-0.129421)--
(0.818912,-0.156216)--
(0.821192,-0.183558)--
(0.82324,-0.211372)--
(0.825701,-0.239888)--
(0.827846,-0.268983)--
(0.829633,-0.298686)--
(0.831676,-0.329284)--
(0.833901,-0.360861)--
(0.835591,-0.393199)--
(0.836689,-0.426314)--
(0.837757,-0.460561)--
(0.837484,-0.495287)--
(0.837633,-0.531578)--
(0.836313,-0.568358)--
(0.834638,-0.6064)--
(0.832499,-0.645752)--
(0.829239,-0.686006)--
(0.825846,-0.728081)--
(0.821126,-0.771088)--
(0.8155,-0.8155)--
(0.80836,-0.860816)--
(0.800562,-0.90806)--
(0.791476,-0.956731)--
(0.780537,-1.00626)--
(0.768079,-1.05717)--
(0.753571,-1.10885)--
(0.736935,-1.16122)--
(0.718094,-1.21423)--
(0.696291,-1.26655)--
(0.672215,-1.3193)--
(0.644895,-1.37047)--
(0.615009,-1.4212)--
(0.582038,-1.47006)--
(0.546114,-1.51689)--
(0.507376,-1.56154)--
(0.465967,-1.60387)--
(0.421689,-1.64237)--
(0.375137,-1.67827)--
(0.326079,-1.70937)--
(0.275102,-1.73692)--
(0.222269,-1.75944)--
(0.168025,-1.77752)--
(0.112642,-1.79039)--
(0.0565264,-1.7987)--
(8.21861e-15,-1.801)--
(-0.0565264,-1.7987)--
(-0.112642,-1.79039)--
(-0.168025,-1.77752)--
(-0.222269,-1.75944)--
(-0.275102,-1.73692)--
(-0.326079,-1.70937)--
(-0.375137,-1.67827)--
(-0.421689,-1.64237)--
(-0.465967,-1.60387)--
(-0.507376,-1.56154)--
(-0.546114,-1.51689)--
(-0.582038,-1.47006)--
(-0.615009,-1.4212)--
(-0.644895,-1.37047)--
(-0.672215,-1.3193)--
(-0.696291,-1.26655)--
(-0.718094,-1.21423)--
(-0.736935,-1.16122)--
(-0.753571,-1.10885)--
(-0.768079,-1.05717)--
(-0.780537,-1.00626)--
(-0.791476,-0.956731)--
(-0.800562,-0.90806)--
(-0.80836,-0.860816)--
(-0.8155,-0.8155)--
(-0.821126,-0.771088)--
(-0.825846,-0.728081)--
(-0.829239,-0.686006)--
(-0.832499,-0.645752)--
(-0.834638,-0.6064)--
(-0.836313,-0.568358)--
(-0.837633,-0.531578)--
(-0.837484,-0.495287)--
(-0.837757,-0.460561)--
(-0.836689,-0.426314)--
(-0.835591,-0.393199)--
(-0.833901,-0.360861)--
(-0.831676,-0.329284)--
(-0.829633,-0.298686)--
(-0.827846,-0.268983)--
(-0.825701,-0.239888)--
(-0.82324,-0.211372)--
(-0.821192,-0.183558)--
(-0.818912,-0.156216)--
(-0.817129,-0.129421)--
(-0.814478,-0.102892)--
(-0.811676,-0.076726)--
(-0.808745,-0.0508819)--
(-0.805704,-0.0253203)--
(-0.802566,-1.08398e-14)--
(-0.798636,0.0250982)--
(-0.794631,0.049994)--
(-0.789853,0.0746631)--
(-0.785715,0.0992589)--
(-0.780812,0.123669)--
(-0.776543,0.148133)--
(-0.772197,0.172606)--
(-0.767764,0.197128)--
(-0.763909,0.221936)--
(-0.759923,0.246914)--
(-0.755784,0.272099)--
(-0.751467,0.297527)--
(-0.747591,0.323512)--
(-0.743458,0.349845)--
(-0.738403,0.376235)--
(-0.733657,0.403331)--
(-0.728538,0.430856)--
(-0.722406,0.458453)--
(-0.716422,0.48688)--
(-0.709928,0.515793)--
(-0.702875,0.545206)--
(-0.695754,0.575578)--
(-0.68794,0.606501)--
(-0.680405,0.638943)--
(-0.672,0.672)--
(-0.66363,0.706694)--
(-0.654665,0.742572)--
(-0.64499,0.779659)--
(-0.634484,0.817972)--
(-0.62344,0.858092)--
(-0.61168,0.90006)--
(-0.59902,0.943904)--
(-0.584913,0.989034)--
(-0.569569,1.03604)--
(-0.552796,1.08492)--
(-0.534402,1.13566)--
(-0.514192,1.18823)--
(-0.491973,1.24258)--
(-0.467551,1.29867)--
(-0.440731,1.35643)--
(-0.411322,1.41578)--
(-0.378782,1.47526)--
(-0.343053,1.53473)--
(-0.303952,1.59337)--
(-0.261053,1.64823)--
(-0.21447,1.69771)--
(-0.164498,1.74021)--
(-0.111487,1.77204)--
(-0.0563487,1.79304)--
(-4.03996e-14,1.79959);

	\draw[very thick, rotate=-10] (-0.5,0) .. controls (-0.166666666,2) and (0.166666666,2) .. (0.5,0) .. controls (0.833333333,-2) and (-0.8333333333,-2) .. (-0.5,0) ;
    \draw[orange,dashed,thick,rotate=-10] (0,1.79959)--
(0.0563487,1.79304)--
(0.111487,1.77204)--
(0.164498,1.74021)--
(0.21447,1.69771)--
(0.261053,1.64823)--
(0.303952,1.59337)--
(0.343053,1.53473)--
(0.378782,1.47526)--
(0.411322,1.41578)--
(0.440731,1.35643)--
(0.467551,1.29867)--
(0.491973,1.24258)--
(0.514192,1.18823)--
(0.534402,1.13566)--
(0.552796,1.08492)--
(0.569569,1.03604)--
(0.584913,0.989034)--
(0.59902,0.943904)--
(0.61168,0.90006)--
(0.62344,0.858092)--
(0.634484,0.817972)--
(0.64499,0.779659)--
(0.654665,0.742572)--
(0.66363,0.706694)--
(0.672,0.672)--
(0.680405,0.638943)--
(0.68794,0.606501)--
(0.695754,0.575578)--
(0.702875,0.545206)--
(0.709928,0.515793)--
(0.716422,0.48688)--
(0.722406,0.458453)--
(0.728538,0.430856)--
(0.733657,0.403331)--
(0.738403,0.376235)--
(0.743458,0.349845)--
(0.747591,0.323512)--
(0.751467,0.297527)--
(0.755784,0.272099)--
(0.759923,0.246914)--
(0.763909,0.221936)--
(0.767764,0.197128)--
(0.772197,0.172606)--
(0.776543,0.148133)--
(0.780812,0.123669)--
(0.785715,0.0992589)--
(0.789853,0.0746631)--
(0.794631,0.049994)--
(0.798636,0.0250982)--
(0.802566,-1.02009e-15)--
(0.805704,-0.0253203)--
(0.808745,-0.0508819)--
(0.811676,-0.076726)--
(0.814478,-0.102892)--
(0.817129,-0.129421)--
(0.818912,-0.156216)--
(0.821192,-0.183558)--
(0.82324,-0.211372)--
(0.825701,-0.239888)--
(0.827846,-0.268983)--
(0.829633,-0.298686)--
(0.831676,-0.329284)--
(0.833901,-0.360861)--
(0.835591,-0.393199)--
(0.836689,-0.426314)--
(0.837757,-0.460561)--
(0.837484,-0.495287)--
(0.837633,-0.531578)--
(0.836313,-0.568358)--
(0.834638,-0.6064)--
(0.832499,-0.645752)--
(0.829239,-0.686006)--
(0.825846,-0.728081)--
(0.821126,-0.771088)--
(0.8155,-0.8155)--
(0.80836,-0.860816)--
(0.800562,-0.90806)--
(0.791476,-0.956731)--
(0.780537,-1.00626)--
(0.768079,-1.05717)--
(0.753571,-1.10885)--
(0.736935,-1.16122)--
(0.718094,-1.21423)--
(0.696291,-1.26655)--
(0.672215,-1.3193)--
(0.644895,-1.37047)--
(0.615009,-1.4212)--
(0.582038,-1.47006)--
(0.546114,-1.51689)--
(0.507376,-1.56154)--
(0.465967,-1.60387)--
(0.421689,-1.64237)--
(0.375137,-1.67827)--
(0.326079,-1.70937)--
(0.275102,-1.73692)--
(0.222269,-1.75944)--
(0.168025,-1.77752)--
(0.112642,-1.79039)--
(0.0565264,-1.7987)--
(8.21861e-15,-1.801)--
(-0.0565264,-1.7987)--
(-0.112642,-1.79039)--
(-0.168025,-1.77752)--
(-0.222269,-1.75944)--
(-0.275102,-1.73692)--
(-0.326079,-1.70937)--
(-0.375137,-1.67827)--
(-0.421689,-1.64237)--
(-0.465967,-1.60387)--
(-0.507376,-1.56154)--
(-0.546114,-1.51689)--
(-0.582038,-1.47006)--
(-0.615009,-1.4212)--
(-0.644895,-1.37047)--
(-0.672215,-1.3193)--
(-0.696291,-1.26655)--
(-0.718094,-1.21423)--
(-0.736935,-1.16122)--
(-0.753571,-1.10885)--
(-0.768079,-1.05717)--
(-0.780537,-1.00626)--
(-0.791476,-0.956731)--
(-0.800562,-0.90806)--
(-0.80836,-0.860816)--
(-0.8155,-0.8155)--
(-0.821126,-0.771088)--
(-0.825846,-0.728081)--
(-0.829239,-0.686006)--
(-0.832499,-0.645752)--
(-0.834638,-0.6064)--
(-0.836313,-0.568358)--
(-0.837633,-0.531578)--
(-0.837484,-0.495287)--
(-0.837757,-0.460561)--
(-0.836689,-0.426314)--
(-0.835591,-0.393199)--
(-0.833901,-0.360861)--
(-0.831676,-0.329284)--
(-0.829633,-0.298686)--
(-0.827846,-0.268983)--
(-0.825701,-0.239888)--
(-0.82324,-0.211372)--
(-0.821192,-0.183558)--
(-0.818912,-0.156216)--
(-0.817129,-0.129421)--
(-0.814478,-0.102892)--
(-0.811676,-0.076726)--
(-0.808745,-0.0508819)--
(-0.805704,-0.0253203)--
(-0.802566,-1.08398e-14)--
(-0.798636,0.0250982)--
(-0.794631,0.049994)--
(-0.789853,0.0746631)--
(-0.785715,0.0992589)--
(-0.780812,0.123669)--
(-0.776543,0.148133)--
(-0.772197,0.172606)--
(-0.767764,0.197128)--
(-0.763909,0.221936)--
(-0.759923,0.246914)--
(-0.755784,0.272099)--
(-0.751467,0.297527)--
(-0.747591,0.323512)--
(-0.743458,0.349845)--
(-0.738403,0.376235)--
(-0.733657,0.403331)--
(-0.728538,0.430856)--
(-0.722406,0.458453)--
(-0.716422,0.48688)--
(-0.709928,0.515793)--
(-0.702875,0.545206)--
(-0.695754,0.575578)--
(-0.68794,0.606501)--
(-0.680405,0.638943)--
(-0.672,0.672)--
(-0.66363,0.706694)--
(-0.654665,0.742572)--
(-0.64499,0.779659)--
(-0.634484,0.817972)--
(-0.62344,0.858092)--
(-0.61168,0.90006)--
(-0.59902,0.943904)--
(-0.584913,0.989034)--
(-0.569569,1.03604)--
(-0.552796,1.08492)--
(-0.534402,1.13566)--
(-0.514192,1.18823)--
(-0.491973,1.24258)--
(-0.467551,1.29867)--
(-0.440731,1.35643)--
(-0.411322,1.41578)--
(-0.378782,1.47526)--
(-0.343053,1.53473)--
(-0.303952,1.59337)--
(-0.261053,1.64823)--
(-0.21447,1.69771)--
(-0.164498,1.74021)--
(-0.111487,1.77204)--
(-0.0563487,1.79304)--
(-4.03996e-14,1.79959);

\end{scope}
	
\end{tikzpicture}

%% file: images/ExclVolSph.tex
\begin{tikzpicture}
   \node at (0,-0.2) {\includegraphics[height=6.1cm]{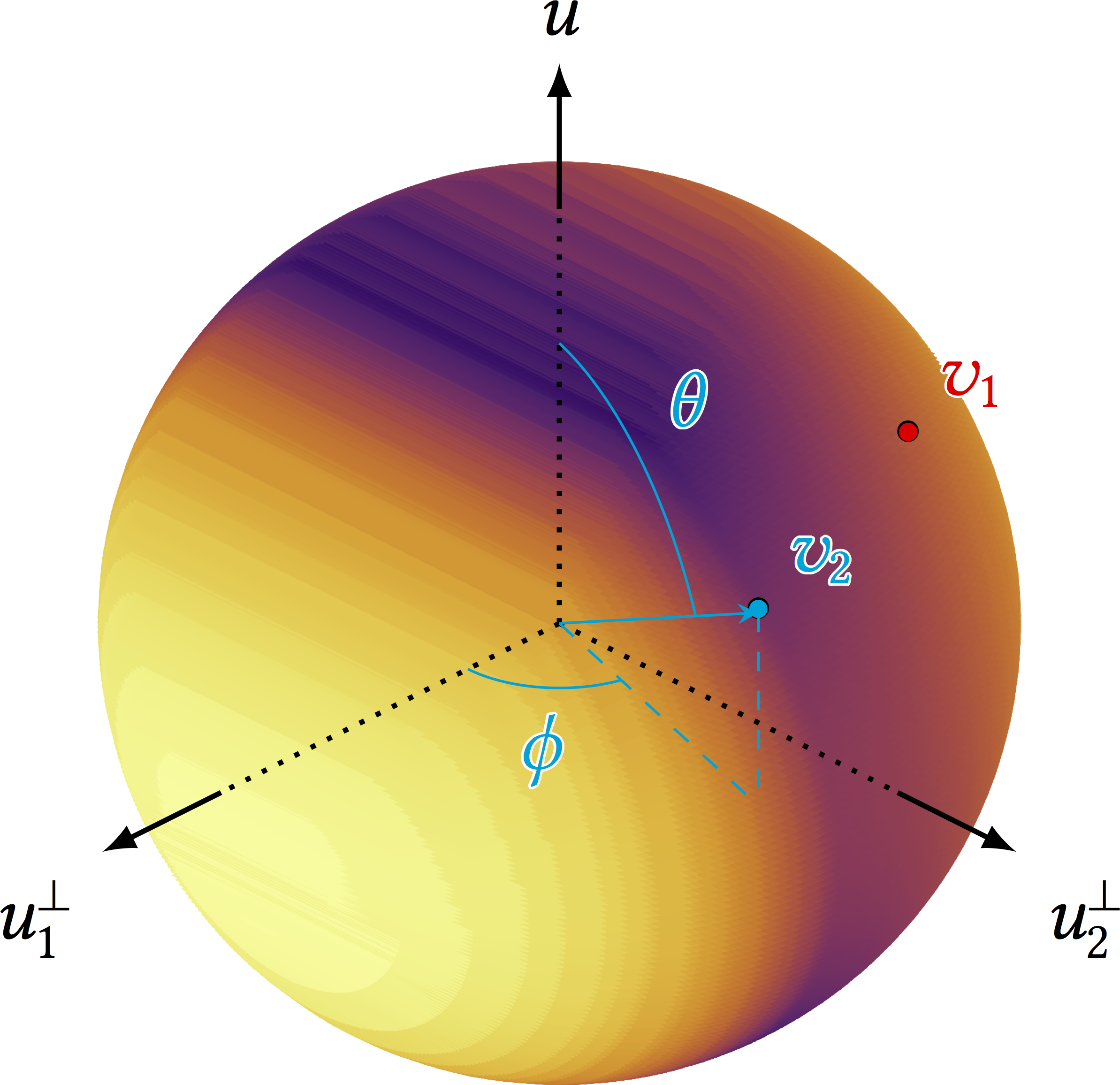}};
   \node at (-0.68,3.8) {\includegraphics[width=6cm]{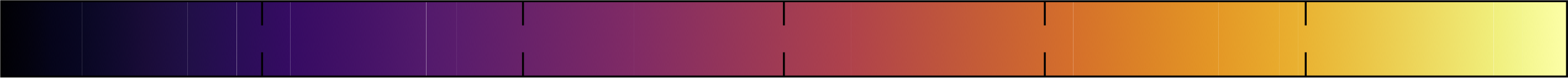}};
   \node at (2.32,3.4) {30};
   \node at (1.32,3.4) {29.5};
   \node at (0.32,3.4) {29};
   \node at (-0.68,3.4) {28.5};
   \node at (-1.68,3.4) {28};
   \node at (-2.68,3.4) {27.5};
   \node at (-3.68,3.4) {27};
   \node at (-0.68,4.25) {excluded volume $V_{\text{excl}}$ $\lbrack \sigma^3_w\rbrack$};
	\node at (-4,4.15) {\textbf{(c)}}; 
     \begin{scope}[shift={(5,0)}]
     	\node at (0,0) {\includegraphics[height=8cm]{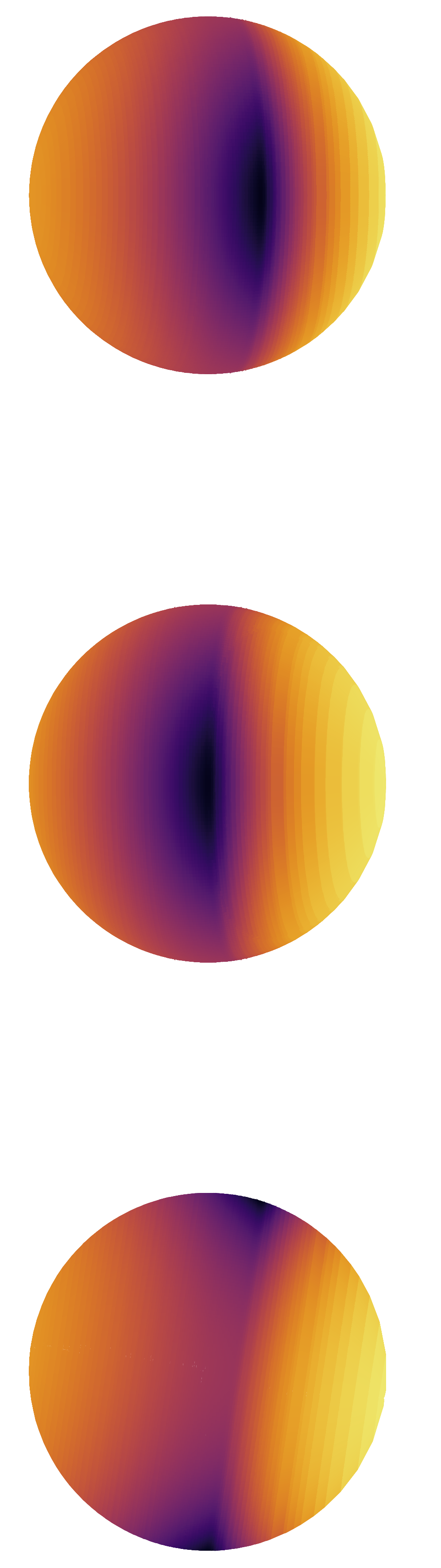}};
	
	\draw[thick,->] (-1.06,1) -- (-1.06,-1) node[anchor=west] {$u^{\perp}_2$};
	\draw[thick,->] (-1.06,1) -- (0.94,1) node[anchor=north] {$u^{\perp}_1$};	
	\node at (-0.06,1.35) {Bottom View}; 
	\node at (-1.5,1.35) {\textbf{(e)}}; 
	
	\draw[thick,->] (-1.06,2) -- (-1.06,4) node[anchor=west] {$u^{\perp}_2$};
	\draw[thick,->] (-1.06,2) -- (0.94,2) node[anchor=south] {$u^{\perp}_1$};	
	\node at (-0.06,4.35) {Top View}; 
	\node at (-1.5,4.35) {\textbf{(d)}}; 

	\draw[thick,->] (-1.06,-4) -- (-1.06,-2) node[anchor=west] {$u$};
	\draw[thick,->] (-1.06,-4) -- (0.94,-4) node[anchor=south] {$u^{\perp}_1$};	
	\node at (-0.06,-1.65)  {Side View}; 
	\node at (-1.5,-1.65) {\textbf{(f)}}; 
    \end{scope}
    
\node[opacity=0.8] at (-6.0,-2.35) {\includegraphics[scale=0.2]{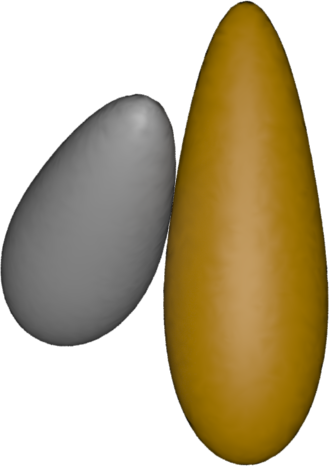}};
\node[opacity=0.8] at (-6.28,2) {\includegraphics[scale=0.2]{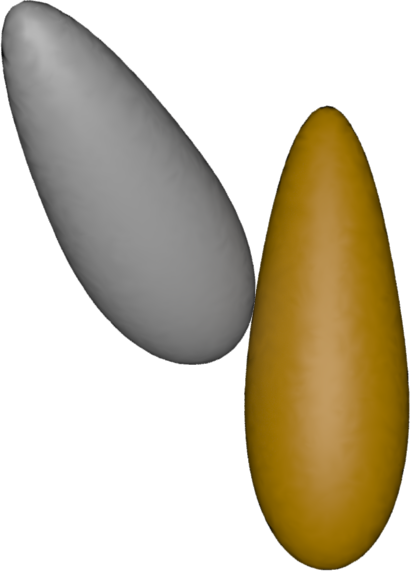}};
\node[black] at (-5.94,1.8) {\textbullet};
\node[black,anchor=west] at (-5.94,1.8) {$\tilde{p}_c$};
\draw[thick,->] (-4.7,1) -- (-4.7,3) node [midway,anchor=west] {$u$};
\draw[thick,->] (-7.1,-0.1) -- (-5.1,-0.1) node [midway,anchor=south] {$u^{\perp}_1$};
\node at (-4.7,-0.1) {$\otimes$};
\node[anchor=south]  at (-4.7,-0.1) {$u^{\perp}_2$};
\node at (-8,4) {\textbf{(a)}}; 

\draw[red,thick,->] (-7.2,1.65) --  ([shift=(119.8:0.8174667902275907*2)]-7.2,1.65) node [midway,anchor=north east] {$v_1$};

\node[black] at (-5.94,-2.2) {\textbullet};
\node[black,anchor=west] at (-5.94,-2.2) {$\tilde{p}_c$};
\draw[thick,->] (-4.7,-3) -- (-4.7,-1) node [midway,anchor=west] {$u$};
\draw[thick,->] (-7.1,-4.1) -- (-5.1,-4.1) node [midway,anchor=south] {$u^{\perp}_1$};
\node at (-4.7,-4.1) {$\otimes$};
\node[anchor=south]  at (-4.7,-4.1) {$u^{\perp}_2$};
\draw[cyan,thick,->] (-7.2,-2.2) --  ([shift=(60:0.47347696443046405*2)]-7.2,-2.2) node [midway,anchor=east] {$v_2$};
\node at (-8,-1) {\textbf{(b)}}; 

   \end{tikzpicture}

%% file: images/excluded_V.tex
    \begin{tikzpicture}
    \begin{axis}[
        ylabel=excl. vol. $V_{\text{excl}}$ $\lbrack \sigma^3_w \rbrack$,
        xlabel=$\beta$,
        ytick pos=left,
        xmin = 0,
        xmax = 180,
        xtick={0,45,90,135,180},
        xticklabels={0,$\frac{\pi}{4}$,$\frac{\pi}{2}$,$\frac{3\pi}{4}$,$\pi$},
        ymin = 27,
        ymax = 30.5,
        width=0.9\textwidth,
        height=0.3\textwidth,
            ]
	\addplot table[x expr= \thisrow{alpha}, y expr= \thisrow{norm}*2.828427125, col sep=comma] {images/Data/ExcludedVolume_II_0.08.dat};
	\addplot table[x expr= \thisrow{alpha}, y expr= \thisrow{norm}*2.828427125, col sep=comma] {images/Data/ExcludedVolume_69_0.08.dat};
    \end{axis}

\begin{scope}[shift={(0,0.252\textwidth)},scale=0.35,rotate=90]
            \draw[very thick, dotted] (0,3.3) -- (0,-3.3);
            
    \begin{scope}[shift={(0,1.5)},rotate=180]
        \draw[ball color=cyan,shading=ball] (-0.5,0) .. controls (-0.236842105,-2) and (0.236842105,-2) .. (0.5,0) .. controls (0.763157894,2) and (-0.763157894,2) .. (-0.5,0);
        \draw[ultra thick, ->] (0,1.5) -- (0,-1.5);
    \end{scope}

    \begin{scope}[ shift={(0,-1.5)}]
        \draw[ball color=orange,shading=ball] (-0.5,0) .. controls (-0.236842105,-2) and (0.236842105,-2) .. (0.5,0) .. controls (0.763157894,2) and (-0.763157894,2) .. (-0.5,0);
        \draw[ultra thick, ->] (0,1.5) -- (0,-1.5);
    \end{scope}
\end{scope}

\begin{scope}[shift={(0,0.3608\textwidth)},scale=0.35,rotate=90]
            \draw[very thick, dotted] (0,3.3) -- (0,-3.3);
            
    \begin{scope}[shift={(0,1.5)},rotate=180]
        \draw[ball color=cyan,shading=ball] (-0.5,0) .. controls (-0.236842105,-2) and (0.236842105,-2) .. (0.5,0) .. controls (0.763157894,2) and (-0.763157894,2) .. (-0.5,0);
        \draw[ultra thick, ->] (0,1.5) -- (0,-1.5);
    \end{scope}

    \begin{scope}[ shift={(0,-1.5)}]
        \draw[ball color=orange,shading=ball] (-0.5,0) .. controls (-0.236842105,-2) and (0.236842105,-2) .. (0.5,0) .. controls (0.763157894,2) and (-0.763157894,2) .. (-0.5,0);
        \draw[ultra thick, ->] (0,1.5) -- (0,-1.5);
    \end{scope}
\end{scope}

\begin{scope}[shift={(0.2029\textwidth,0.2682\textwidth)},scale=0.35,rotate=45]
    \begin{scope}[shift={(0,1.17)},rotate=135]
        \draw[ball color=cyan,shading=ball] (-0.5,0) .. controls (-0.236842105,-2) and (0.236842105,-2) .. (0.5,0) .. controls (0.763157894,2) and (-0.763157894,2) .. (-0.5,0);
        \draw[ultra thick, ->] (0,1.5) -- (0,-1.5);
         \draw[very thick,dashed] ([shift=(270:1.5)]0,0) arc (270:315:1.5);
    \end{scope}

    \begin{scope}[ shift={(0,-1.17)},rotate=45]
        \draw[ball color=orange,shading=ball] (-0.5,0) .. controls (-0.236842105,-2) and (0.236842105,-2) .. (0.5,0) .. controls (0.763157894,2) and (-0.763157894,2) .. (-0.5,0);
        \draw[ultra thick, ->] (0,1.5) -- (0,-1.5);
        \draw[very thick,dashed] ([shift=(-90:1.5)]0,0) arc (-90:-135:1.5);
    \end{scope}
            \draw[very thick, dotted] (0,2.8) -- (0,-2.8);
\end{scope}

\begin{scope}[shift={(0.2029\textwidth,0.3664\textwidth)},scale=0.35,rotate=45]
    \begin{scope}[shift={(-0.8,1.17)},rotate=-135]
        \draw[ball color=cyan,shading=ball] (-0.5,0) .. controls (-0.236842105,-2) and (0.236842105,-2) .. (0.5,0) .. controls (0.763157894,2) and (-0.763157894,2) .. (-0.5,0);
        \draw[ultra thick, ->] (0,1.5) -- (0,-1.5);
         \draw[very thick,dashed] ([shift=(270:1.5)]0,0) arc (270:235:1.5);
    \end{scope}

    \begin{scope}[ shift={(0.8,-1.17)},rotate=45]
        \draw[ball color=orange,shading=ball] (-0.5,0) .. controls (-0.236842105,-2) and (0.236842105,-2) .. (0.5,0) .. controls (0.763157894,2) and (-0.763157894,2) .. (-0.5,0);
        \draw[ultra thick, ->] (0,1.5) -- (0,-1.5);
        \draw[very thick,dashed] ([shift=(-90:1.5)]0,0) arc (-90:-135:1.5);
    \end{scope}
            \draw[dotted] (0.8,1.2) -- (0.8,-2.8);
            \draw[very thick, dotted] (0,2) -- (0,-2);
            \draw[dotted] (-0.8,2.8) -- (-0.8,-1.2);
\end{scope}

\begin{scope}[shift={(0.4058\textwidth,0.2628\textwidth)},scale=0.35,rotate=0]
    \begin{scope}[shift={(0,0.55)},rotate=90]
        \draw[ball color=cyan,shading=ball] (-0.5,0) .. controls (-0.236842105,-2) and (0.236842105,-2) .. (0.5,0) .. controls (0.763157894,2) and (-0.763157894,2) .. (-0.5,0);
        \draw[ultra thick, ->] (0,1.5) -- (0,-1.5);
     \draw[very thick,dashed] ([shift=(270:1.5)]0,0) arc (270:360:1.5);
    \end{scope}

    \begin{scope}[ shift={(0,-0.55)},rotate=90]
        \draw[ball color=orange,shading=ball] (-0.5,0) .. controls (-0.236842105,-2) and (0.236842105,-2) .. (0.5,0) .. controls (0.763157894,2) and (-0.763157894,2) .. (-0.5,0);
        \draw[ultra thick, ->] (0,1.5) -- (0,-1.5);
         \draw[very thick,dashed] ([shift=(-90:1.5)]0,0) arc (-90:-180:1.5);
    \end{scope}
            \draw[very thick, dotted] (0,2.2) -- (0,-2.2);
            \node at (0.6,-1.4) {$\beta$};   
            \node at (0.6,1.4) {$\beta$};        
\end{scope}

\begin{scope}[shift={(0.4058\textwidth,0.3698\textwidth)},scale=0.35,rotate=0]
    \begin{scope}[shift={(-0.4,0.55)},rotate=-90]
        \draw[ball color=cyan,shading=ball] (-0.5,0) .. controls (-0.236842105,-2) and (0.236842105,-2) .. (0.5,0) .. controls (0.763157894,2) and (-0.763157894,2) .. (-0.5,0);
        \draw[ultra thick, ->] (0,1.5) -- (0,-1.5);
     \draw[very thick,dashed] ([shift=(270:1.5)]0,0) arc (270:180:1.5);
    \end{scope}

    \begin{scope}[ shift={(0.4,-0.55)},rotate=90]
        \draw[ball color=orange,shading=ball] (-0.5,0) .. controls (-0.236842105,-2) and (0.236842105,-2) .. (0.5,0) .. controls (0.763157894,2) and (-0.763157894,2) .. (-0.5,0);
        \draw[ultra thick, ->] (0,1.5) -- (0,-1.5);
         \draw[very thick,dashed] ([shift=(-90:1.5)]0,0) arc (-90:-180:1.5);
    \end{scope}
            \draw[very thick, dotted] (0,2.2) -- (0,-2.2);
            \draw[dotted] (-0.4,2.2) -- (-0.4,-2.2);
            \draw[dotted] (0.4,2.2) -- (0.4,-2.2);
            \node at (1,-1.4) {$\beta$};   
            \node at (-1,1.4) {$\beta$};        
\end{scope}

\begin{scope}[shift={(0.6087\textwidth,0.2674\textwidth)},scale=0.35,rotate=-45]

    \begin{scope}[shift={(0,1.11)},rotate=45]
        \draw[ball color=cyan,shading=ball] (-0.5,0) .. controls (-0.236842105,-2) and (0.236842105,-2) .. (0.5,0) .. controls (0.763157894,2) and (-0.763157894,2) .. (-0.5,0);
        \draw[ultra thick, ->] (0,1.5) -- (0,-1.5);
         \draw[very thick,dashed] ([shift=(270:1.5)]0,0) arc (270:405:1.5);
    \end{scope}

    \begin{scope}[ shift={(0,-1.11)},rotate=135]
        \draw[ball color=orange,shading=ball] (-0.5,0) .. controls (-0.236842105,-2) and (0.236842105,-2) .. (0.5,0) .. controls (0.763157894,2) and (-0.763157894,2) .. (-0.5,0);
        \draw[ultra thick, ->] (0,1.5) -- (0,-1.5);
       \draw[very thick,dashed] ([shift=(-90:1.5)]0,0) arc (-90:-225:1.5);
    \end{scope}
            \draw[very thick, dotted] (0,2.8) -- (0,-2.8);            
\end{scope}

\begin{scope}[shift={(0.6087\textwidth,0.363\textwidth)},scale=0.35,rotate=-45]
    \begin{scope}[shift={(0.95,1.11)},rotate=-45]
        \draw[ball color=cyan,shading=ball] (-0.5,0) .. controls (-0.236842105,-2) and (0.236842105,-2) .. (0.5,0) .. controls (0.763157894,2) and (-0.763157894,2) .. (-0.5,0);
        \draw[ultra thick, ->] (0,1.5) -- (0,-1.5);
         \draw[very thick,dashed] ([shift=(270:1.5)]0,0) arc (270:135:1.5);
    \end{scope}

    \begin{scope}[ shift={(-0.95,-1.11)},rotate=135]
        \draw[ball color=orange,shading=ball] (-0.5,0) .. controls (-0.236842105,-2) and (0.236842105,-2) .. (0.5,0) .. controls (0.763157894,2) and (-0.763157894,2) .. (-0.5,0);
        \draw[ultra thick, ->] (0,1.5) -- (0,-1.5);
       \draw[very thick,dashed] ([shift=(-90:1.5)]0,0) arc (-90:-225:1.5);
    \end{scope}
            \draw[very thick, dotted] (0,1.9) -- (0,-1.9);            
            \draw[dotted] (0.95,2.8) -- (0.95,-1);            
            \draw[dotted] (-0.95,1) -- (-0.95,-2.8);            
\end{scope}

\begin{scope}[shift={(0.8116\textwidth,0.252\textwidth)},scale=0.35,rotate=-90]
    \begin{scope}[shift={(0,-1.5)},rotate=180]
        \draw[ball color=orange,shading=ball] (-0.5,0) .. controls (-0.236842105,-2) and (0.236842105,-2) .. (0.5,0) .. controls (0.763157894,2) and (-0.763157894,2) .. (-0.5,0);
        \draw[ultra thick, ->] (0,1.5) -- (0,-1.5);
        \draw[very thick,dashed] ([shift=(270:1.5)]0,0) arc (270:90:1.5);
    \end{scope}

    \begin{scope}[ shift={(0,1.5)}]
        \draw[ball color=cyan,shading=ball] (-0.5,0) .. controls (-0.236842105,-2) and (0.236842105,-2) .. (0.5,0) .. controls (0.763157894,2) and (-0.763157894,2) .. (-0.5,0);
        \draw[ultra thick, ->] (0,1.5) -- (0,-1.5);
       \draw[very thick,dashed] ([shift=(-90:1.5)]0,0) arc (-90:90:1.5);
    \end{scope}
            \draw[very thick, dotted] (0,3.3) -- (0,-3.3);
\end{scope}

\begin{scope}[shift={(0.8116\textwidth,0.3608\textwidth)},scale=0.35,rotate=-90]
    \begin{scope}[shift={(0,-1.5)},rotate=180]
        \draw[ball color=orange,shading=ball] (-0.5,0) .. controls (-0.236842105,-2) and (0.236842105,-2) .. (0.5,0) .. controls (0.763157894,2) and (-0.763157894,2) .. (-0.5,0);
        \draw[ultra thick, ->] (0,1.5) -- (0,-1.5);
        \draw[very thick,dashed] ([shift=(270:1.5)]0,0) arc (270:90:1.5);
    \end{scope}

    \begin{scope}[ shift={(0,1.5)}]
        \draw[ball color=cyan,shading=ball] (-0.5,0) .. controls (-0.236842105,-2) and (0.236842105,-2) .. (0.5,0) .. controls (0.763157894,2) and (-0.763157894,2) .. (-0.5,0);
        \draw[ultra thick, ->] (0,1.5) -- (0,-1.5);
       \draw[very thick,dashed] ([shift=(-90:1.5)]0,0) arc (-90:90:1.5);
    \end{scope}
            \draw[very thick, dotted] (0,3.3) -- (0,-3.3);
\end{scope}

\draw[blue,very thick] (0.8886\textwidth,0.2185\textwidth) rectangle (-0.077\textwidth,0.3179\textwidth);
\node[circle, draw=blue,fill=blue,regular polygon sides=4,scale=0.3] at (-0.0658\textwidth,0.3039\textwidth) {};
\node[anchor=west] at (-0.0658\textwidth,0.3039\textwidth) {Roll};
\draw[red,very thick] (0.8886\textwidth,0.3246\textwidth) rectangle (-0.077\textwidth,0.415\textwidth);
\node[regular polygon, draw=red,fill=red,regular polygon sides=4,scale=0.3] at (-0.0658\textwidth,0.401\textwidth) { };
\node[anchor=west] at (-0.0658\textwidth,0.401\textwidth) {Slide};

\begin{scope}[shift={(0.279\textwidth,0.0558\textwidth)},scale=0.35,rotate=90]
    \begin{scope}[shift={(0,0.51)},rotate=82.5]
        \draw[ball color=cyan,shading=ball] (-0.5,0) .. controls (-0.236842105,-2) and (0.236842105,-2) .. (0.5,0) .. controls (0.763157894,2) and (-0.763157894,2) .. (-0.5,0);
    \end{scope}

    \begin{scope}[ shift={(0,-0.51)},rotate=97.5]
        \draw[ball color=cyan,shading=ball] (-0.5,0) .. controls (-0.236842105,-2) and (0.236842105,-2) .. (0.5,0) .. controls (0.763157894,2) and (-0.763157894,2) .. (-0.5,0);
    \end{scope}
            \draw[blue,very thick] (1.8,1.8) rectangle (-1.8,-1.8);            
\end{scope}

\begin{scope}[shift={(0.558\textwidth,0.0558\textwidth)},scale=0.35,rotate=90]
    \begin{scope}[shift={(0,0.51)},rotate=-82.5]
        \draw[ball color=cyan,shading=ball] (-0.5,0) .. controls (-0.236842105,-2) and (0.236842105,-2) .. (0.5,0) .. controls (0.763157894,2) and (-0.763157894,2) .. (-0.5,0);
    \end{scope}

    \begin{scope}[ shift={(0,-0.51)},rotate=97.5]
        \draw[ball color=cyan,shading=ball] (-0.5,0) .. controls (-0.236842105,-2) and (0.236842105,-2) .. (0.5,0) .. controls (0.763157894,2) and (-0.763157894,2) .. (-0.5,0);
    \end{scope}
            \draw[red,very thick] (1.8,1.8) rectangle (-1.8,-1.8);            
\end{scope}

\draw[->,blue,very thick] (0.3147\textwidth,0.0558\textwidth) -- (0.43\textwidth,0.02\textwidth);
\draw[->,red,very thick] (0.5223\textwidth,0.0558\textwidth) -- (0.46\textwidth,0.02\textwidth);

\end{tikzpicture}

%% file: images/separation.tex
\begin{tikzpicture}
\begin{axis}[
    xlabel={step $n$ $\lbrack 10^6\rbrack$},
    ylabel= separation $R$ $\lbrack\sigma_w\rbrack$,
    xtick pos=left,
    ytick pos=left,
    xmin = 0,
    xmax = 5,
    ymin = 0.8,
    ymax= 6.3,
    width=14cm,
    height=6cm,
    x tick label style={
    /pgf/number format/.cd,
        fixed,
        fixed zerofill,
        precision=1,
    /tikz/.cd
    },
    scaled x ticks=false,
    legend style={draw=none}
    ]
    
   \addplot[mark=none,red,ultra thick] table[x expr= \thisrow{step}/1000000, y expr= \thisrow{r}, col sep=comma] {images/Data/separationNc2Ns1498Vr0_08rho0_45Hard.csv}; 
   \addplot[mark=none,blue,very thick] table[x expr= \thisrow{step}/1000000, y expr= \thisrow{r}, col sep=comma] {images/Data/separationNc2Ns1498Vr0_08rho0_45GO_2.csv}; 
   \addplot[mark=none,orange,thick] table[x expr= \thisrow{step}/1000000, y expr= \thisrow{r}, col sep=comma] {images/Data/separationNc2Ns1498Vr0_08rho0_45NonAdd.csv};
    \addplot[name path=ub,draw=none, mark=none,domain=0.0:5.0] {1};  
    \addplot[name path=lb,draw=none, mark=none,domain=0.0:5.0] {2.6};  
    \addplot[cyan,opacity=0.3] fill between [ of=lb and ub];
    \legend{HPR,PHGO,NAHPR}

    \node at (axis cs:1.25,1.6) {zone of influence};    

\end{axis}
\end{tikzpicture}

%% file: images/ori_hard_go.tex
    \begin{tikzpicture}
    \node[] at (0.1\textwidth,-0.45\textwidth){\includegraphics[height=0.19\textwidth]{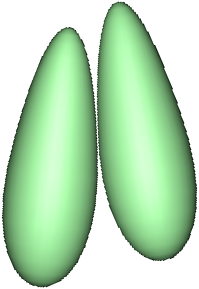}};
    \draw[blue,ultra thick,densely dashed] (0\textwidth,-0.35\textwidth) rectangle (0.2\textwidth,-0.55\textwidth);  
    \node[blue] at (0.025\textwidth,-0.37\textwidth) {\textbf{(I)}};
    \node[] at (0.3116\textwidth,-0.45\textwidth){\includegraphics[height=0.19\textwidth]{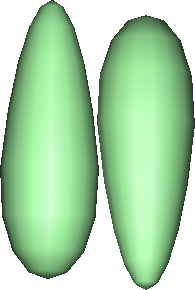}};
    \draw[cyan,ultra thick,densely dashdotted] (0.4116\textwidth,-0.35\textwidth) rectangle (0.2116\textwidth,-0.55\textwidth);  
    \node[cyan] at (0.2366\textwidth,-0.37\textwidth) {\textbf{(II)}};

    \node at (0.2058\textwidth,0.33\textwidth) {\Large \textbf{HPR}};

    \node[] at (0.55\textwidth,-0.45\textwidth){\includegraphics[height=0.19\textwidth]{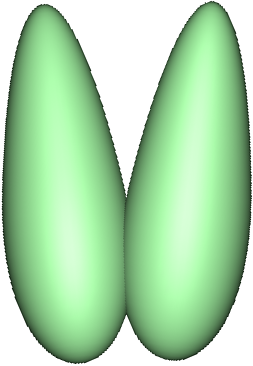}};
    \draw[red,ultra thick,densely dotted] (0.45\textwidth,-0.35\textwidth) rectangle (0.65\textwidth,-0.55\textwidth);  
    \node[red] at (0.475\textwidth,-0.37\textwidth) {\textbf{(III)}};

    \node[] at (0.7616\textwidth,-0.45\textwidth){\includegraphics[height=0.19\textwidth]{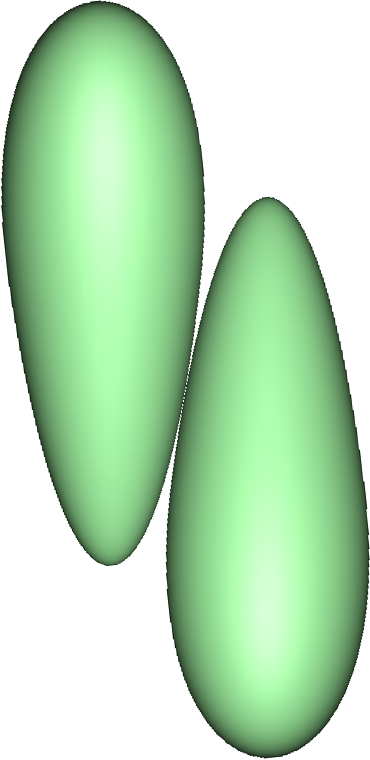}};
    \draw[cyan,ultra thick,densely dashdotted] (0.6616\textwidth,-0.35\textwidth) rectangle (0.8616\textwidth,-0.55\textwidth);  
    \node[cyan] at (0.6866\textwidth,-0.37\textwidth) {\textbf{(IV)}};

    \node at (0.6558\textwidth,0.33\textwidth) {\Large \textbf{PHGO}};

    \draw[very thick,dashed] (0.4308\textwidth,0.31\textwidth) -- (0.4308\textwidth,-0.55\textwidth);
    \begin{axis}[
        ylabel=ocurrence $\lbrack a.u. \rbrack$,
        xlabel=relative angle $\alpha$,
        xtick pos=left,
        ytick pos=left,
        xmin = -3.15,
        xmax = 3.15,
        ymin = 0.0,
        ymax = 1,
        ybar interval,
        width=0.5\textwidth,
        height=0.3\textwidth,
        xtick={-3.14159,-2.356,-1.5708,-0.7854,0,0.7854,1.5708,2.356,3.14159},
        xticklabels={,,,,,,,,},
        y tick label style={
        /pgf/number format/.cd,
            fixed,
            fixed zerofill,
            precision=1,
        /tikz/.cd
        },
        ylabel absolute,
        ylabel near ticks
            ]
    \addplot table[x expr= \thisrow{alpha}, y expr= \thisrow{norm}/0.085, col sep=comma] {images/Data/oriNc2Ns1498Vr0_08rho0_45Hard.csv}; 
    \draw[cyan,ultra thick,densely dashdotted] (axis cs:-3.1,0) -- (axis cs:-3.1,1.2);  
    \node[cyan] at (axis cs:-2.8,0.75) {(II)};
    \draw[cyan,ultra thick,densely dashdotted] (axis cs:3.1,0) -- (axis cs:3.1,1.2);  
    \node[cyan] at (axis cs:2.8,0.75) {(II)};
    \draw[blue,ultra thick,densely dashed] (axis cs:-0.26,0) -- (axis cs:-0.26,1.2);  
    \node[blue] at (axis cs:-0.56,0.75) {(I)};
    \draw[red,ultra thick,densely dotted] (axis cs:0.32,0) -- (axis cs:0.32,1.2);  
    \end{axis}
    \node at (0,-0.0223\textwidth) {-$\pi$}; 
    \node at (0.05145\textwidth,-0.0223\textwidth) {-$\frac{3\pi}{4}$}; 
    \node at (0.1029\textwidth,-0.0223\textwidth) {-$\frac{\pi}{2}$}; 
    \node at (0.15435\textwidth,-0.0223\textwidth) {-$\frac{\pi}{4}$}; 
    \node at (0.2058\textwidth,-0.0223\textwidth) {$0$}; 
    \node at (0.25725\textwidth,-0.0223\textwidth) {$\frac{\pi}{4}$}; 
    \node at (0.3087\textwidth,-0.0223\textwidth) {$\frac{\pi}{2}$}; 
    \node at (0.36015\textwidth,-0.0223\textwidth) {$\frac{3\pi}{4}$}; 
    \node at (0.4116\textwidth,-0.0223\textwidth) {$\pi$}; 
    \node at (0.023\textwidth,0.1952\textwidth) {\textbf{(a)}};

    \draw[ultra thick,<->] (0,0.2232\textwidth) -- (0.2058\textwidth,0.2232\textwidth);  
    \node[anchor=west] at (-0.0021\textwidth,0.2957\textwidth) {``A-config.''};
    \draw[ultra thick,<->] (0.4116\textwidth,0.2232\textwidth) -- (0.2058\textwidth,0.2232\textwidth);  
    \node[anchor=east] at (0.4199\textwidth,0.2957\textwidth) {``V-config.''};
\begin{scope}[shift={(0.2709\textwidth,0.3292\textwidth)},scale=0.5,rotate=90]
    \begin{scope}[shift={(-3,0)},rotate=180]
        \draw[ball color=cyan,shading=ball] (-0.5,0) .. controls (-0.1,-2) and (0.1,-2) .. (0.5,0) .. controls (0.9,2) and (-0.9,2) .. (-0.5,0) ;
        \draw[ultra thick, ->] (0,1.5) -- (0,-1.5);
        \draw[very thick, dotted] (0,1.5) -- (0,4.1);
        \draw[very thick] ([shift=(270:2)]0,4.1) arc (270:250:2);
    \end{scope}

    \node at (-2.7,-2.1) [anchor=west] {$\alpha$}; 

    \begin{scope}[ shift={(-1.6,-0.3)},rotate=160]
        \draw[ball color=cyan,shading=ball] (-0.5,0) .. controls (-0.1,-2) and (0.1,-2) .. (0.5,0) .. controls (0.9,2) and (-0.9,2) .. (-0.5,0) ;
        \draw[ultra thick, ->] (0,1.5) -- (0,-1.5);
        \draw[very thick, dotted] (0,1.5) -- (0,4);
    \end{scope}
\end{scope}

\begin{scope}[shift={(0.1374\textwidth,0.3292\textwidth)},scale=0.5,rotate=90]
    \begin{scope}[shift={(-3,0)},rotate=180]
        \draw[ball color=cyan,shading=ball] (-0.5,0) .. controls (-0.1,-2) and (0.1,-2) .. (0.5,0) .. controls (0.9,2) and (-0.9,2) .. (-0.5,0) ;
        \draw[ultra thick, ->] (0,1.5) -- (0,-1.5);
        \draw[very thick, dotted] (0,-1.5) -- (0,-4.1);
        \draw[very thick] ([shift=(-270:2)]0,-4.1) arc (-270:-250:2);
    \end{scope}

    \node at (-2.7,3.2) [anchor=west] {$\alpha$}; 

    \begin{scope}[ shift={(-1.6,0.3)},rotate=-160]
        \draw[ball color=cyan,shading=ball] (-0.5,0) .. controls (-0.1,-2) and (0.1,-2) .. (0.5,0) .. controls (0.9,2) and (-0.9,2) .. (-0.5,0) ;
        \draw[ultra thick, ->] (0,1.5) -- (0,-1.5);
        \draw[very thick, dotted] (0,-1.5) -- (0,-4);
    \end{scope}
\end{scope}

    \begin{axis}[
        yshift =-0.28\textwidth,
        ylabel=ocurrence $\lbrack a.u. \rbrack$,
        xlabel=lateral distance $z$ $\lbrack \sigma_w \rbrack$,
        xtick pos=left,
        ytick pos=left,
        xmin = -3,
        xmax = 3,
        ymin = 0.0,
        ymax = 1,
        ybar interval,
        width=0.5\textwidth,
        height=0.3\textwidth,
        xtick={-3,-2,-1,0,1,2,3},
        xticklabels={,,,,,,},
        y tick label style={
        /pgf/number format/.cd,
            fixed,
            fixed zerofill,
            precision=1,
        /tikz/.cd
        },
        ylabel absolute,
        ylabel near ticks
            ]
   \addplot table[x expr= \thisrow{r}*sqrt(2), y expr= \thisrow{N}/75, col sep=comma] {images/Data/distNc2Ns1498Vr0_08rho0_45Hard.csv};
    \end{axis}
    \node at (0,-0.3023\textwidth) {-3}; 
    \node at (0.0686\textwidth,-0.3023\textwidth) {-2}; 
    \node at (0.1372\textwidth,-0.3023\textwidth) {-1}; 
    \node at (0.2058\textwidth,-0.3023\textwidth) {$0$}; 
    \node at (0.2744\textwidth,-0.3023\textwidth) {1}; 
    \node at (0.343\textwidth,-0.3023\textwidth) {2}; 
    \node at (0.4116\textwidth,-0.3023\textwidth) {3}; 
    \node at (0.023\textwidth,-0.0848\textwidth) {\textbf{(c)}};

    \begin{scope}[shift={(0.2058\textwidth,-0.28\textwidth)},scale=1.23]
    \clip (-1.56,0) rectangle (1.56,0.6);
    \draw[fill=orange,thick,rotate=-90,opacity=0.2] (-0.5,0) .. controls (-0.236842105,2) and (0.236842105,2) .. (0.5,0) .. controls (0.763157894,-2) and (-0.763157894,-2) .. (-0.5,0);
    \end{scope}

    \begin{axis}[
        xshift =0.45\textwidth,
        ylabel=ocurrence $\lbrack a.u. \rbrack$,
        xlabel=relative angle $\alpha$,
        xtick pos=left,
        ytick pos=right,
        xmin = -3.15,
        xmax = 3.15,
        ymin = 0.0,
        ymax = 1,
        ybar interval,
        width=0.5\textwidth,
        height=0.3\textwidth,
        xtick={-3.14159,-2.356,-1.5708,-0.7854,0,0.7854,1.5708,2.356,3.14159},
        xticklabels={,,,,,,,,},
        y tick label style={
        /pgf/number format/.cd,
            fixed,
            fixed zerofill,
            precision=1,
        /tikz/.cd
        },
        ylabel absolute,
        ylabel near ticks,
        yticklabel pos=right
            ]
   \addplot table[x expr= \thisrow{alpha}, y expr= \thisrow{norm}/0.19, col sep=comma] {images/Data/oriNc2Ns1498Vr0_08rho0_45GO_2.csv}; 
    \draw[cyan,ultra thick,densely dashdotted] (axis cs:-3.1,0) -- (axis cs:-3.1,1.2);  
    \node[cyan] at (axis cs:-2.75,0.75) {(IV)};
    \draw[cyan,ultra thick,densely dashdotted] (axis cs:3.1,0) -- (axis cs:3.1,1.2);  
    \node[cyan] at (axis cs:2.75,0.75) {(IV)};
    \draw[blue,ultra thick,densely dashed] (axis cs:-0.26,0) -- (axis cs:-0.26,1.2);  
    \draw[red,ultra thick,densely dotted] (axis cs:0.32,0) -- (axis cs:0.32,1.2);  
    \node[red] at (axis cs:0.72,0.75) {(III)};
    \end{axis}
    \node at (0.45\textwidth,-0.0223\textwidth) {-$\pi$}; 
    \node at (0.50145\textwidth,-0.0223\textwidth) {-$\frac{3\pi}{4}$}; 
    \node at (0.5529\textwidth,-0.0223\textwidth) {-$\frac{\pi}{2}$}; 
    \node at (0.60435\textwidth,-0.0223\textwidth) {-$\frac{\pi}{4}$}; 
    \node at (0.6558\textwidth,-0.0223\textwidth) {$0$}; 
    \node at (0.70725\textwidth,-0.0223\textwidth) {$\frac{\pi}{4}$}; 
    \node at (0.7587\textwidth,-0.0223\textwidth) {$\frac{\pi}{2}$}; 
    \node at (0.81015\textwidth,-0.0223\textwidth) {$\frac{3\pi}{4}$}; 
    \node at (0.8616\textwidth,-0.0223\textwidth) {$\pi$}; 
    \node at (0.473\textwidth,0.1952\textwidth) {\textbf{(b)}};

    \draw[ultra thick,<->] (0.46\textwidth,0.2232\textwidth) -- (0.6558\textwidth,0.2232\textwidth);  
    \node[anchor=west] at (0.4579\textwidth,0.2957\textwidth) {``A-config.''};
    \draw[ultra thick,<->] (0.8616\textwidth,0.2232\textwidth) -- (0.6558\textwidth,0.2232\textwidth);  
    \node[anchor=east] at (0.8699\textwidth,0.2957\textwidth) {``V-config.''};
\begin{scope}[shift={(0.7209\textwidth,0.3292\textwidth)},scale=0.5,rotate=90]
    \begin{scope}[shift={(-3,0)},rotate=180]
        \draw[ball color=cyan,shading=ball] (-0.5,0) .. controls (-0.1,-2) and (0.1,-2) .. (0.5,0) .. controls (0.9,2) and (-0.9,2) .. (-0.5,0) ;
        \draw[ultra thick, ->] (0,1.5) -- (0,-1.5);
        \draw[very thick, dotted] (0,1.5) -- (0,4.1);
        \draw[very thick] ([shift=(270:2)]0,4.1) arc (270:250:2);
    \end{scope}

    \node at (-2.7,-2.1) [anchor=west] {$\alpha$}; 

    \begin{scope}[ shift={(-1.6,-0.3)},rotate=160]
        \draw[ball color=cyan,shading=ball] (-0.5,0) .. controls (-0.1,-2) and (0.1,-2) .. (0.5,0) .. controls (0.9,2) and (-0.9,2) .. (-0.5,0) ;
        \draw[ultra thick, ->] (0,1.5) -- (0,-1.5);
        \draw[very thick, dotted] (0,1.5) -- (0,4);
    \end{scope}
\end{scope}

\begin{scope}[shift={(0.5874\textwidth,0.3292\textwidth)},scale=0.5,rotate=90]
    \begin{scope}[shift={(-3,0)},rotate=180]
        \draw[ball color=cyan,shading=ball] (-0.5,0) .. controls (-0.1,-2) and (0.1,-2) .. (0.5,0) .. controls (0.9,2) and (-0.9,2) .. (-0.5,0) ;
        \draw[ultra thick, ->] (0,1.5) -- (0,-1.5);
        \draw[very thick, dotted] (0,-1.5) -- (0,-4.1);
        \draw[very thick] ([shift=(-270:2)]0,-4.1) arc (-270:-250:2);
    \end{scope}

    \node at (-2.7,3.2) [anchor=west] {$\alpha$}; 

    \begin{scope}[ shift={(-1.6,0.3)},rotate=-160]
        \draw[ball color=cyan,shading=ball] (-0.5,0) .. controls (-0.1,-2) and (0.1,-2) .. (0.5,0) .. controls (0.9,2) and (-0.9,2) .. (-0.5,0) ;
        \draw[ultra thick, ->] (0,1.5) -- (0,-1.5);
        \draw[very thick, dotted] (0,-1.5) -- (0,-4);
    \end{scope}
\end{scope}

    \begin{axis}[
        yshift =-0.28\textwidth,
        xshift =0.45\textwidth,
        ylabel=ocurrence $\lbrack a.u. \rbrack$,
        xlabel=lateral distance $z$ $\lbrack \sigma_w \rbrack$,
        xtick pos=left,
        ytick pos=right,
        xmin = -3,
        xmax = 3,
        ymin = 0.0,
        ymax = 1,
        ybar interval,
        width=0.5\textwidth,
        height=0.3\textwidth,
        xtick={-3,-2,-1,0,1,2,3},
        xticklabels={,,,,,,},
        y tick label style={
        /pgf/number format/.cd,
            fixed,
            fixed zerofill,
            precision=1,
        /tikz/.cd
        },
        ylabel absolute,
        ylabel near ticks,
        yticklabel pos=right
            ]
   \addplot table[x expr= \thisrow{r}*sqrt(2), y expr= \thisrow{N}/175, col sep=comma] {images/Data/distNc2Ns1498Vr0_08rho0_45GO_2.csv};
    \end{axis}
    \node at (0.45\textwidth,-0.3023\textwidth) {-3}; 
    \node at (0.5186\textwidth,-0.3023\textwidth) {-2}; 
    \node at (0.5872\textwidth,-0.3023\textwidth) {-1}; 
    \node at (0.6558\textwidth,-0.3023\textwidth) {$0$}; 
    \node at (0.7244\textwidth,-0.3023\textwidth) {1}; 
    \node at (0.793\textwidth,-0.3023\textwidth) {2}; 
    \node at (0.8616\textwidth,-0.3023\textwidth) {3}; 
    \node at (0.473\textwidth,-0.0848\textwidth) {\textbf{(d)}};

    \begin{scope}[shift={(0.6558\textwidth,-0.28\textwidth)},scale=1.23]
    \clip (-1.56,0) rectangle (1.56,0.6);
    \draw[fill=orange,thick,rotate=-90,opacity=0.2] (-0.5,0) .. controls (-0.236842105,2) and (0.236842105,2) .. (0.5,0) .. controls (0.763157894,-2) and (-0.763157894,-2) .. (-0.5,0);
    \end{scope}
\end{tikzpicture}

%% file: images/pricklypear.tex
\begin{tikzpicture}
\node at (-0.2,-0.5) {(b)};
\node at (-5.3,-0.5) {(a)};
\begin{scope}[shift={(-3.4,0.5)},scale=1.5,rotate=90]

\begin{scope}[rotate=15]

\draw[thick,blue] (0.2636795,-1.2959707) -- ([shift=(-15:0.055)]0.2636795,-1.2959707);
\draw[thick,blue] (0.29314,-1.1899514) -- ([shift=(-15:0.09)]0.29314,-1.1899514);
\draw[thick,blue] (0.3226005,-1.0839321) -- ([shift=(-15:0.107)]0.3226005,-1.0839321);
\draw[thick,blue] (0.352061,-0.9779128) -- ([shift=(-15:0.11)]0.352061,-0.9779128);
\draw[thick,blue] (0.3815215,-0.8718935) -- ([shift=(-15:0.107)]0.3815215,-0.8718935);
\draw[thick,blue] (0.410982,-0.7658742) -- ([shift=(-15:0.1)]0.410982,-0.7658742);
\draw[thick,blue] (0.4404425,-0.6598549) -- ([shift=(-15:0.085)]0.4404425,-0.6598549);
\draw[thick,blue] (0.469903,-0.5538356) -- ([shift=(-15:0.06)]0.469903,-0.5538356);
\draw[thick,blue] (0.4993635,-0.4478163) -- ([shift=(-15:0.035)]0.4993635,-0.4478163);
\draw[thick,blue] (0.24894925,-1.34898035) -- ([shift=(-15:0.03)]0.24894925,-1.34898035);
\draw[thick,blue] (0.27840975,-1.24296105) -- ([shift=(-15:0.075)]0.27840975,-1.24296105);
\draw[thick,blue] (0.30787025,-1.13694175) -- ([shift=(-15:0.1)]0.30787025,-1.13694175);
\draw[thick,blue] (0.33733075,-1.03092245) -- ([shift=(-15:0.11)]0.33733075,-1.03092245);
\draw[thick,blue] (0.36679125,-0.92490315) -- ([shift=(-15:0.11)]0.36679125,-0.92490315);
\draw[thick,blue] (0.39625175,-0.81888385) -- ([shift=(-15:0.105)]0.39625175,-0.81888385);
\draw[thick,blue] (0.42571225,-0.71286455) -- ([shift=(-15:0.093)]0.42571225,-0.71286455);
\draw[thick,blue] (0.45517275,-0.60684525) -- ([shift=(-15:0.075)]0.45517275,-0.60684525);
\draw[thick,blue] (0.48463325,-0.50082595) -- ([shift=(-15:0.05)]0.48463325,-0.50082595);
\draw[thick,blue] (0.51409375,-0.39480665) -- ([shift=(-15:0.02)]0.51409375,-0.39480665);
\draw[thick,blue] (-0.256314375,-1.322475525) -- ([shift=(195:0.042)]-0.256314375,-1.322475525);
\draw[thick,blue] (-0.271044625,-1.269465875) -- ([shift=(195:0.063)]-0.271044625,-1.269465875);
\draw[thick,blue] (-0.285774875,-1.216456225) -- ([shift=(195:0.08)]-0.285774875,-1.216456225);
\draw[thick,blue] (-0.300505125,-1.163446575) -- ([shift=(195:0.09)]-0.300505125,-1.163446575);
\draw[thick,blue] (-0.315235375,-1.110436925) -- ([shift=(195:0.1)]-0.315235375,-1.110436925);
\draw[thick,blue] (-0.329965625,-1.057427275) -- ([shift=(195:0.108)]-0.329965625,-1.057427275);
\draw[thick,blue] (-0.344695875,-1.004417625) -- ([shift=(195:0.11)]-0.344695875,-1.004417625);
\draw[thick,blue] (-0.359426125,-0.951407975) -- ([shift=(195:0.112)]-0.359426125,-0.951407975);
\draw[thick,blue] (-0.374156375,-0.898398325) -- ([shift=(195:0.107)]-0.374156375,-0.898398325);
\draw[thick,blue] (-0.388886625,-0.845388675) -- ([shift=(195:0.105)]-0.388886625,-0.845388675);
\draw[thick,blue] (-0.403616875,-0.792379025) -- ([shift=(195:0.102)]-0.403616875,-0.792379025);
\draw[thick,blue] (-0.418347125,-0.739369375) -- ([shift=(195:0.094)]-0.418347125,-0.739369375);
\draw[thick,blue] (-0.433077375,-0.686359725) -- ([shift=(195:0.086)]-0.433077375,-0.686359725);
\draw[thick,blue] (-0.447807625,-0.633350075) -- ([shift=(195:0.077)]-0.447807625,-0.633350075);
\draw[thick,blue] (-0.462537875,-0.580340425) -- ([shift=(195:0.066)]-0.462537875,-0.580340425);
\draw[thick,blue] (-0.477268125,-0.527330775) -- ([shift=(195:0.054)]-0.477268125,-0.527330775);
\draw[thick,blue] (-0.491998375,-0.474321125) -- ([shift=(195:0.04)]-0.491998375,-0.474321125);
\draw[thick,blue] (-0.506728625,-0.421311475) -- ([shift=(195:0.027)]-0.506728625,-0.421311475);

 \draw[thick]
(0,-1.5) --
(0.00997031,-1.49945) --
(0.0199406,-1.49891) --
(0.0299095,-1.49834) --
(0.0397871,-1.49688) --
(0.0496634,-1.49541) --
(0.0595326,-1.49389) --
(0.0692274,-1.4915) --
(0.0789175,-1.48909) --
(0.088591,-1.48661) --
(0.0980227,-1.48334) --
(0.107446,-1.48003) --
(0.116842,-1.47665) --
(0.125961,-1.47259) --
(0.135067,-1.46849) --
(0.144132,-1.4643) --
(0.152886,-1.4595) --
(0.161622,-1.45466) --
(0.170308,-1.44974) --
(0.178681,-1.4443) --
(0.187034,-1.43883) --
(0.195324,-1.43326) --
(0.20329,-1.42724) --
(0.211229,-1.42119) --
(0.219102,-1.41504) --
(0.226675,-1.40854) --
(0.234219,-1.40199) --
(0.528824,-0.341797) --
(0.528487,-0.331818) --
(0.528094,-0.32184) --
(0.527659,-0.311865) --
(0.527206,-0.30189) --
(0.526727,-0.291916) --
(0.526225,-0.281944) --
(0.525697,-0.271972) --
(0.525122,-0.262004) --
(0.524513,-0.252037) --
(0.523882,-0.242072) --
(0.523216,-0.232109) --
(0.522524,-0.222148) --
(0.521812,-0.212188) --
(0.521065,-0.202231) --
(0.520292,-0.192276) --
(0.519494,-0.182322) --
(0.518651,-0.172373) --
(0.517779,-0.162426) --
(0.516892,-0.15248) --
(0.515981,-0.142537) --
(0.515049,-0.132595) --
(0.51409,-0.122656) --
(0.513078,-0.112722) --
(0.512036,-0.102792) --
(0.510984,-0.0928619) --
(0.509914,-0.0829343) --
(0.508827,-0.0730085) --
(0.507711,-0.0630859) --
(0.506539,-0.0531695) --
(0.505337,-0.0432569) --
(0.504129,-0.0333453) --
(0.502906,-0.0234353) --
(0.501671,-0.0135268) --
(0.500416,-0.00362088) --
(0.499126,0.00628066) --
(0.497817,0.0161798) --
(0.496503,0.026078) --
(0.49518,0.0359752) --
(0.493851,0.0458715) --
(0.492513,0.0557667) --
(0.491157,0.0656594) --
(0.489792,0.0755508) --
(0.488425,0.0854419) --
(0.487052,0.0953322) --
(0.485675,0.105222) --
(0.484287,0.11511) --
(0.482881,0.124996) --
(0.481467,0.134881) --
(0.48005,0.144765) --
(0.478627,0.154648) --
(0.477199,0.16453) --
(0.47576,0.174411) --
(0.474302,0.18429) --
(0.472836,0.194166) --
(0.471366,0.204043) --
(0.469891,0.213918) --
(0.468411,0.223793) --
(0.466918,0.233666) --
(0.465404,0.243536) --
(0.463881,0.253404) --
(0.462353,0.263272) --
(0.460819,0.273139) --
(0.45928,0.283005) --
(0.457728,0.292868) --
(0.456159,0.302729) --
(0.454581,0.312589) --
(0.452995,0.322448) --
(0.451398,0.332305) --
(0.449794,0.34216) --
(0.448181,0.352014) --
(0.446554,0.361866) --
(0.44492,0.371716) --
(0.443273,0.381565) --
(0.441608,0.39141) --
(0.439936,0.401254) --
(0.438257,0.411097) --
(0.436568,0.420938) --
(0.434872,0.430779) --
(0.433159,0.440616) --
(0.431424,0.450449) --
(0.429679,0.46028) --
(0.427927,0.470111) --
(0.426164,0.479939) --
(0.424395,0.489766) --
(0.422611,0.499591) --
(0.420808,0.509412) --
(0.418995,0.519232) --
(0.417168,0.529048) --
(0.415324,0.538861) --
(0.413472,0.548673) --
(0.411611,0.558484) --
(0.409739,0.568292) --
(0.407857,0.578098) --
(0.405954,0.5879) --
(0.404022,0.597697) --
(0.40208,0.607491) --
(0.400127,0.617284) --
(0.39816,0.627073) --
(0.396185,0.636861) --
(0.394197,0.646646) --
(0.392192,0.656428) --
(0.390176,0.666207) --
(0.388137,0.675982) --
(0.386069,0.685751) --
(0.383992,0.695517) --
(0.381899,0.705281) --
(0.37979,0.715041) --
(0.377674,0.724799) --
(0.375541,0.734554) --
(0.37339,0.744305) --
(0.371228,0.754053) --
(0.369034,0.763794) --
(0.366809,0.773529) --
(0.364573,0.78326) --
(0.362317,0.792987) --
(0.360042,0.80271) --
(0.357758,0.81243) --
(0.355452,0.822145) --
(0.353124,0.831855) --
(0.350786,0.841563) --
(0.348419,0.851263) --
(0.346024,0.860957) --
(0.343617,0.870648) --
(0.341176,0.880331) --
(0.338705,0.890005) --
(0.336223,0.899677) --
(0.333711,0.90934) --
(0.331174,0.918998) --
(0.328626,0.928653) --
(0.326045,0.938298) --
(0.323438,0.947937) --
(0.320818,0.957573) --
(0.31816,0.967198) --
(0.315473,0.976815) --
(0.312773,0.986428) --
(0.310027,0.996028) --
(0.307248,1.00562) --
(0.304456,1.01521) --
(0.301613,1.02478) --
(0.298736,1.03434) --
(0.295845,1.0439) --
(0.292903,1.05344) --
(0.289928,1.06297) --
(0.28694,1.0725) --
(0.283894,1.08201) --
(0.280812,1.0915) --
(0.277716,1.101) --
(0.274555,1.11047) --
(0.271357,1.11993) --
(0.268143,1.12938) --
(0.264856,1.13881) --
(0.26153,1.14823) --
(0.258187,1.15763) --
(0.25476,1.16701) --
(0.251293,1.17638) --
(0.247808,1.18573) --
(0.244228,1.19506) --
(0.240603,1.20436) --
(0.236961,1.21366) --
(0.233211,1.22291) --
(0.229413,1.23215) --
(0.225598,1.24137) --
(0.221657,1.25055) --
(0.217668,1.2597) --
(0.213659,1.26885) --
(0.2095,1.27792) --
(0.205287,1.28698) --
(0.201053,1.29602) --
(0.196639,1.30498) --
(0.192167,1.31391) --
(0.187673,1.32282) --
(0.18297,1.33163) --
(0.178207,1.34041) --
(0.17342,1.34917) --
(0.168381,1.35779) --
(0.163278,1.36637) --
(0.158149,1.37494) --
(0.152704,1.38331) --
(0.147188,1.39163) --
(0.141647,1.39994) --
(0.13571,1.40797) --
(0.129698,1.41594) --
(0.123659,1.42389) --
(0.117121,1.43144) --
(0.110502,1.43892) --
(0.103856,1.44637) --
(0.0965752,1.4532) --
(0.0892133,1.45995) --
(0.0818262,1.46666) --
(0.0736564,1.47241) --
(0.065414,1.47804) --
(0.05715,1.48365) --
(0.0480044,1.48765) --
(0.0388073,1.49154) --
(0.0296006,1.49541) --
(0.0197427,1.49699) --
(0.00987107,1.4985) --
(0,1.5) --
(-0.00987107,1.4985) --
(-0.0197427,1.49699) --
(-0.0296006,1.49541) --
(-0.0388073,1.49154) --
(-0.0480044,1.48765) --
(-0.05715,1.48365) --
(-0.065414,1.47804) --
(-0.0736564,1.47241) --
(-0.0818262,1.46666) --
(-0.0892133,1.45995) --
(-0.0965752,1.4532) --
(-0.103856,1.44637) --
(-0.110502,1.43892) --
(-0.117121,1.43144) --
(-0.123659,1.42389) --
(-0.129698,1.41594) --
(-0.13571,1.40797) --
(-0.141647,1.39994) --
(-0.147188,1.39163) --
(-0.152704,1.38331) --
(-0.158149,1.37494) --
(-0.163278,1.36637) --
(-0.168381,1.35779) --
(-0.17342,1.34917) --
(-0.178207,1.34041) --
(-0.18297,1.33163) --
(-0.187673,1.32282) --
(-0.192167,1.31391) --
(-0.196639,1.30498) --
(-0.201053,1.29602) --
(-0.205287,1.28698) --
(-0.2095,1.27792) --
(-0.213659,1.26885) --
(-0.217668,1.2597) --
(-0.221657,1.25055) --
(-0.225598,1.24137) --
(-0.229413,1.23215) --
(-0.233211,1.22291) --
(-0.236961,1.21366) --
(-0.240603,1.20436) --
(-0.244228,1.19506) --
(-0.247808,1.18573) --
(-0.251293,1.17638) --
(-0.25476,1.16701) --
(-0.258187,1.15763) --
(-0.26153,1.14823) --
(-0.264856,1.13881) --
(-0.268143,1.12938) --
(-0.271357,1.11993) --
(-0.274555,1.11047) --
(-0.277716,1.101) --
(-0.280812,1.0915) --
(-0.283894,1.08201) --
(-0.28694,1.0725) --
(-0.289928,1.06297) --
(-0.292903,1.05344) --
(-0.295845,1.0439) --
(-0.298736,1.03434) --
(-0.301613,1.02478) --
(-0.304456,1.01521) --
(-0.307248,1.00562) --
(-0.310027,0.996028) --
(-0.312773,0.986428) --
(-0.315473,0.976815) --
(-0.31816,0.967198) --
(-0.320818,0.957573) --
(-0.323438,0.947937) --
(-0.326045,0.938298) --
(-0.328626,0.928653) --
(-0.331174,0.918998) --
(-0.333711,0.90934) --
(-0.336223,0.899677) --
(-0.338705,0.890005) --
(-0.341176,0.880331) --
(-0.343617,0.870648) --
(-0.346024,0.860957) --
(-0.348419,0.851263) --
(-0.350786,0.841563) --
(-0.353124,0.831855) --
(-0.355452,0.822145) --
(-0.357758,0.81243) --
(-0.360042,0.80271) --
(-0.362317,0.792987) --
(-0.364573,0.78326) --
(-0.366809,0.773529) --
(-0.369034,0.763794) --
(-0.371228,0.754053) --
(-0.37339,0.744305) --
(-0.375541,0.734554) --
(-0.377674,0.724799) --
(-0.37979,0.715041) --
(-0.381899,0.705281) --
(-0.383992,0.695517) --
(-0.386069,0.685751) --
(-0.388137,0.675982) --
(-0.390176,0.666207) --
(-0.392192,0.656428) --
(-0.394197,0.646646) --
(-0.396185,0.636861) --
(-0.39816,0.627073) --
(-0.400127,0.617284) --
(-0.40208,0.607491) --
(-0.404022,0.597697) --
(-0.405954,0.5879) --
(-0.407857,0.578098) --
(-0.409739,0.568292) --
(-0.411611,0.558484) --
(-0.413472,0.548673) --
(-0.415324,0.538861) --
(-0.417168,0.529048) --
(-0.418995,0.519232) --
(-0.420808,0.509412) --
(-0.422611,0.499591) --
(-0.424395,0.489766) --
(-0.426164,0.479939) --
(-0.427927,0.470111) --
(-0.429679,0.46028) --
(-0.431424,0.450449) --
(-0.433159,0.440616) --
(-0.434872,0.430779) --
(-0.436568,0.420938) --
(-0.438257,0.411097) --
(-0.439936,0.401254) --
(-0.441608,0.39141) --
(-0.443273,0.381565) --
(-0.44492,0.371716) --
(-0.446554,0.361866) --
(-0.448181,0.352014) --
(-0.449794,0.34216) --
(-0.451398,0.332305) --
(-0.452995,0.322448) --
(-0.454581,0.312589) --
(-0.456159,0.302729) --
(-0.457728,0.292868) --
(-0.45928,0.283005) --
(-0.460819,0.273139) --
(-0.462353,0.263272) --
(-0.463881,0.253404) --
(-0.465404,0.243536) --
(-0.466918,0.233666) --
(-0.468411,0.223793) --
(-0.469891,0.213918) --
(-0.471366,0.204043) --
(-0.472836,0.194166) --
(-0.474302,0.18429) --
(-0.47576,0.174411) --
(-0.477199,0.16453) --
(-0.478627,0.154648) --
(-0.48005,0.144765) --
(-0.481467,0.134881) --
(-0.482881,0.124996) --
(-0.484287,0.11511) --
(-0.485675,0.105222) --
(-0.487052,0.0953322) --
(-0.488425,0.0854419) --
(-0.489792,0.0755508) --
(-0.491157,0.0656594) --
(-0.492513,0.0557667) --
(-0.493851,0.0458715) --
(-0.49518,0.0359752) --
(-0.496503,0.026078) --
(-0.497817,0.0161798) --
(-0.499126,0.00628066) --
(-0.500416,-0.00362088) --
(-0.501671,-0.0135268) --
(-0.502906,-0.0234353) --
(-0.504129,-0.0333453) --
(-0.505337,-0.0432569) --
(-0.506539,-0.0531695) --
(-0.507711,-0.0630859) --
(-0.508827,-0.0730085) --
(-0.509914,-0.0829343) --
(-0.510984,-0.0928619) --
(-0.512036,-0.102792) --
(-0.513078,-0.112722) --
(-0.51409,-0.122656) --
(-0.515049,-0.132595) --
(-0.515981,-0.142537) --
(-0.516892,-0.15248) --
(-0.517779,-0.162426) --
(-0.518651,-0.172373) --
(-0.519494,-0.182322) --
(-0.520292,-0.192276) --
(-0.521065,-0.202231) --
(-0.521812,-0.212188) --
(-0.522524,-0.222148) --
(-0.523216,-0.232109) --
(-0.523882,-0.242072) --
(-0.524513,-0.252037) --
(-0.525122,-0.262004) --
(-0.525697,-0.271972) --
(-0.526225,-0.281944) --
(-0.526727,-0.291916) --
(-0.527206,-0.30189) --
(-0.527659,-0.311865) --
(-0.528094,-0.32184) --
(-0.528487,-0.331818) --
(-0.528824,-0.341797) --
(-0.234219,-1.40199) --
(-0.226675,-1.40854) --
(-0.219102,-1.41504) --
(-0.211229,-1.42119) --
(-0.20329,-1.42724) --
(-0.195324,-1.43326) --
(-0.187034,-1.43883) --
(-0.178681,-1.4443) --
(-0.170308,-1.44974) --
(-0.161622,-1.45466) --
(-0.152886,-1.4595) --
(-0.144132,-1.4643) --
(-0.135067,-1.46849) --
(-0.125961,-1.47259) --
(-0.116842,-1.47665) --
(-0.107446,-1.48003) --
(-0.0980227,-1.48334) --
(-0.088591,-1.48661) --
(-0.0789175,-1.48909) --
(-0.0692274,-1.4915) --
(-0.0595326,-1.49389) --
(-0.0496634,-1.49541) --
(-0.0397871,-1.49688) --
(-0.0299095,-1.49834) --
(-0.0199406,-1.49891) --
(-0.00997031,-1.49945) --
(0,-1.5);
\end{scope}

\begin{scope}[shift={(1.33,0)},rotate=-15]

\draw[thick,blue] (0.2636795,-1.2959707) -- ([shift=(-15:0.055)]0.2636795,-1.2959707);
\draw[thick,blue] (0.29314,-1.1899514) -- ([shift=(-15:0.09)]0.29314,-1.1899514);
\draw[thick,blue] (0.3226005,-1.0839321) -- ([shift=(-15:0.107)]0.3226005,-1.0839321);
\draw[thick,blue] (0.352061,-0.9779128) -- ([shift=(-15:0.11)]0.352061,-0.9779128);
\draw[thick,blue] (0.3815215,-0.8718935) -- ([shift=(-15:0.107)]0.3815215,-0.8718935);
\draw[thick,blue] (0.410982,-0.7658742) -- ([shift=(-15:0.1)]0.410982,-0.7658742);
\draw[thick,blue] (0.4404425,-0.6598549) -- ([shift=(-15:0.085)]0.4404425,-0.6598549);
\draw[thick,blue] (0.469903,-0.5538356) -- ([shift=(-15:0.06)]0.469903,-0.5538356);
\draw[thick,blue] (0.4993635,-0.4478163) -- ([shift=(-15:0.035)]0.4993635,-0.4478163);
\draw[thick,blue] (0.24894925,-1.34898035) -- ([shift=(-15:0.03)]0.24894925,-1.34898035);
\draw[thick,blue] (0.27840975,-1.24296105) -- ([shift=(-15:0.075)]0.27840975,-1.24296105);
\draw[thick,blue] (0.30787025,-1.13694175) -- ([shift=(-15:0.1)]0.30787025,-1.13694175);
\draw[thick,blue] (0.33733075,-1.03092245) -- ([shift=(-15:0.11)]0.33733075,-1.03092245);
\draw[thick,blue] (0.36679125,-0.92490315) -- ([shift=(-15:0.11)]0.36679125,-0.92490315);
\draw[thick,blue] (0.39625175,-0.81888385) -- ([shift=(-15:0.105)]0.39625175,-0.81888385);
\draw[thick,blue] (0.42571225,-0.71286455) -- ([shift=(-15:0.093)]0.42571225,-0.71286455);
\draw[thick,blue] (0.45517275,-0.60684525) -- ([shift=(-15:0.075)]0.45517275,-0.60684525);
\draw[thick,blue] (0.48463325,-0.50082595) -- ([shift=(-15:0.05)]0.48463325,-0.50082595);
\draw[thick,blue] (0.51409375,-0.39480665) -- ([shift=(-15:0.02)]0.51409375,-0.39480665);
\draw[thick,blue] (-0.256314375,-1.322475525) -- ([shift=(195:0.042)]-0.256314375,-1.322475525);
\draw[thick,blue] (-0.271044625,-1.269465875) -- ([shift=(195:0.063)]-0.271044625,-1.269465875);
\draw[thick,blue] (-0.285774875,-1.216456225) -- ([shift=(195:0.08)]-0.285774875,-1.216456225);
\draw[thick,blue] (-0.300505125,-1.163446575) -- ([shift=(195:0.09)]-0.300505125,-1.163446575);
\draw[thick,blue] (-0.315235375,-1.110436925) -- ([shift=(195:0.1)]-0.315235375,-1.110436925);
\draw[thick,blue] (-0.329965625,-1.057427275) -- ([shift=(195:0.108)]-0.329965625,-1.057427275);
\draw[thick,blue] (-0.344695875,-1.004417625) -- ([shift=(195:0.11)]-0.344695875,-1.004417625);
\draw[thick,blue] (-0.359426125,-0.951407975) -- ([shift=(195:0.112)]-0.359426125,-0.951407975);
\draw[thick,blue] (-0.374156375,-0.898398325) -- ([shift=(195:0.107)]-0.374156375,-0.898398325);
\draw[thick,blue] (-0.388886625,-0.845388675) -- ([shift=(195:0.105)]-0.388886625,-0.845388675);
\draw[thick,blue] (-0.403616875,-0.792379025) -- ([shift=(195:0.102)]-0.403616875,-0.792379025);
\draw[thick,blue] (-0.418347125,-0.739369375) -- ([shift=(195:0.094)]-0.418347125,-0.739369375);
\draw[thick,blue] (-0.433077375,-0.686359725) -- ([shift=(195:0.086)]-0.433077375,-0.686359725);
\draw[thick,blue] (-0.447807625,-0.633350075) -- ([shift=(195:0.077)]-0.447807625,-0.633350075);
\draw[thick,blue] (-0.462537875,-0.580340425) -- ([shift=(195:0.066)]-0.462537875,-0.580340425);
\draw[thick,blue] (-0.477268125,-0.527330775) -- ([shift=(195:0.054)]-0.477268125,-0.527330775);
\draw[thick,blue] (-0.491998375,-0.474321125) -- ([shift=(195:0.04)]-0.491998375,-0.474321125);
\draw[thick,blue] (-0.506728625,-0.421311475) -- ([shift=(195:0.027)]-0.506728625,-0.421311475);

 \draw[thick]
(0,-1.5) --
(0.00997031,-1.49945) --
(0.0199406,-1.49891) --
(0.0299095,-1.49834) --
(0.0397871,-1.49688) --
(0.0496634,-1.49541) --
(0.0595326,-1.49389) --
(0.0692274,-1.4915) --
(0.0789175,-1.48909) --
(0.088591,-1.48661) --
(0.0980227,-1.48334) --
(0.107446,-1.48003) --
(0.116842,-1.47665) --
(0.125961,-1.47259) --
(0.135067,-1.46849) --
(0.144132,-1.4643) --
(0.152886,-1.4595) --
(0.161622,-1.45466) --
(0.170308,-1.44974) --
(0.178681,-1.4443) --
(0.187034,-1.43883) --
(0.195324,-1.43326) --
(0.20329,-1.42724) --
(0.211229,-1.42119) --
(0.219102,-1.41504) --
(0.226675,-1.40854) --
(0.234219,-1.40199) --
(0.528824,-0.341797) --
(0.528487,-0.331818) --
(0.528094,-0.32184) --
(0.527659,-0.311865) --
(0.527206,-0.30189) --
(0.526727,-0.291916) --
(0.526225,-0.281944) --
(0.525697,-0.271972) --
(0.525122,-0.262004) --
(0.524513,-0.252037) --
(0.523882,-0.242072) --
(0.523216,-0.232109) --
(0.522524,-0.222148) --
(0.521812,-0.212188) --
(0.521065,-0.202231) --
(0.520292,-0.192276) --
(0.519494,-0.182322) --
(0.518651,-0.172373) --
(0.517779,-0.162426) --
(0.516892,-0.15248) --
(0.515981,-0.142537) --
(0.515049,-0.132595) --
(0.51409,-0.122656) --
(0.513078,-0.112722) --
(0.512036,-0.102792) --
(0.510984,-0.0928619) --
(0.509914,-0.0829343) --
(0.508827,-0.0730085) --
(0.507711,-0.0630859) --
(0.506539,-0.0531695) --
(0.505337,-0.0432569) --
(0.504129,-0.0333453) --
(0.502906,-0.0234353) --
(0.501671,-0.0135268) --
(0.500416,-0.00362088) --
(0.499126,0.00628066) --
(0.497817,0.0161798) --
(0.496503,0.026078) --
(0.49518,0.0359752) --
(0.493851,0.0458715) --
(0.492513,0.0557667) --
(0.491157,0.0656594) --
(0.489792,0.0755508) --
(0.488425,0.0854419) --
(0.487052,0.0953322) --
(0.485675,0.105222) --
(0.484287,0.11511) --
(0.482881,0.124996) --
(0.481467,0.134881) --
(0.48005,0.144765) --
(0.478627,0.154648) --
(0.477199,0.16453) --
(0.47576,0.174411) --
(0.474302,0.18429) --
(0.472836,0.194166) --
(0.471366,0.204043) --
(0.469891,0.213918) --
(0.468411,0.223793) --
(0.466918,0.233666) --
(0.465404,0.243536) --
(0.463881,0.253404) --
(0.462353,0.263272) --
(0.460819,0.273139) --
(0.45928,0.283005) --
(0.457728,0.292868) --
(0.456159,0.302729) --
(0.454581,0.312589) --
(0.452995,0.322448) --
(0.451398,0.332305) --
(0.449794,0.34216) --
(0.448181,0.352014) --
(0.446554,0.361866) --
(0.44492,0.371716) --
(0.443273,0.381565) --
(0.441608,0.39141) --
(0.439936,0.401254) --
(0.438257,0.411097) --
(0.436568,0.420938) --
(0.434872,0.430779) --
(0.433159,0.440616) --
(0.431424,0.450449) --
(0.429679,0.46028) --
(0.427927,0.470111) --
(0.426164,0.479939) --
(0.424395,0.489766) --
(0.422611,0.499591) --
(0.420808,0.509412) --
(0.418995,0.519232) --
(0.417168,0.529048) --
(0.415324,0.538861) --
(0.413472,0.548673) --
(0.411611,0.558484) --
(0.409739,0.568292) --
(0.407857,0.578098) --
(0.405954,0.5879) --
(0.404022,0.597697) --
(0.40208,0.607491) --
(0.400127,0.617284) --
(0.39816,0.627073) --
(0.396185,0.636861) --
(0.394197,0.646646) --
(0.392192,0.656428) --
(0.390176,0.666207) --
(0.388137,0.675982) --
(0.386069,0.685751) --
(0.383992,0.695517) --
(0.381899,0.705281) --
(0.37979,0.715041) --
(0.377674,0.724799) --
(0.375541,0.734554) --
(0.37339,0.744305) --
(0.371228,0.754053) --
(0.369034,0.763794) --
(0.366809,0.773529) --
(0.364573,0.78326) --
(0.362317,0.792987) --
(0.360042,0.80271) --
(0.357758,0.81243) --
(0.355452,0.822145) --
(0.353124,0.831855) --
(0.350786,0.841563) --
(0.348419,0.851263) --
(0.346024,0.860957) --
(0.343617,0.870648) --
(0.341176,0.880331) --
(0.338705,0.890005) --
(0.336223,0.899677) --
(0.333711,0.90934) --
(0.331174,0.918998) --
(0.328626,0.928653) --
(0.326045,0.938298) --
(0.323438,0.947937) --
(0.320818,0.957573) --
(0.31816,0.967198) --
(0.315473,0.976815) --
(0.312773,0.986428) --
(0.310027,0.996028) --
(0.307248,1.00562) --
(0.304456,1.01521) --
(0.301613,1.02478) --
(0.298736,1.03434) --
(0.295845,1.0439) --
(0.292903,1.05344) --
(0.289928,1.06297) --
(0.28694,1.0725) --
(0.283894,1.08201) --
(0.280812,1.0915) --
(0.277716,1.101) --
(0.274555,1.11047) --
(0.271357,1.11993) --
(0.268143,1.12938) --
(0.264856,1.13881) --
(0.26153,1.14823) --
(0.258187,1.15763) --
(0.25476,1.16701) --
(0.251293,1.17638) --
(0.247808,1.18573) --
(0.244228,1.19506) --
(0.240603,1.20436) --
(0.236961,1.21366) --
(0.233211,1.22291) --
(0.229413,1.23215) --
(0.225598,1.24137) --
(0.221657,1.25055) --
(0.217668,1.2597) --
(0.213659,1.26885) --
(0.2095,1.27792) --
(0.205287,1.28698) --
(0.201053,1.29602) --
(0.196639,1.30498) --
(0.192167,1.31391) --
(0.187673,1.32282) --
(0.18297,1.33163) --
(0.178207,1.34041) --
(0.17342,1.34917) --
(0.168381,1.35779) --
(0.163278,1.36637) --
(0.158149,1.37494) --
(0.152704,1.38331) --
(0.147188,1.39163) --
(0.141647,1.39994) --
(0.13571,1.40797) --
(0.129698,1.41594) --
(0.123659,1.42389) --
(0.117121,1.43144) --
(0.110502,1.43892) --
(0.103856,1.44637) --
(0.0965752,1.4532) --
(0.0892133,1.45995) --
(0.0818262,1.46666) --
(0.0736564,1.47241) --
(0.065414,1.47804) --
(0.05715,1.48365) --
(0.0480044,1.48765) --
(0.0388073,1.49154) --
(0.0296006,1.49541) --
(0.0197427,1.49699) --
(0.00987107,1.4985) --
(0,1.5) --
(-0.00987107,1.4985) --
(-0.0197427,1.49699) --
(-0.0296006,1.49541) --
(-0.0388073,1.49154) --
(-0.0480044,1.48765) --
(-0.05715,1.48365) --
(-0.065414,1.47804) --
(-0.0736564,1.47241) --
(-0.0818262,1.46666) --
(-0.0892133,1.45995) --
(-0.0965752,1.4532) --
(-0.103856,1.44637) --
(-0.110502,1.43892) --
(-0.117121,1.43144) --
(-0.123659,1.42389) --
(-0.129698,1.41594) --
(-0.13571,1.40797) --
(-0.141647,1.39994) --
(-0.147188,1.39163) --
(-0.152704,1.38331) --
(-0.158149,1.37494) --
(-0.163278,1.36637) --
(-0.168381,1.35779) --
(-0.17342,1.34917) --
(-0.178207,1.34041) --
(-0.18297,1.33163) --
(-0.187673,1.32282) --
(-0.192167,1.31391) --
(-0.196639,1.30498) --
(-0.201053,1.29602) --
(-0.205287,1.28698) --
(-0.2095,1.27792) --
(-0.213659,1.26885) --
(-0.217668,1.2597) --
(-0.221657,1.25055) --
(-0.225598,1.24137) --
(-0.229413,1.23215) --
(-0.233211,1.22291) --
(-0.236961,1.21366) --
(-0.240603,1.20436) --
(-0.244228,1.19506) --
(-0.247808,1.18573) --
(-0.251293,1.17638) --
(-0.25476,1.16701) --
(-0.258187,1.15763) --
(-0.26153,1.14823) --
(-0.264856,1.13881) --
(-0.268143,1.12938) --
(-0.271357,1.11993) --
(-0.274555,1.11047) --
(-0.277716,1.101) --
(-0.280812,1.0915) --
(-0.283894,1.08201) --
(-0.28694,1.0725) --
(-0.289928,1.06297) --
(-0.292903,1.05344) --
(-0.295845,1.0439) --
(-0.298736,1.03434) --
(-0.301613,1.02478) --
(-0.304456,1.01521) --
(-0.307248,1.00562) --
(-0.310027,0.996028) --
(-0.312773,0.986428) --
(-0.315473,0.976815) --
(-0.31816,0.967198) --
(-0.320818,0.957573) --
(-0.323438,0.947937) --
(-0.326045,0.938298) --
(-0.328626,0.928653) --
(-0.331174,0.918998) --
(-0.333711,0.90934) --
(-0.336223,0.899677) --
(-0.338705,0.890005) --
(-0.341176,0.880331) --
(-0.343617,0.870648) --
(-0.346024,0.860957) --
(-0.348419,0.851263) --
(-0.350786,0.841563) --
(-0.353124,0.831855) --
(-0.355452,0.822145) --
(-0.357758,0.81243) --
(-0.360042,0.80271) --
(-0.362317,0.792987) --
(-0.364573,0.78326) --
(-0.366809,0.773529) --
(-0.369034,0.763794) --
(-0.371228,0.754053) --
(-0.37339,0.744305) --
(-0.375541,0.734554) --
(-0.377674,0.724799) --
(-0.37979,0.715041) --
(-0.381899,0.705281) --
(-0.383992,0.695517) --
(-0.386069,0.685751) --
(-0.388137,0.675982) --
(-0.390176,0.666207) --
(-0.392192,0.656428) --
(-0.394197,0.646646) --
(-0.396185,0.636861) --
(-0.39816,0.627073) --
(-0.400127,0.617284) --
(-0.40208,0.607491) --
(-0.404022,0.597697) --
(-0.405954,0.5879) --
(-0.407857,0.578098) --
(-0.409739,0.568292) --
(-0.411611,0.558484) --
(-0.413472,0.548673) --
(-0.415324,0.538861) --
(-0.417168,0.529048) --
(-0.418995,0.519232) --
(-0.420808,0.509412) --
(-0.422611,0.499591) --
(-0.424395,0.489766) --
(-0.426164,0.479939) --
(-0.427927,0.470111) --
(-0.429679,0.46028) --
(-0.431424,0.450449) --
(-0.433159,0.440616) --
(-0.434872,0.430779) --
(-0.436568,0.420938) --
(-0.438257,0.411097) --
(-0.439936,0.401254) --
(-0.441608,0.39141) --
(-0.443273,0.381565) --
(-0.44492,0.371716) --
(-0.446554,0.361866) --
(-0.448181,0.352014) --
(-0.449794,0.34216) --
(-0.451398,0.332305) --
(-0.452995,0.322448) --
(-0.454581,0.312589) --
(-0.456159,0.302729) --
(-0.457728,0.292868) --
(-0.45928,0.283005) --
(-0.460819,0.273139) --
(-0.462353,0.263272) --
(-0.463881,0.253404) --
(-0.465404,0.243536) --
(-0.466918,0.233666) --
(-0.468411,0.223793) --
(-0.469891,0.213918) --
(-0.471366,0.204043) --
(-0.472836,0.194166) --
(-0.474302,0.18429) --
(-0.47576,0.174411) --
(-0.477199,0.16453) --
(-0.478627,0.154648) --
(-0.48005,0.144765) --
(-0.481467,0.134881) --
(-0.482881,0.124996) --
(-0.484287,0.11511) --
(-0.485675,0.105222) --
(-0.487052,0.0953322) --
(-0.488425,0.0854419) --
(-0.489792,0.0755508) --
(-0.491157,0.0656594) --
(-0.492513,0.0557667) --
(-0.493851,0.0458715) --
(-0.49518,0.0359752) --
(-0.496503,0.026078) --
(-0.497817,0.0161798) --
(-0.499126,0.00628066) --
(-0.500416,-0.00362088) --
(-0.501671,-0.0135268) --
(-0.502906,-0.0234353) --
(-0.504129,-0.0333453) --
(-0.505337,-0.0432569) --
(-0.506539,-0.0531695) --
(-0.507711,-0.0630859) --
(-0.508827,-0.0730085) --
(-0.509914,-0.0829343) --
(-0.510984,-0.0928619) --
(-0.512036,-0.102792) --
(-0.513078,-0.112722) --
(-0.51409,-0.122656) --
(-0.515049,-0.132595) --
(-0.515981,-0.142537) --
(-0.516892,-0.15248) --
(-0.517779,-0.162426) --
(-0.518651,-0.172373) --
(-0.519494,-0.182322) --
(-0.520292,-0.192276) --
(-0.521065,-0.202231) --
(-0.521812,-0.212188) --
(-0.522524,-0.222148) --
(-0.523216,-0.232109) --
(-0.523882,-0.242072) --
(-0.524513,-0.252037) --
(-0.525122,-0.262004) --
(-0.525697,-0.271972) --
(-0.526225,-0.281944) --
(-0.526727,-0.291916) --
(-0.527206,-0.30189) --
(-0.527659,-0.311865) --
(-0.528094,-0.32184) --
(-0.528487,-0.331818) --
(-0.528824,-0.341797) --
(-0.234219,-1.40199) --
(-0.226675,-1.40854) --
(-0.219102,-1.41504) --
(-0.211229,-1.42119) --
(-0.20329,-1.42724) --
(-0.195324,-1.43326) --
(-0.187034,-1.43883) --
(-0.178681,-1.4443) --
(-0.170308,-1.44974) --
(-0.161622,-1.45466) --
(-0.152886,-1.4595) --
(-0.144132,-1.4643) --
(-0.135067,-1.46849) --
(-0.125961,-1.47259) --
(-0.116842,-1.47665) --
(-0.107446,-1.48003) --
(-0.0980227,-1.48334) --
(-0.088591,-1.48661) --
(-0.0789175,-1.48909) --
(-0.0692274,-1.4915) --
(-0.0595326,-1.49389) --
(-0.0496634,-1.49541) --
(-0.0397871,-1.49688) --
(-0.0299095,-1.49834) --
(-0.0199406,-1.49891) --
(-0.00997031,-1.49945) --
(0,-1.5);
\end{scope}

\begin{scope}[shift={(2.65,-0.4)},rotate=130]

\draw[thick,blue] (0.2636795,-1.2959707) -- ([shift=(-15:0.055)]0.2636795,-1.2959707);
\draw[thick,blue] (0.29314,-1.1899514) -- ([shift=(-15:0.09)]0.29314,-1.1899514);
\draw[thick,blue] (0.3226005,-1.0839321) -- ([shift=(-15:0.107)]0.3226005,-1.0839321);
\draw[thick,blue] (0.352061,-0.9779128) -- ([shift=(-15:0.11)]0.352061,-0.9779128);
\draw[thick,blue] (0.3815215,-0.8718935) -- ([shift=(-15:0.107)]0.3815215,-0.8718935);
\draw[thick,blue] (0.410982,-0.7658742) -- ([shift=(-15:0.1)]0.410982,-0.7658742);
\draw[thick,blue] (0.4404425,-0.6598549) -- ([shift=(-15:0.085)]0.4404425,-0.6598549);
\draw[thick,blue] (0.469903,-0.5538356) -- ([shift=(-15:0.06)]0.469903,-0.5538356);
\draw[thick,blue] (0.4993635,-0.4478163) -- ([shift=(-15:0.035)]0.4993635,-0.4478163);
\draw[thick,blue] (0.24894925,-1.34898035) -- ([shift=(-15:0.03)]0.24894925,-1.34898035);
\draw[thick,blue] (0.27840975,-1.24296105) -- ([shift=(-15:0.075)]0.27840975,-1.24296105);
\draw[thick,blue] (0.30787025,-1.13694175) -- ([shift=(-15:0.1)]0.30787025,-1.13694175);
\draw[thick,blue] (0.33733075,-1.03092245) -- ([shift=(-15:0.11)]0.33733075,-1.03092245);
\draw[thick,blue] (0.36679125,-0.92490315) -- ([shift=(-15:0.11)]0.36679125,-0.92490315);
\draw[thick,blue] (0.39625175,-0.81888385) -- ([shift=(-15:0.105)]0.39625175,-0.81888385);
\draw[thick,blue] (0.42571225,-0.71286455) -- ([shift=(-15:0.093)]0.42571225,-0.71286455);
\draw[thick,blue] (0.45517275,-0.60684525) -- ([shift=(-15:0.075)]0.45517275,-0.60684525);
\draw[thick,blue] (0.48463325,-0.50082595) -- ([shift=(-15:0.05)]0.48463325,-0.50082595);
\draw[thick,blue] (0.51409375,-0.39480665) -- ([shift=(-15:0.02)]0.51409375,-0.39480665);

\draw[thick,blue] (-0.256314375,-1.322475525) -- ([shift=(195:0.042)]-0.256314375,-1.322475525);
\draw[thick,blue] (-0.271044625,-1.269465875) -- ([shift=(195:0.063)]-0.271044625,-1.269465875);
\draw[thick,blue] (-0.285774875,-1.216456225) -- ([shift=(195:0.08)]-0.285774875,-1.216456225);
\draw[thick,blue] (-0.300505125,-1.163446575) -- ([shift=(195:0.09)]-0.300505125,-1.163446575);
\draw[thick,blue] (-0.315235375,-1.110436925) -- ([shift=(195:0.1)]-0.315235375,-1.110436925);
\draw[thick,blue] (-0.329965625,-1.057427275) -- ([shift=(195:0.108)]-0.329965625,-1.057427275);
\draw[thick,blue] (-0.344695875,-1.004417625) -- ([shift=(195:0.11)]-0.344695875,-1.004417625);
\draw[thick,blue] (-0.359426125,-0.951407975) -- ([shift=(195:0.112)]-0.359426125,-0.951407975);
\draw[thick,blue] (-0.374156375,-0.898398325) -- ([shift=(195:0.107)]-0.374156375,-0.898398325);
\draw[thick,blue] (-0.388886625,-0.845388675) -- ([shift=(195:0.105)]-0.388886625,-0.845388675);
\draw[thick,blue] (-0.403616875,-0.792379025) -- ([shift=(195:0.102)]-0.403616875,-0.792379025);
\draw[thick,blue] (-0.418347125,-0.739369375) -- ([shift=(195:0.094)]-0.418347125,-0.739369375);
\draw[thick,blue] (-0.433077375,-0.686359725) -- ([shift=(195:0.086)]-0.433077375,-0.686359725);
\draw[thick,blue] (-0.447807625,-0.633350075) -- ([shift=(195:0.077)]-0.447807625,-0.633350075);
\draw[thick,blue] (-0.462537875,-0.580340425) -- ([shift=(195:0.066)]-0.462537875,-0.580340425);
\draw[thick,blue] (-0.477268125,-0.527330775) -- ([shift=(195:0.054)]-0.477268125,-0.527330775);
\draw[thick,blue] (-0.491998375,-0.474321125) -- ([shift=(195:0.04)]-0.491998375,-0.474321125);
\draw[thick,blue] (-0.506728625,-0.421311475) -- ([shift=(195:0.027)]-0.506728625,-0.421311475);

 \draw[thick]
(0,-1.5) --
(0.00997031,-1.49945) --
(0.0199406,-1.49891) --
(0.0299095,-1.49834) --
(0.0397871,-1.49688) --
(0.0496634,-1.49541) --
(0.0595326,-1.49389) --
(0.0692274,-1.4915) --
(0.0789175,-1.48909) --
(0.088591,-1.48661) --
(0.0980227,-1.48334) --
(0.107446,-1.48003) --
(0.116842,-1.47665) --
(0.125961,-1.47259) --
(0.135067,-1.46849) --
(0.144132,-1.4643) --
(0.152886,-1.4595) --
(0.161622,-1.45466) --
(0.170308,-1.44974) --
(0.178681,-1.4443) --
(0.187034,-1.43883) --
(0.195324,-1.43326) --
(0.20329,-1.42724) --
(0.211229,-1.42119) --
(0.219102,-1.41504) --
(0.226675,-1.40854) --
(0.234219,-1.40199) --
(0.528824,-0.341797) --
(0.528487,-0.331818) --
(0.528094,-0.32184) --
(0.527659,-0.311865) --
(0.527206,-0.30189) --
(0.526727,-0.291916) --
(0.526225,-0.281944) --
(0.525697,-0.271972) --
(0.525122,-0.262004) --
(0.524513,-0.252037) --
(0.523882,-0.242072) --
(0.523216,-0.232109) --
(0.522524,-0.222148) --
(0.521812,-0.212188) --
(0.521065,-0.202231) --
(0.520292,-0.192276) --
(0.519494,-0.182322) --
(0.518651,-0.172373) --
(0.517779,-0.162426) --
(0.516892,-0.15248) --
(0.515981,-0.142537) --
(0.515049,-0.132595) --
(0.51409,-0.122656) --
(0.513078,-0.112722) --
(0.512036,-0.102792) --
(0.510984,-0.0928619) --
(0.509914,-0.0829343) --
(0.508827,-0.0730085) --
(0.507711,-0.0630859) --
(0.506539,-0.0531695) --
(0.505337,-0.0432569) --
(0.504129,-0.0333453) --
(0.502906,-0.0234353) --
(0.501671,-0.0135268) --
(0.500416,-0.00362088) --
(0.499126,0.00628066) --
(0.497817,0.0161798) --
(0.496503,0.026078) --
(0.49518,0.0359752) --
(0.493851,0.0458715) --
(0.492513,0.0557667) --
(0.491157,0.0656594) --
(0.489792,0.0755508) --
(0.488425,0.0854419) --
(0.487052,0.0953322) --
(0.485675,0.105222) --
(0.484287,0.11511) --
(0.482881,0.124996) --
(0.481467,0.134881) --
(0.48005,0.144765) --
(0.478627,0.154648) --
(0.477199,0.16453) --
(0.47576,0.174411) --
(0.474302,0.18429) --
(0.472836,0.194166) --
(0.471366,0.204043) --
(0.469891,0.213918) --
(0.468411,0.223793) --
(0.466918,0.233666) --
(0.465404,0.243536) --
(0.463881,0.253404) --
(0.462353,0.263272) --
(0.460819,0.273139) --
(0.45928,0.283005) --
(0.457728,0.292868) --
(0.456159,0.302729) --
(0.454581,0.312589) --
(0.452995,0.322448) --
(0.451398,0.332305) --
(0.449794,0.34216) --
(0.448181,0.352014) --
(0.446554,0.361866) --
(0.44492,0.371716) --
(0.443273,0.381565) --
(0.441608,0.39141) --
(0.439936,0.401254) --
(0.438257,0.411097) --
(0.436568,0.420938) --
(0.434872,0.430779) --
(0.433159,0.440616) --
(0.431424,0.450449) --
(0.429679,0.46028) --
(0.427927,0.470111) --
(0.426164,0.479939) --
(0.424395,0.489766) --
(0.422611,0.499591) --
(0.420808,0.509412) --
(0.418995,0.519232) --
(0.417168,0.529048) --
(0.415324,0.538861) --
(0.413472,0.548673) --
(0.411611,0.558484) --
(0.409739,0.568292) --
(0.407857,0.578098) --
(0.405954,0.5879) --
(0.404022,0.597697) --
(0.40208,0.607491) --
(0.400127,0.617284) --
(0.39816,0.627073) --
(0.396185,0.636861) --
(0.394197,0.646646) --
(0.392192,0.656428) --
(0.390176,0.666207) --
(0.388137,0.675982) --
(0.386069,0.685751) --
(0.383992,0.695517) --
(0.381899,0.705281) --
(0.37979,0.715041) --
(0.377674,0.724799) --
(0.375541,0.734554) --
(0.37339,0.744305) --
(0.371228,0.754053) --
(0.369034,0.763794) --
(0.366809,0.773529) --
(0.364573,0.78326) --
(0.362317,0.792987) --
(0.360042,0.80271) --
(0.357758,0.81243) --
(0.355452,0.822145) --
(0.353124,0.831855) --
(0.350786,0.841563) --
(0.348419,0.851263) --
(0.346024,0.860957) --
(0.343617,0.870648) --
(0.341176,0.880331) --
(0.338705,0.890005) --
(0.336223,0.899677) --
(0.333711,0.90934) --
(0.331174,0.918998) --
(0.328626,0.928653) --
(0.326045,0.938298) --
(0.323438,0.947937) --
(0.320818,0.957573) --
(0.31816,0.967198) --
(0.315473,0.976815) --
(0.312773,0.986428) --
(0.310027,0.996028) --
(0.307248,1.00562) --
(0.304456,1.01521) --
(0.301613,1.02478) --
(0.298736,1.03434) --
(0.295845,1.0439) --
(0.292903,1.05344) --
(0.289928,1.06297) --
(0.28694,1.0725) --
(0.283894,1.08201) --
(0.280812,1.0915) --
(0.277716,1.101) --
(0.274555,1.11047) --
(0.271357,1.11993) --
(0.268143,1.12938) --
(0.264856,1.13881) --
(0.26153,1.14823) --
(0.258187,1.15763) --
(0.25476,1.16701) --
(0.251293,1.17638) --
(0.247808,1.18573) --
(0.244228,1.19506) --
(0.240603,1.20436) --
(0.236961,1.21366) --
(0.233211,1.22291) --
(0.229413,1.23215) --
(0.225598,1.24137) --
(0.221657,1.25055) --
(0.217668,1.2597) --
(0.213659,1.26885) --
(0.2095,1.27792) --
(0.205287,1.28698) --
(0.201053,1.29602) --
(0.196639,1.30498) --
(0.192167,1.31391) --
(0.187673,1.32282) --
(0.18297,1.33163) --
(0.178207,1.34041) --
(0.17342,1.34917) --
(0.168381,1.35779) --
(0.163278,1.36637) --
(0.158149,1.37494) --
(0.152704,1.38331) --
(0.147188,1.39163) --
(0.141647,1.39994) --
(0.13571,1.40797) --
(0.129698,1.41594) --
(0.123659,1.42389) --
(0.117121,1.43144) --
(0.110502,1.43892) --
(0.103856,1.44637) --
(0.0965752,1.4532) --
(0.0892133,1.45995) --
(0.0818262,1.46666) --
(0.0736564,1.47241) --
(0.065414,1.47804) --
(0.05715,1.48365) --
(0.0480044,1.48765) --
(0.0388073,1.49154) --
(0.0296006,1.49541) --
(0.0197427,1.49699) --
(0.00987107,1.4985) --
(0,1.5) --
(-0.00987107,1.4985) --
(-0.0197427,1.49699) --
(-0.0296006,1.49541) --
(-0.0388073,1.49154) --
(-0.0480044,1.48765) --
(-0.05715,1.48365) --
(-0.065414,1.47804) --
(-0.0736564,1.47241) --
(-0.0818262,1.46666) --
(-0.0892133,1.45995) --
(-0.0965752,1.4532) --
(-0.103856,1.44637) --
(-0.110502,1.43892) --
(-0.117121,1.43144) --
(-0.123659,1.42389) --
(-0.129698,1.41594) --
(-0.13571,1.40797) --
(-0.141647,1.39994) --
(-0.147188,1.39163) --
(-0.152704,1.38331) --
(-0.158149,1.37494) --
(-0.163278,1.36637) --
(-0.168381,1.35779) --
(-0.17342,1.34917) --
(-0.178207,1.34041) --
(-0.18297,1.33163) --
(-0.187673,1.32282) --
(-0.192167,1.31391) --
(-0.196639,1.30498) --
(-0.201053,1.29602) --
(-0.205287,1.28698) --
(-0.2095,1.27792) --
(-0.213659,1.26885) --
(-0.217668,1.2597) --
(-0.221657,1.25055) --
(-0.225598,1.24137) --
(-0.229413,1.23215) --
(-0.233211,1.22291) --
(-0.236961,1.21366) --
(-0.240603,1.20436) --
(-0.244228,1.19506) --
(-0.247808,1.18573) --
(-0.251293,1.17638) --
(-0.25476,1.16701) --
(-0.258187,1.15763) --
(-0.26153,1.14823) --
(-0.264856,1.13881) --
(-0.268143,1.12938) --
(-0.271357,1.11993) --
(-0.274555,1.11047) --
(-0.277716,1.101) --
(-0.280812,1.0915) --
(-0.283894,1.08201) --
(-0.28694,1.0725) --
(-0.289928,1.06297) --
(-0.292903,1.05344) --
(-0.295845,1.0439) --
(-0.298736,1.03434) --
(-0.301613,1.02478) --
(-0.304456,1.01521) --
(-0.307248,1.00562) --
(-0.310027,0.996028) --
(-0.312773,0.986428) --
(-0.315473,0.976815) --
(-0.31816,0.967198) --
(-0.320818,0.957573) --
(-0.323438,0.947937) --
(-0.326045,0.938298) --
(-0.328626,0.928653) --
(-0.331174,0.918998) --
(-0.333711,0.90934) --
(-0.336223,0.899677) --
(-0.338705,0.890005) --
(-0.341176,0.880331) --
(-0.343617,0.870648) --
(-0.346024,0.860957) --
(-0.348419,0.851263) --
(-0.350786,0.841563) --
(-0.353124,0.831855) --
(-0.355452,0.822145) --
(-0.357758,0.81243) --
(-0.360042,0.80271) --
(-0.362317,0.792987) --
(-0.364573,0.78326) --
(-0.366809,0.773529) --
(-0.369034,0.763794) --
(-0.371228,0.754053) --
(-0.37339,0.744305) --
(-0.375541,0.734554) --
(-0.377674,0.724799) --
(-0.37979,0.715041) --
(-0.381899,0.705281) --
(-0.383992,0.695517) --
(-0.386069,0.685751) --
(-0.388137,0.675982) --
(-0.390176,0.666207) --
(-0.392192,0.656428) --
(-0.394197,0.646646) --
(-0.396185,0.636861) --
(-0.39816,0.627073) --
(-0.400127,0.617284) --
(-0.40208,0.607491) --
(-0.404022,0.597697) --
(-0.405954,0.5879) --
(-0.407857,0.578098) --
(-0.409739,0.568292) --
(-0.411611,0.558484) --
(-0.413472,0.548673) --
(-0.415324,0.538861) --
(-0.417168,0.529048) --
(-0.418995,0.519232) --
(-0.420808,0.509412) --
(-0.422611,0.499591) --
(-0.424395,0.489766) --
(-0.426164,0.479939) --
(-0.427927,0.470111) --
(-0.429679,0.46028) --
(-0.431424,0.450449) --
(-0.433159,0.440616) --
(-0.434872,0.430779) --
(-0.436568,0.420938) --
(-0.438257,0.411097) --
(-0.439936,0.401254) --
(-0.441608,0.39141) --
(-0.443273,0.381565) --
(-0.44492,0.371716) --
(-0.446554,0.361866) --
(-0.448181,0.352014) --
(-0.449794,0.34216) --
(-0.451398,0.332305) --
(-0.452995,0.322448) --
(-0.454581,0.312589) --
(-0.456159,0.302729) --
(-0.457728,0.292868) --
(-0.45928,0.283005) --
(-0.460819,0.273139) --
(-0.462353,0.263272) --
(-0.463881,0.253404) --
(-0.465404,0.243536) --
(-0.466918,0.233666) --
(-0.468411,0.223793) --
(-0.469891,0.213918) --
(-0.471366,0.204043) --
(-0.472836,0.194166) --
(-0.474302,0.18429) --
(-0.47576,0.174411) --
(-0.477199,0.16453) --
(-0.478627,0.154648) --
(-0.48005,0.144765) --
(-0.481467,0.134881) --
(-0.482881,0.124996) --
(-0.484287,0.11511) --
(-0.485675,0.105222) --
(-0.487052,0.0953322) --
(-0.488425,0.0854419) --
(-0.489792,0.0755508) --
(-0.491157,0.0656594) --
(-0.492513,0.0557667) --
(-0.493851,0.0458715) --
(-0.49518,0.0359752) --
(-0.496503,0.026078) --
(-0.497817,0.0161798) --
(-0.499126,0.00628066) --
(-0.500416,-0.00362088) --
(-0.501671,-0.0135268) --
(-0.502906,-0.0234353) --
(-0.504129,-0.0333453) --
(-0.505337,-0.0432569) --
(-0.506539,-0.0531695) --
(-0.507711,-0.0630859) --
(-0.508827,-0.0730085) --
(-0.509914,-0.0829343) --
(-0.510984,-0.0928619) --
(-0.512036,-0.102792) --
(-0.513078,-0.112722) --
(-0.51409,-0.122656) --
(-0.515049,-0.132595) --
(-0.515981,-0.142537) --
(-0.516892,-0.15248) --
(-0.517779,-0.162426) --
(-0.518651,-0.172373) --
(-0.519494,-0.182322) --
(-0.520292,-0.192276) --
(-0.521065,-0.202231) --
(-0.521812,-0.212188) --
(-0.522524,-0.222148) --
(-0.523216,-0.232109) --
(-0.523882,-0.242072) --
(-0.524513,-0.252037) --
(-0.525122,-0.262004) --
(-0.525697,-0.271972) --
(-0.526225,-0.281944) --
(-0.526727,-0.291916) --
(-0.527206,-0.30189) --
(-0.527659,-0.311865) --
(-0.528094,-0.32184) --
(-0.528487,-0.331818) --
(-0.528824,-0.341797) --
(-0.234219,-1.40199) --
(-0.226675,-1.40854) --
(-0.219102,-1.41504) --
(-0.211229,-1.42119) --
(-0.20329,-1.42724) --
(-0.195324,-1.43326) --
(-0.187034,-1.43883) --
(-0.178681,-1.4443) --
(-0.170308,-1.44974) --
(-0.161622,-1.45466) --
(-0.152886,-1.4595) --
(-0.144132,-1.4643) --
(-0.135067,-1.46849) --
(-0.125961,-1.47259) --
(-0.116842,-1.47665) --
(-0.107446,-1.48003) --
(-0.0980227,-1.48334) --
(-0.088591,-1.48661) --
(-0.0789175,-1.48909) --
(-0.0692274,-1.4915) --
(-0.0595326,-1.49389) --
(-0.0496634,-1.49541) --
(-0.0397871,-1.49688) --
(-0.0299095,-1.49834) --
(-0.0199406,-1.49891) --
(-0.00997031,-1.49945) --
(0,-1.5);
\end{scope}
\end{scope}

    \begin{axis}[
    scale only axis=true,
        ylabel=$x$,
        xlabel=$z$,
        xtick pos=left,
        ytick pos=left,
        xmin = -1.7,
        xmax = 1.7,
        ymin = 0,
        ymax = 0.7,
        width=0.75*0.9\textwidth,
        height=0.75*0.1845\textwidth,
        x tick label style={
        /pgf/number format/.cd,
            fixed,
            fixed zerofill,
            precision=1,
        /tikz/.cd
        },
        ylabel near ticks,
        ylabel style={rotate=-90},
        y tick label style={
        /pgf/number format/.cd,
            fixed,
            fixed zerofill,
            precision=1,
        /tikz/.cd
        },
        legend columns=2,
        legend style={
at={(0.5,1)},
anchor=south,
/tikz/every even column/.append style={column sep=1.656cm},
}
        ]
   \addplot[draw,dashed,thick] table[x expr= \thisrow{y}, y expr= \thisrow{x}, col sep=comma] {images/Data/Bezier3-3.8.txt}; 
    \addplot[draw,dotted,thick] table[x expr= \thisrow{y}, y expr= \thisrow{x}, col sep=comma] {images/Data/BezierD.txt}; 
    \addplot[draw,dashdotted,thick] table[x expr= \thisrow{y}, y expr= \thisrow{x}, col sep=comma] {images/Data/BezierN.txt}; 
    \addplot[draw,thick] table[x expr= \thisrow{y}, y expr= \thisrow{x}, col sep=comma] {images/Data/BezierH.txt}; 
    \legend{HPR,\quad PHGO-based overlap \quad,\quad equivalent self-non-additive shape \qquad, NAHPR}
    \end{axis}

\end{tikzpicture}

%% file: images/ori_nonadd.tex
    \begin{tikzpicture}
    \node[] at (0.62735\textwidth,0.11305\textwidth){\includegraphics[height=0.19\textwidth]{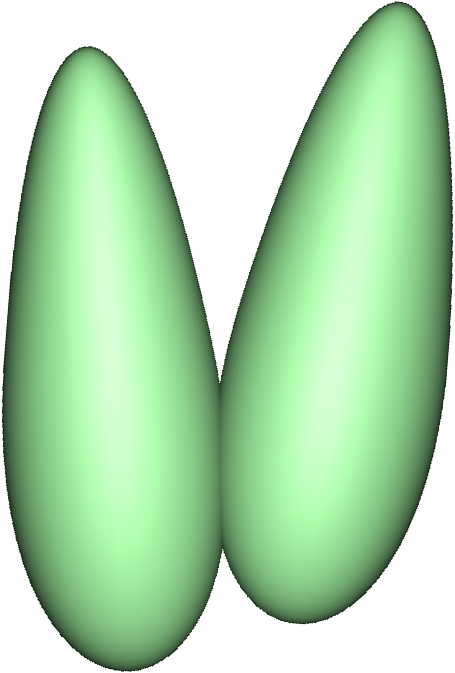}};
    \draw[red,ultra thick,densely dotted] (0.5301\textwidth,0.0158\textwidth) rectangle (0.7246\textwidth,0.2103\textwidth);  
    \node[] at (0.62735\textwidth,-0.10145\textwidth){\includegraphics[height=0.19\textwidth]{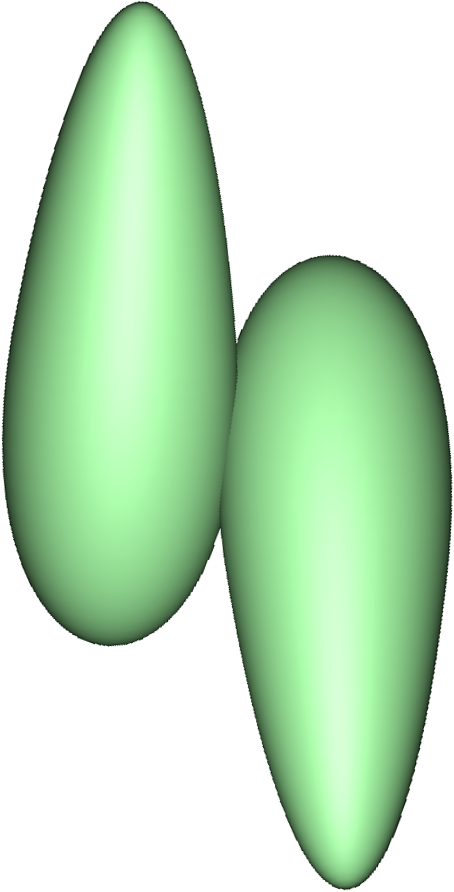}};
    \draw[cyan,ultra thick,densely dashdotted] (0.5301\textwidth,-0.1987\textwidth) rectangle (0.7246\textwidth,-0.0042\textwidth);  

    \draw[ultra thick] (0.5301\textwidth,-0.2187\textwidth) rectangle (0.7246\textwidth,-0.2787\textwidth);  
    \node at (0.62735\textwidth,-0.2487\textwidth) {\Large \textbf{NAHPR}};

    \begin{axis}[
        ylabel=ocurrence $\lbrack a.u. \rbrack$,
        xlabel=relative angle $\alpha$,
        xtick pos=left,
        ytick pos=left,
        xmin = -3.15,
        xmax = 3.15,
        ymin = 0.0,
        ymax = 1,
        ybar interval,
        width=0.6\textwidth,
        height=0.3\textwidth,
        xtick={-3.14159,-2.356,-1.5708,-0.7854,0,0.7854,1.5708,2.356,3.14159},
        xticklabels={,,,,,,,,},
        y tick label style={
        /pgf/number format/.cd,
            fixed,
            fixed zerofill,
            precision=1,
        /tikz/.cd
        },
        ylabel absolute,
        ]
    \addplot table[x expr= \thisrow{alpha}, y expr= \thisrow{norm}/0.1, col sep=comma] {images/Data/oriNc2Ns1498Vr0_08rho0_45NonAdd.csv}; 
    \draw[cyan,ultra thick,densely dashdotted] (axis cs:-3.1,0) -- (axis cs:-3.1,1.2);  
    \draw[cyan,ultra thick,densely dashdotted] (axis cs:3.1,0) -- (axis cs:3.1,1.2);  
    \draw[blue,ultra thick,densely dashed] (axis cs:-0.26,0) -- (axis cs:-0.26,1.2);  
    \draw[red,ultra thick,densely dotted] (axis cs:0.36,0) -- (axis cs:0.36,1.2);  
    \end{axis}
    \node at (0,-0.0223\textwidth) {-$\pi$}; 
    \node at (0.06395\textwidth,-0.0223\textwidth) {-$\frac{3\pi}{4}$}; 
    \node at (0.1279\textwidth,-0.0223\textwidth) {-$\frac{\pi}{2}$}; 
    \node at (0.19185\textwidth,-0.0223\textwidth) {-$\frac{\pi}{4}$}; 
    \node at (0.2558\textwidth,-0.0223\textwidth) {$0$}; 
    \node at (0.31975\textwidth,-0.0223\textwidth) {$\frac{\pi}{4}$}; 
    \node at (0.3837\textwidth,-0.0223\textwidth) {$\frac{\pi}{2}$}; 
    \node at (0.44765\textwidth,-0.0223\textwidth) {$\frac{3\pi}{4}$}; 
    \node at (0.5116\textwidth,-0.0223\textwidth) {$\pi$}; 

    \draw[ultra thick,<->] (0,0.2232\textwidth) -- (0.2558\textwidth,0.2232\textwidth);  
    \node[anchor=west] at (0.0279\textwidth,0.2957\textwidth) {``A-config.''};
    \draw[ultra thick,<->] (0.5116\textwidth,0.2232\textwidth) -- (0.2558\textwidth,0.2232\textwidth);  
    \node[anchor=east] at (0.4799\textwidth,0.2957\textwidth) {``V-config.''};
\begin{scope}[shift={(0.3459\textwidth,0.3292\textwidth)},scale=0.5,rotate=90]
    \begin{scope}[shift={(-3,0)},rotate=180]
        \draw[ball color=cyan,shading=ball] (-0.5,0) .. controls (-0.1,-2) and (0.1,-2) .. (0.5,0) .. controls (0.9,2) and (-0.9,2) .. (-0.5,0) ;
        \draw[ultra thick, ->] (0,1.5) -- (0,-1.5);
        \draw[very thick, dotted] (0,1.5) -- (0,4.1);
        \draw[very thick] ([shift=(270:2)]0,4.1) arc (270:250:2);
    \end{scope}

    \node at (-2.7,-2.1) [anchor=west] {$\alpha$}; 

    \begin{scope}[ shift={(-1.6,-0.3)},rotate=160]
        \draw[ball color=cyan,shading=ball] (-0.5,0) .. controls (-0.1,-2) and (0.1,-2) .. (0.5,0) .. controls (0.9,2) and (-0.9,2) .. (-0.5,0) ;
        \draw[ultra thick, ->] (0,1.5) -- (0,-1.5);
        \draw[very thick, dotted] (0,1.5) -- (0,4);
    \end{scope}
\end{scope}

\begin{scope}[shift={(0.1674\textwidth,0.3292\textwidth)},scale=0.5,rotate=90]
    \begin{scope}[shift={(-3,0)},rotate=180]
        \draw[ball color=cyan,shading=ball] (-0.5,0) .. controls (-0.1,-2) and (0.1,-2) .. (0.5,0) .. controls (0.9,2) and (-0.9,2) .. (-0.5,0) ;
        \draw[ultra thick, ->] (0,1.5) -- (0,-1.5);
        \draw[very thick, dotted] (0,-1.5) -- (0,-4.1);
        \draw[very thick] ([shift=(-270:2)]0,-4.1) arc (-270:-250:2);
    \end{scope}

    \node at (-2.7,3.2) [anchor=west] {$\alpha$}; 

    \begin{scope}[ shift={(-1.6,0.3)},rotate=-160]
        \draw[ball color=cyan,shading=ball] (-0.5,0) .. controls (-0.1,-2) and (0.1,-2) .. (0.5,0) .. controls (0.9,2) and (-0.9,2) .. (-0.5,0) ;
        \draw[ultra thick, ->] (0,1.5) -- (0,-1.5);
        \draw[very thick, dotted] (0,-1.5) -- (0,-4);
    \end{scope}
\end{scope}

    \begin{axis}[
        yshift =-0.28\textwidth,
        ylabel=ocurrence $\lbrack a.u. \rbrack$,
        xlabel=lateral distance $z$ $\lbrack \sigma_w \rbrack$,
        xtick pos=left,
        ytick pos=left,
        xmin = -3,
        xmax = 3,
        ymin = 0.0,
        ymax = 1,
        ybar interval,
        width=0.6\textwidth,
        height=0.3\textwidth,
        xtick={-3,-2,-1,0,1,2,3},
        xticklabels={,,,,,,},
        y tick label style={
        /pgf/number format/.cd,
            fixed,
            fixed zerofill,
            precision=1,
        /tikz/.cd
        },
        ylabel absolute,
            ]
   \addplot table[x expr= \thisrow{r}*sqrt(2), y expr= \thisrow{N}/160, col sep=comma] {images/Data/distNc2Ns1498Vr0_08rho0_45NonAdd.csv};
    \end{axis}
    \node at (0,-0.3023\textwidth) {-3}; 
    \node at (0.0853\textwidth,-0.3023\textwidth) {-2}; 
    \node at (0.1705\textwidth,-0.3023\textwidth) {-1}; 
    \node at (0.2558\textwidth,-0.3023\textwidth) {$0$}; 
    \node at (0.3411\textwidth,-0.3023\textwidth) {1}; 
    \node at (0.4263\textwidth,-0.3023\textwidth) {2}; 
    \node at (0.5116\textwidth,-0.3023\textwidth) {3}; 

    \begin{scope}[shift={(0.2558\textwidth,-0.28\textwidth)},scale=1.53]
    \clip (-1.56,0) rectangle (1.56,0.6);
    \draw[fill=orange,thick,rotate=-90,opacity=0.2] (-0.5,0) .. controls (-0.236842105,2) and (0.236842105,2) .. (0.5,0) .. controls (0.763157894,-2) and (-0.763157894,-2) .. (-0.5,0);
    \end{scope}

    \node at (0.023\textwidth,0.1952\textwidth) {\textbf{(a)}};
    \node at (0.023\textwidth,-0.0848\textwidth) {\textbf{(b)}};

    \node at (0.5452\textwidth,0.1952\textwidth) {\textbf{(c)}};
    \node at (0.5452\textwidth,-0.0193\textwidth) {\textbf{(d)}};

\end{tikzpicture}

%% file: images/ExclVolAlg.tex
\definecolor{exclVolColor}{rgb}{0,0.9,0.9}
\definecolor{exclVolColor1}{rgb}{0.9,0.45,0}
\definecolor{exclVolColor2}{rgb}{0.75,0.75,0.75}
\begin{tikzpicture}

\node at (-5.122,2.9) {\textbf{Step 1+2:}};
\node at (-5.122,2.4) {Surface meshes};

\node[] at (-5.122,0){\includegraphics[angle=-90,scale=0.2]{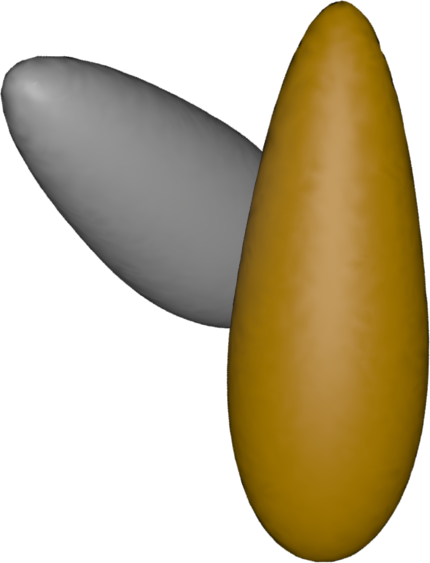}};
\node at (-5,1.2) {\textcolor{exclVolColor2}{$B_2$}};
\node at (-6.3,0.1) {\textcolor{exclVolColor1}{$B_1$}};

\node at (0,2.9) {\textbf{Step 3:}};
\node at (0,2.4) {Parallel surface construction};

\node[] at (0,0){\includegraphics[angle=-90,scale=0.2]{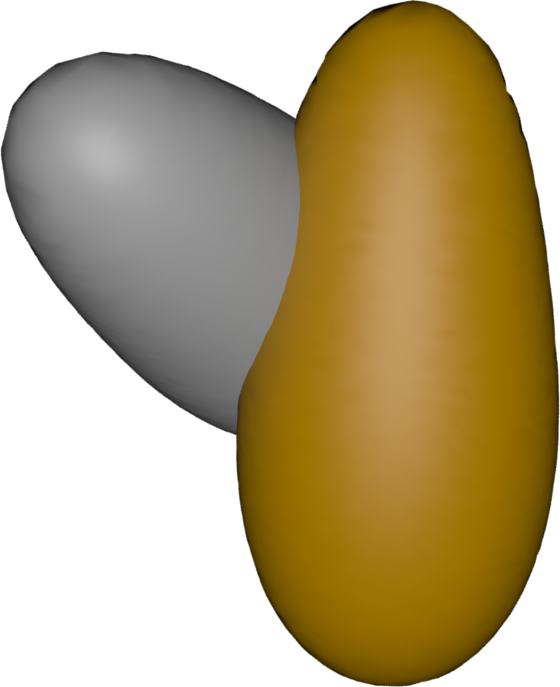}};
\node at (-0.6,1.2) {\textcolor{exclVolColor2}{$B'_2$}};
\node at (-1.5,0.5) {\textcolor{exclVolColor1}{$B'_1$}};

\node at (5.122,2.9) {\textbf{Step 4:}};
\node at (5.122,2.4) {Excluded volume merging};

\node[] at (5.122,0){\includegraphics[angle=-90,scale=0.2]{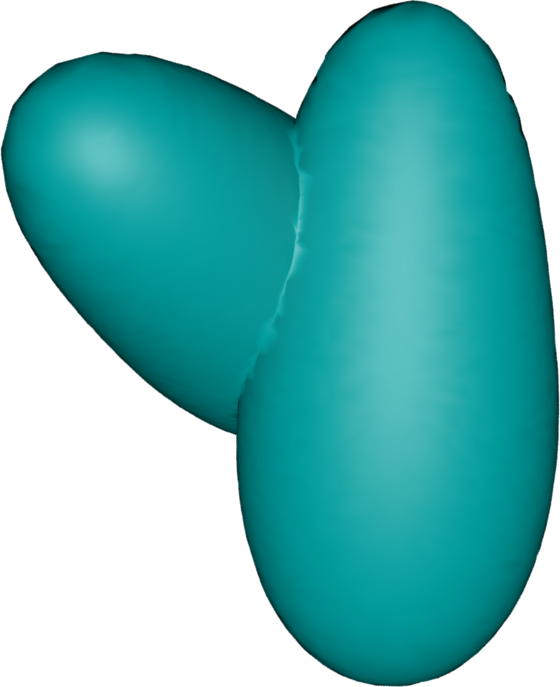}};

\node at (4.1,0.8) {\textcolor{exclVolColor}{$V_{\text{excl}}$}};
\end{tikzpicture}